\pdfoutput=1
\documentclass[12pt,letterpaper]{imamat}
\usepackage{latexsym,amssymb,lastpage,subcaption}
\usepackage{graphicx,amsfonts}
\usepackage{epstopdf}
\usepackage{pgfplots}
\usepackage{times,mathptmx,bm,amsmath,amsthm}
\usepackage{dcolumn}

\usepackage{marginnote}
\newcommand{\lbl}[1]{\label{#1}}
\def\Hmi{H_{\max,i}}
\def\Hma{H_{\max,1}}
\def\Hmb{H_{\max,2}}
\def\hci{H_{\mathrm{c},i}}
\def\hsi{H_{\mathrm{s},i}}
\def\hsa{H_{\mathrm{s},1}}
\def\hsb{H_{\mathrm{s},2}}

\def\pmax{p_{\max}}
\def\hmax{h_{\max}}
\def\hmin{h_{\min}}
\def\pbar{\bar{p}}
\def\hb{\bar{h}}
\def\hpeak{h_{\mathrm{peak}}}
\def\eps{\epsilon}
\def\hstab{h_*}
\def\Lh{\mathcal{L}_{\hstab}}

\begin{document}

\title{Steady-states of thin film droplets on chemically heterogeneous substrates}

\shorttitle{Droplets on heterogeneous substrates} 
\shortauthorlist{Liu and Witelski} 

\author{
\name{Weifan Liu}
\address{Department of Mathematics, Syracuse University}
\name{Thomas P. Witelski}
\address{Department of Mathematics, Duke University}
}
\maketitle

\begin{abstract}
{We study steady-state thin films on a chemically heterogeneous substrates of finite size, subject to no-flux boundary conditions. Based on the structure of the bifurcation diagram, we classify the one-dimensional steady-state solutions that exist on such substrates into six different branches and develop asymptotic estimates for the steady-states on each branch. We show using perturbation expansions, that leading order solutions provide good predictions of the steady-state thin films on  stepwise-patterned substrates. The analysis in one dimension can be extended to axisymmetric solutions. We also examine the influence of the wettability contrast on linear stability and dynamics. Results are also applied to describe two-dimensional droplets on hydrophilic square patches and striped regions used in microfluidic applications.}
{thin films, lubrication theory, heterogeneous substrates, disjoining pressure, pinned droplets}
\\
\end{abstract}

\section{Introduction}
\par
Thin liquid films on solid substrates are often seen in nature and engineering applications, for example, as tear films on the eye, lubricating coatings, and functional layers in microfluidic devices (see for example \cite{thiele2003modelling}). Microfluidic systems  manipulate small amounts of fluids, using channels with dimensions at scale of micrometers (\cite{whitesides2006origins}). Microfluidics has found many applications in cell biology and chemical synthesis (\cite{lo2013application,whitesides2006origins}). The effect of substrate wetting properties on the equilibrium liquid droplet formed on a solid, especially features like contact angle, pressure, and mass, has attracted extensive research attention due to applications in liquid coating and inkjet printing (\cite{bhushan2009lotus,dong2006visualization,sakai2008effect,son2008spreading,yuan2013contact}). Specifically, the steady-state thin films have been previously studied through the approach of numerical methods, asymptotic approximations and ellipsoidal droplet approximation (\cite{glasner03coarsening,gomba2009analytical,lubarda2011analysis,mac2016new}). 
\par 
Much theoretical understanding of thin films has been limited to films on homogeneous substrates. Profiles of steady-state solutions under the action of different forms of intermolecular potentials of homogeneous substrates have been previously investigated and described (\cite{bertozzi2001dewetting,glasner03coarsening,gomba2009analytical}). In one study, \cite{glasner03coarsening} considered isolated steady-state droplet parameterized by uniform pressure on an infinite domain, given by the homoclinic solution of the system. Through asymptotic matching, they showed that at leading order, large homoclinic droplets could be well approximated by parabolic profiles. In another study, \cite{bertozzi2001dewetting} performed similar analysis and computations for steady-state thin films on finite domains. Asymptotic analysis for both the bifurcation structure and solution profile of such films were presented. 
\par
However, many naturally occurring surfaces are chemically heterogeneous due to contamination or differentiated structures in biological or other contexts. Designed chemically heterogeneous substrates have been increasingly used for the engineering of micropatterns of thin films and applications that require accurate dispensing and distribution of liquids on solid surfaces (\cite{zope2001self}). One example of such applications is in the design of the chemical patterns of the nozzle plate in inkjet print heads (\cite{Blizynyuk_thesis,kooij2012directional}). Quantifying the characteristics of wetting layer on the nozzle plate and designing suitable chemical patterns to control the motion of the ink are critical to improving the printing quality  (\cite{Blizynyuk_thesis,kooij2012directional}). Another application is microcontact printing where a stamp is used to transfer the material onto a substrate to create a desired pattern. Understanding equilibrium droplet shape on chemically patterned substrate is essential to optimizing the printing process (see \cite{darhuber2000morphology}). Chemically patterned substrates have also been used in the fabrication of polymer field effect transistors where a substrate with a hydrophobic stripe is employed to split  a deposited liquid droplet (\cite{wang2004polymer}).
\par
Previously, \cite{lenz1998morphological} investigated the morphologies of different equilibrium states of liquids on a surface that consists of hydrophilic domains in a hydrophobic matrix. By minimizing the interfacial free energy subject to constant liquid volume, they found that the different morphologies are determined by the liquid volume and the area fraction of the hydrophilic domains. \cite{kavspar2016confinement} explored the effect of alternating hydrophobic and hydrophilic area of a rectangular micro-arrayed surface on the overall confinement and spillover of water droplets. They gave an estimate for the contact angle of the droplet in terms of the height of the spherical cap $h$ and a coefficient $a$ that accounts for the properties of the confining surface.
\par
In the framework of lubrication theory, the evolution of thin liquid films on a homogeneous solid substrate is governed by an equation of the form (\cite{myers,oron1997long,crastermatar})
\begin{equation}\label{eq:het_PDE}
\frac{\partial h}{\partial t}=\frac{\partial}{\partial x}\left(h^3\frac{\partial}{\partial x}\left[\tilde{\Pi}(h)-\frac{\partial^2h}{\partial x^2}\right]\right)
\end{equation}
with $\tilde{\Pi}=A\Pi(h)$ where $A$ is the Hamaker constant and $\Pi(h)$ is the homogeneous disjoining pressure function. The Hamaker constant $A$ determines the equilibrium contact angle formed by the liquid droplet on a substrate. Numerical simulations of lubrication approximation have been presented in \cite{kargupta2000instability, kargupta2001templating, kargupta2002morphological} to inspire experimental studies and illustrate the instability and pattern formation of thin film on chemically heterogeneous substrates with a stepwise pattern. More systematic analytical studies using lubrication approximation were presented in \cite{brusch2002dewetting, kao2006rupture, thiele2003modelling} where a spatially dependent Hamaker coefficient $A(x)$ was introduced in the long-wave equation. A disjoining pressure of the form 
\begin{equation}
\tilde \Pi(h,x)=A(x)\Pi(h)
\end{equation}
was used to model thin films on a domain with periodic boundary conditions. Specifically, \cite{brusch2002dewetting, thiele2003modelling} studied the effect of a smoothly patterned substrate on stationary droplet profiles using wettability as a control parameter. The heterogeneous substrate considered was a small-amplitude sinusoidal modulation of the form $\displaystyle A(x)=1+\delta\cos\left(k_px\right)$ where $k_p$ determines the imposed heterogeneity period and $\delta\ll 1$ describes the amplitude of heterogeneity. The smooth spatial variation and the assumption that $\delta\ll 1$ allow for the analysis of the solutions on heterogeneous substrates through an asymptotic expansion in terms of $\delta$. By varying the amplitude and periodicity of the chemical pattern, they identified the parameter range where the pinning mechanism emerges from coarsening.
\par
However, for an engineered patterned substrate, a piecewise-constant $A(x)$ would be a better description than a sinusoidal. For example, micro-patterned surfaces with alternate hydrophilic and hydrophobic rectangular areas are extensively used in digital microfluidics and high-throughput screening nanoarrays (\cite{kavspar2016confinement}). In such applications, a stepwise Hamaker coefficient is needed to model the chemical properties of the surfaces. \cite{kao2006rupture} studied the stationary states of thin films on substrates with square-wave patterning in both one and two dimensions in addition to those with small-amplitude sinusoidal patterning. Specifically, they considered a piecewise constant $A(x)$ with periodic boundary conditions, given by 
\begin{equation}\label{eq:A_Kao}
A(x)=
\begin{cases}
1+\delta^3 & \frac{3\pi}{2}n\leq k_px \leq \frac{\pi}{2}+\frac{3\pi}{2}n,\\
1-\delta^3&\frac{\pi}{2}<k_px \leq \frac{3\pi}{2},
\end{cases}
\end{equation}
for patterning wavenumber $k_p$ and $n=0,1$ on $x\in[0,2\pi]$. To study the bifurcation of stationary states on substrates with such patterning, they wrote $A(x)$ as a Fourier series. In particular, they performed asymptotic analysis for solutions near the bifurcation point. Imperfect bifurcations were observed for patterning of the form \eqref{eq:A_Kao}. They found that the bifurcations and steady-states resemble those for sinusoidally patterned substrates.
\par
In this paper, we study the steady-state solutions of thin films on a stepwise-patterned substrate over a range of wettability contrast. We classify the steady-state solutions that exist on such substrates into branches. We found new branches of solutions characterizing pinned droplets that arise as a consequence of the heterogeneity of the substrates. 
For each branch of solutions, we present systematic asymptotic analysis of the steady-state profile and the structure of the bifurcation diagram. Through asymptotic analysis and numerical simulations, we determine the dependence of steady-state thin films on parameters such as mass, pressure, and heterogeneity strength. We employ a phase-plane approach, which allows us to perform asymptotic analysis in the limit of large heterogeneity contrast. Increasing heterogeneity contrast has an increasing confining and pinning effect on the film droplet, which prevents fluid film from leaking into the more hydrophobic surrounding region. To quantify this phenomenon, we present an effective measure of the fluid leakage and show that the leakage is inversely proportional to the heterogeneity contrast. In addition, we investigate the stability of the steady-state solutions on heterogeneous substrates and show that the analysis derived for one-dimensional solutions can be extended to axisymmetric solutions and more general two-dimensional solutions. Finally, we illustrate the influence of chemical heterogeneity on the dynamics of thin film evolution.

\section{Problem formulation}
\par
\begin{figure}
\begin{center}
\includegraphics[width=3in]{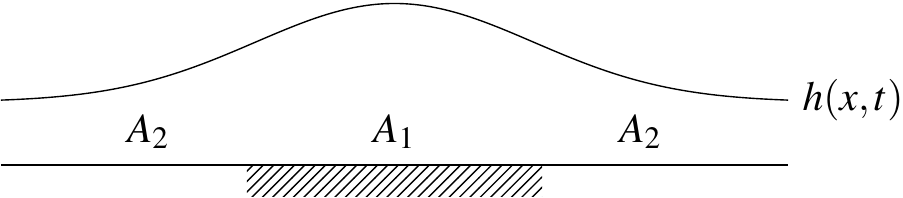}
\end{center}
\caption{Schematic diagram of a thin film  on a heterogeneous substrates with stepwise chemical patterning yielding relatively hydrophilic ($A_1$) and hydrophobic ($A_2> A_1$) regions.} 
\label{fig:schematic}
\end{figure}

 We study thin films on a heterogeneous substrate prescribed with a piecewise chemical patterning;  Figure \ref{fig:schematic} shows a schematic diagram of a thin film on heterogeneous substrate in one dimension with a hydrophilic region on $-s<x<s$ surrounded by hydrophobic regions on the overall domain $-L<x<L$. For convenience we will make use of symmetry to reduce the problem to be on the half-domain $0<x<L$ subject to Neumann boundary conditions (see \cite{LP2000}) and focus primarily on the lowest-order solutions.
\par
 We consider a heterogeneous substrate with a stepwise patterning modeled by a piecewise constant function $A(x)$ where the jumps of $A(x)$ need not be small. In particular, we address analysis of steady-state solutions in the limit of large $A_2$ relative to $A_1$ in
\begin{equation}\lbl{eq:myA}
A(x) =
  \begin{cases}
A_1 & 0\leq x\leq s,\\
A_2 & s<x\leq L.
  \end{cases}
\end{equation}
Here, $L$ is the size of the domain, $s$ is the interface of segmentation and $A_i$ are positive constants. For concreteness, we will normalize relative to the hydrophilic region, generally taking $A_1=1$ and $A_2\ge A_1$. Figure \ref{fig:schematic} shows the schematic diagram of thin film on heterogeneous substrates with such stepwise patterning. Specifically, we consider a disjoining pressure given by a 3-4 inverse power law function which has been used in  \cite{glasner03coarsening,oron1999dewetting,oron2001dynamics,schwartz1998} and others,
\begin{equation}\lbl{eq:dis34}
\Pi(h)=\frac{\epsilon^2}{h^3}-\frac{\epsilon^3}{h^4},
\end{equation}
and the overall representation of wetting effects is given by $\tilde \Pi(h,x)=A(x)\Pi(h)$. The scaling in \eqref{eq:dis34} yields a finite limit for the effective contact angle of droplet solutions as $\epsilon\to 0$, see \cite{glasner2003coarsening}.

We study the thin film on a finite domain subject to no-flux boundary conditions so that the total fluid mass is conserved. The evolution of thin films on chemically heterogeneous substrates of finite-length with chemical patterning $A(x)$ is governed by the partial differential equation for the film height $h(x,t)$ on $0\le x\le L$ (see \cite{o2002theory,oron1997long}),
\begin{subequations}
\lbl{het_prob}
\begin{equation}\lbl{eq:het_PDE}
\frac{\partial h}{\partial t}=\frac{\partial}{\partial x}\left(h^3\frac{\partial}{\partial x}\left[\tilde \Pi(h,x)-\frac{\partial^2h}{\partial x^2}\right]\right),
\end{equation}
subject to no-flux and zero-meniscus boundary conditions,
\begin{equation}\lbl{eq:bc_het}
J(0,t)=0,\qquad J(L,t)=0,\qquad \frac{\partial h}{\partial x}\left(0,t\right)=0,\qquad
\frac{\partial h}{\partial x}\left(L,t\right)=0
\end{equation}
\end{subequations}
where the form of $A(x)$ is given by \eqref{eq:myA}, the dynamic pressure of the thin film is defined by 
\begin{equation}
    p(x,t)\equiv \tilde\Pi(h,x)-{\partial^2 h\over \partial x^2}\qquad \mbox{with}
    \qquad
    \tilde\Pi(h,x)=A(x)\Pi(h),\lbl{defpressure}
\end{equation} 
and $J(x)\equiv h^3\partial_x p$ is the flux. The no-flux boundary conditions \eqref{eq:bc_het} are equivalent to homogeneous Neumann conditions
\begin{equation}\lbl{eq:nbc}
\frac{\partial h}{\partial x}\left(0,t\right)=0,\qquad
\frac{\partial h}{\partial x}\left(L,t\right)=0,\qquad \frac{\partial^3 h}{\partial x^3}\left(0,t\right)=0,\qquad
\frac{\partial^3 h}{\partial x^3}\left(L,t\right)=0,
\end{equation}
yielding reflection symmetry and even extension of solutions with respect to  the boundary, see \cite{LP2000}. These boundary conditions yield solutions that conserve mass, allow us to construct periodic solutions and are consistent with uniform films for the spatially homogeneous case.
\par
This problem has a monotone decreasing energy functional,
\begin{equation}\lbl{eq:E}
E=\int_0^L A(x)U(h)+{\textstyle \frac{1}{2}}(\partial_xh)^2\,dx
\end{equation}
where $U(h)$ is the potential such that $\frac{dU}{dh}=\Pi(h)$. For $\Pi(h)$ of the form \eqref{eq:dis34}, $U(h)$ is given by 
\begin{equation}
U(h)=-\frac{\epsilon^2}{2h^2}+\frac{\epsilon^3}{3h^3}\, .
\lbl{Ueqn}
\end{equation}
This energy was used in \cite{bertozzi2001dewetting,glasner2003} and other papers for the homogeneous case, $A(x)\equiv 1$,
and in \cite{brusch2002dewetting} for the heterogeneous case.
In both cases, $E$ is monotonically decreasing with the same form for the rate of dissipation,
\begin{equation} 
{dE\over dt}= -\int_0^L h^3 (\partial_x p)^2\,dx\le 0,
\lbl{eq:dEdt}
\end{equation}
showing that the dynamics of \eqref{het_prob} follow a gradient flow.
\par
We seek solutions $h(x,t)$ that are continuous and whose first derivative is continuous at $x=s$, i.e.
\begin{subequations}
\lbl{eq:sBCs}
\begin{alignat}{2}
\displaystyle\lim_{x\to s^-}h(x,t)&=\lim_{x\to s^+}h(x,t)\lbl{eq:cont}\\
\displaystyle\lim_{x\to s^-}\frac{\partial h}{\partial x}(x,t)&=\lim_{x\to s^+}\frac{\partial h}{\partial x}(x,t)\lbl{eq:diff}
\end{alignat}
\end{subequations}
and locally conserve mass across the wettability jump at $x=s$.
These conditions yield that solutions will have a continuous pressure \eqref{defpressure} but must have a jump in the curvature at $x=s$, i.e.
\begin{equation}\lbl{eq:jump}
\frac{\partial^2 h}{\partial x^2}(s^+,t)-\frac{\partial^2 h}{\partial x^2}(s^-,t)=(A_2-A_1)\Pi(h(s,t)).
\end{equation}

Steady-state solutions on homogeneous substrates have been previously analyzed in \cite{LP2000,LP2000a,bertozzi2001dewetting,glasner03coarsening,hutchinson2013numerical,pahlavan2018thin} and many other papers. From \eqref{het_prob} and \eqref{eq:dEdt}, it can be seen that all positive steady-state solutions subject to no-flux boundary conditions have uniform constant pressure, i.e. $p\equiv \bar p$. 
This is still true for heterogeneous substrates where $A(x)$ makes $\Pi$ spatially dependent. It follows that the steady-state solutions of \eqref{eq:het_PDE} subject to \eqref{eq:bc_het} satisfy
\begin{subequations}
\lbl{eq:ss_het}
\begin{equation}\lbl{eq:ss_het_intro}
{d^2 h\over dx^2}=A(x)\Pi(h)-\bar p,
\end{equation}
\begin{equation}\lbl{eq:ss_het_intro_bc}
h_x(0)=0,\qquad h_x(L)=0.
\end{equation}
\end{subequations}
For the homogeneous case, all steady states can be described with respect to the range of the function $\Pi(h)$; spatially uniform solutions (``flat films'', $h(x)\equiv\bar{h}$) exist for any positive thickness and correspond to $-\infty <\bar{p}\le \pmax$, where $p_{\max}=27/(256\epsilon)$ is the maximum of $\Pi(\bar{h})$, attained at $\bar{h}=\hpeak=4\epsilon/3$. Nontrivial steady solutions exist for $0< \bar{p}<\pmax$. We will see that the situation with heterogeneous substrates is more complicated.
\par 
For the heterogeneous case where $A(x)$ is a step function with $A_1\neq A_2$, for a steady-state solution to be a flat film and satisfy equation \eqref{eq:jump}, the only option is to have $\Pi(\bar h)=0$, yielding $\bar h=\epsilon$. Hence, $h(x)\equiv\epsilon$ is the only possible flat film solution on a heterogeneous substrate, with corresponding pressure $\bar{p}= 0$.
\par
For the stepwise $A(x)$, analysis of \eqref{eq:ss_het} follows from piecewise-defined autonomous phase plane analysis on $0\le x\le s$ and $s\le x\le L$ with constant $A=A_i$ for $i=1,2$ respectively. From the analysis in \cite{bertozzi2001dewetting} for the phase plane for \eqref{eq:ss_het_intro} with $A\equiv 1$, for $0<\bar{p}<p_{\max}$ then the problem
has two fixed points, a hyperbolic saddle $h=\hsi$ (with $\hsi<\hpeak$) and an elliptic center point $h=\hci$ (with $\hci>\hpeak$), each satisfying 
\begin{equation}
\Pi(H_i)={\bar{p}\over A_i}\, .
\lbl{HsHcEqn}
\end{equation}
There is a homoclinic orbit that passes through the saddle point, defining a single maximal-amplitude droplet on $-\infty<x<\infty$. This solution has $\hsi$ as its global minimum and its corresponding maximum $\Hmi$ is obtained from a first integral, as in \cite{bertozzi2001dewetting}. In the phase plane, the
homoclinic orbit encloses a continuous family of periodic solutions, each having its minimum in the
range $\hsi< \hmin\le \hci$ and corresponding maximum in $\hci\le \hmax< \Hmi$.
\par
\begin{figure}
\centering
 \begin{subfigure}[b]{0.45\linewidth}
 \centering
 \includegraphics[width=2.3in,height=1.7in]{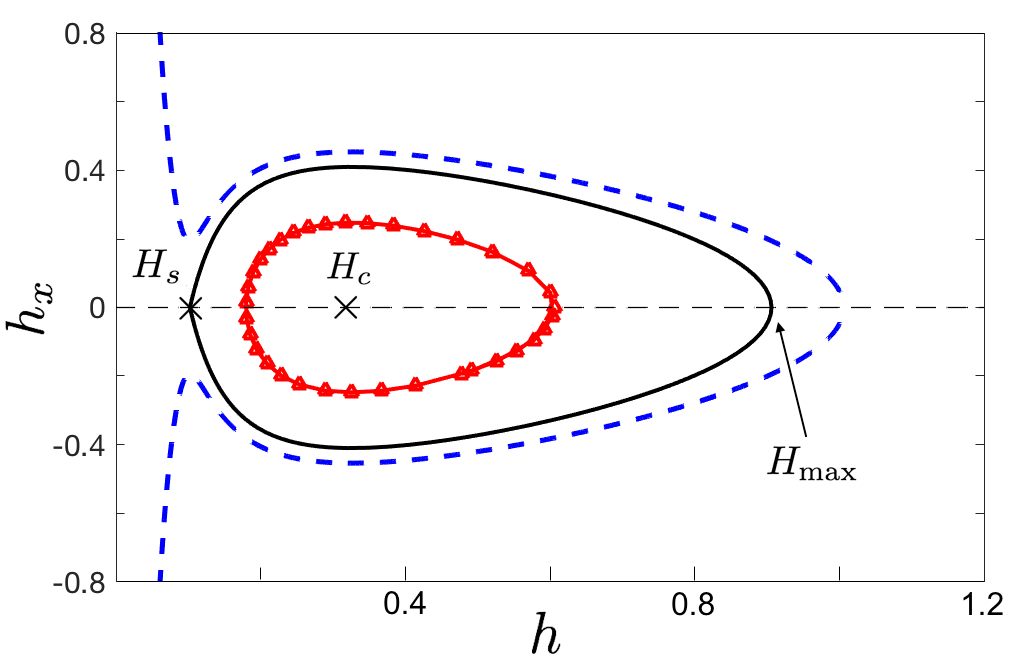}
 \caption{}
\end{subfigure}
\qquad
 \begin{subfigure}[b]{0.45\linewidth}
 \centering
\includegraphics[width=2.4in,height=1.7in]{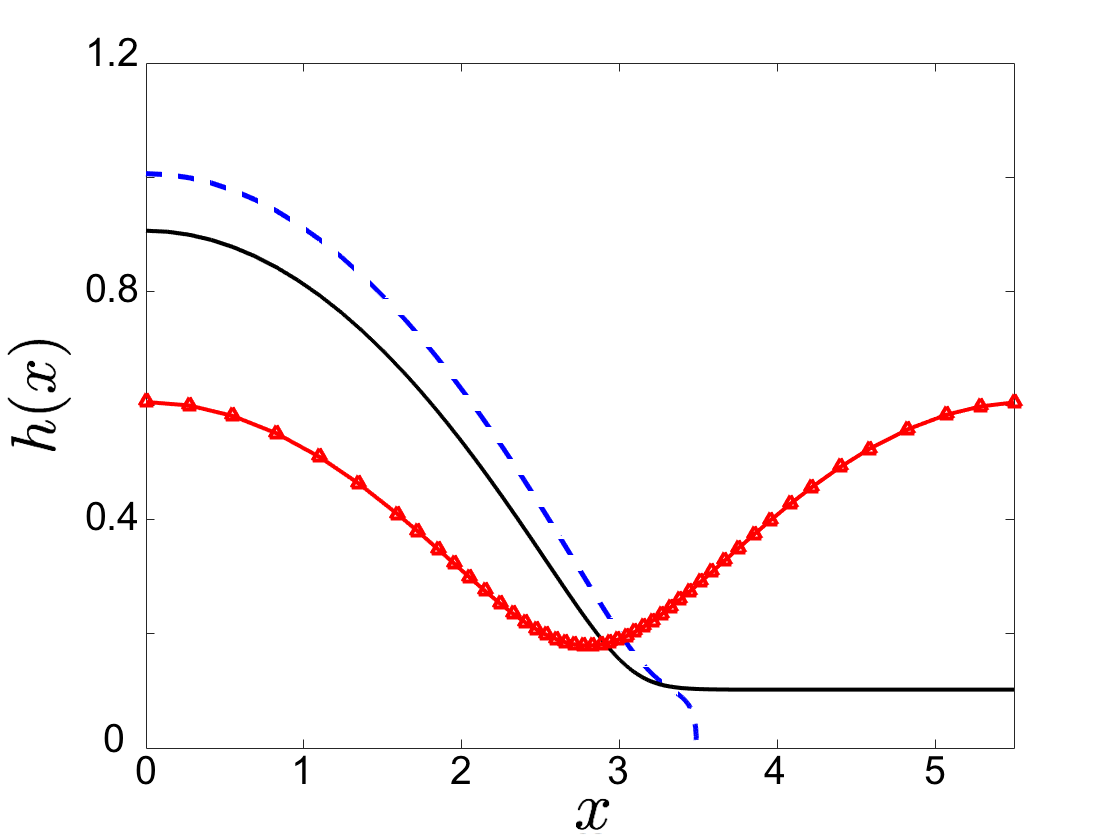}
 \caption{}
\end{subfigure}
\caption{(a) Phase plane for the homogeneous substrate case, with parameters $\bar{p}=0.2, \epsilon=0.1$, showing trajectories for the homoclinic orbit (solid black curve), a periodic solution (red dotted curve) and a solution that lies outside of the homoclinic orbit (dashed blue curve). (b) Profiles of the three solutions corresponding to the trajectories shown in (a).}
\label{fig:pp_whole}
\end{figure}

 
Figure~\ref{fig:pp_whole}(a) illustrates the trajectories in the phase plane: the homoclinic orbit with $h_x\to0$ as $h\to \hsi$ (solid black curve), a periodic solution bounded inside the homoclinic orbit (red dotted curve), and a typical solution lying entirely outside of the homoclinic orbit with $|h_{x}|\to\infty$ and $h\to 0$ at finite $x$ (dashed blue curve) (also see \cite{gomba2017}). Figure~\ref{fig:pp_whole}(b) shows the profiles corresponding to the three trajectories.
On homogeneous substrates, only trajectories that lie inside of the homoclinic orbit yield acceptable steady solutions of \eqref{het_prob}. In this paper, we will show that trajectories that lie outside of the homoclinic orbit will be used to construct steady-states of thin films on heterogeneous substrates. 
\begin{figure}
 \begin{subfigure}{0.5\linewidth}
\centering
  \includegraphics[width=2.5in]{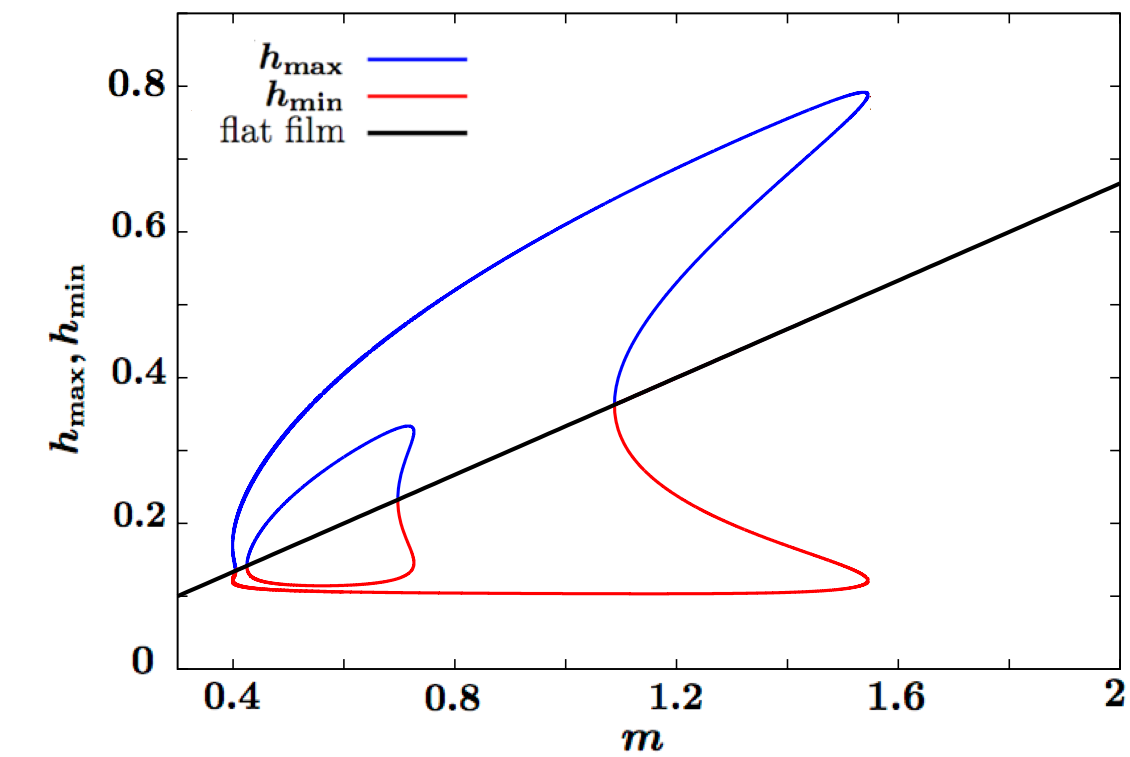}
 \caption{}
\label{fig:mha}
\end{subfigure}
 \begin{subfigure}{0.5\linewidth}
\centering
\includegraphics[width=2.4in]{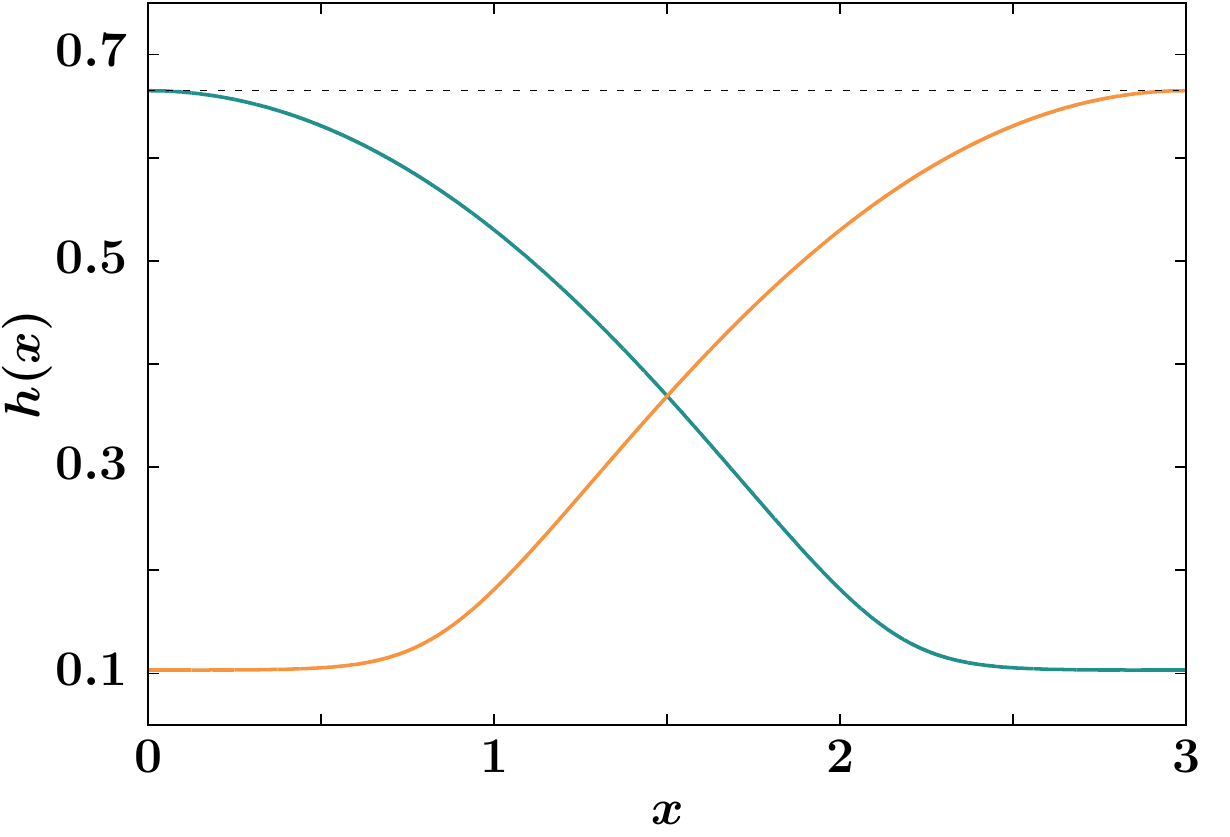}
 \caption{}
\label{fig:mhb}
\end{subfigure}\\\vspace{.1in}
\begin{subfigure}{.5\linewidth}
\centering
\includegraphics[width=2.5in]{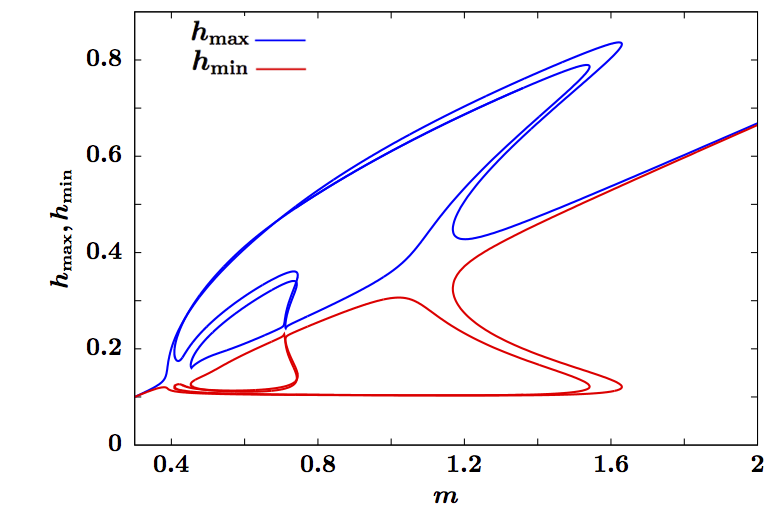}
\caption{}
\label{fig:drop_lra}
\end{subfigure}%
\begin{subfigure}{.5\linewidth}
\centering
\includegraphics[width=2.4in]{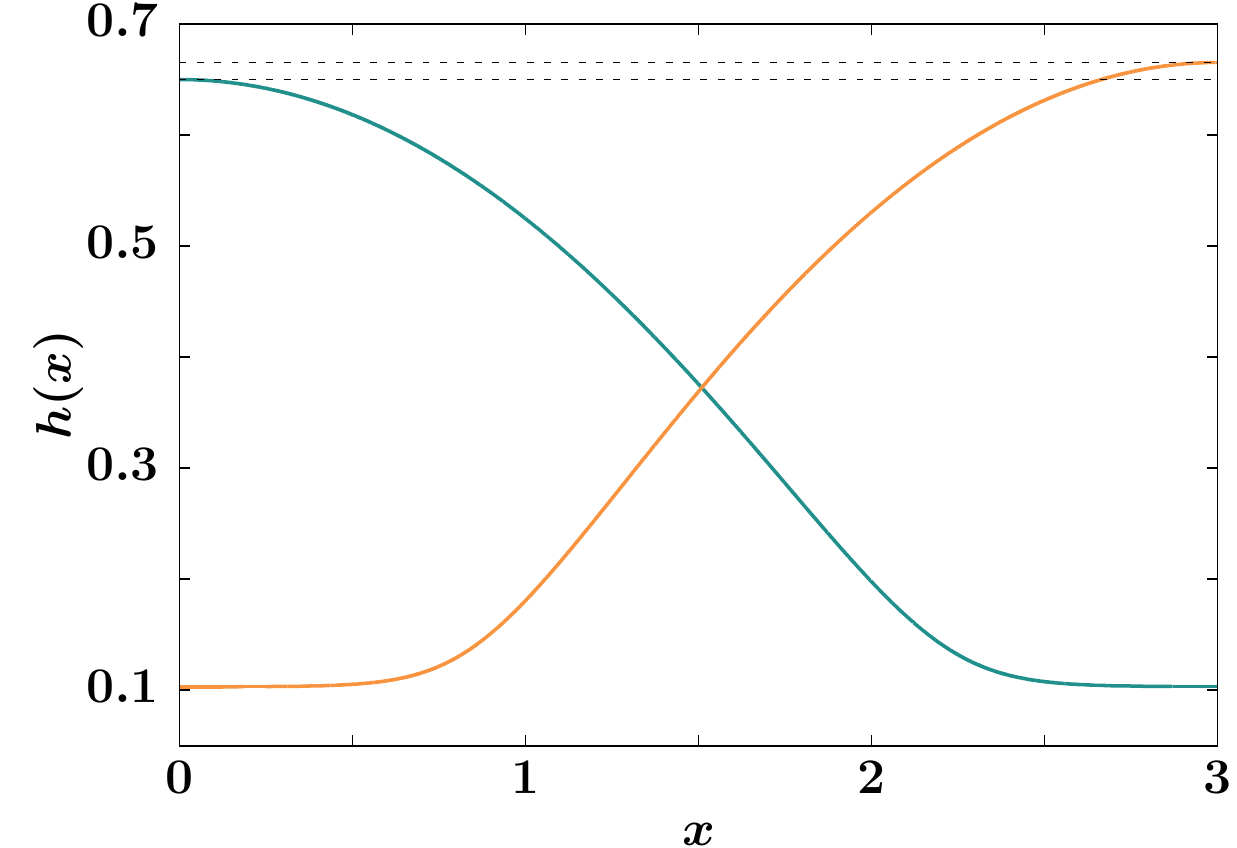}
\caption{}
\label{fig:drop_lrb}
\end{subfigure}\\[1ex]
\begin{subfigure}{\linewidth}
\centering
\includegraphics[width=2.5in]{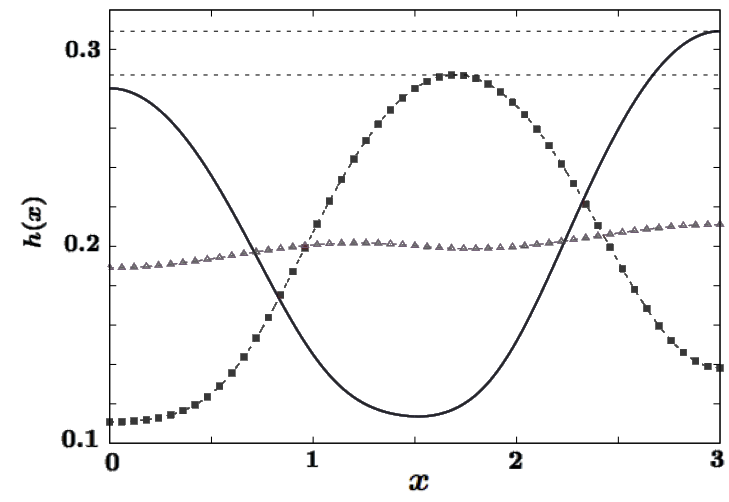}
\caption{}
\label{fig:drop_lrc}
\end{subfigure}
\caption{Steady solutions with $\epsilon=0.1$ on a domain of length $L=3$: (a) Bifurcation diagram of $m$ vs. $h_{\min},h_{\max}$ for solutions on homogeneous substrate, $A(x)\equiv 1$.
(b) Profile of two steady-state droplets from (a) with $m=1.1$ centered at $x=0$ and $x=L$. (c) Bifurcation diagram of $m$ vs. $h_{\min},h_{\max}$ for steady-state solutions on stepwise-patterned substrate with $A_1=1$, $A_2=1.1$ and $s=L/2$. (d) Asymmetric steady-state droplets centered at $x=0$ and $x=L$ both with mass $m=1.1$. The difference in the maximum film thicknesses is highlighted by the two dashed lines. (e) Profiles of three distinct inner loop steady-state solutions on stepwise-patterned substrate, all with mass $m=0.6$.}
\lbl{fig:mh}
\end{figure}
\par
Figure~\ref{fig:mh}(a) shows the numerically computed bifurcation diagram for film mass vs.\  the maximum and minimum film thickness, denoted by $h_{\max}$ and $h_{\min}$, for steady solutions on a homogeneous substrate with length $L=3$. This type of bifurcation diagram has been previously studied in detail by \cite{bertozzi2001dewetting}.
Continuous families of nontrivial (periodic) solutions branch-off from the set of flat films (represented by the diagonal line in Fig.~\ref{fig:mh}(a)) at pairs of pitchfork bifurcation points, $\bar{h}=\bar{h}_{k,\pm}$. The number of loops of solutions, $N$, depends on the domain size $L$ and the derivative of the disjoining pressure through
\begin{equation}
\Pi'(\bar h_k)=-k^2\pi^2/L^2, \qquad k=1,2,\cdots, N\,,
\end{equation} 
see \cite{bertozzi2001dewetting}. For the $(\epsilon, L)$ used here, $N=2$ yielding two loops corresponding to half- and whole-droplets on $[0,L]$.
Figure~\ref{fig:mh}(b) shows the profiles of two droplet solutions with mass $m=1.1$ centered at opposite ends of the domain. Because of the reflection symmetry under $x\to L-x$ for the homogeneous problem, both of these solutions are given by the same state from the bifurcation diagram. 
\par
Figure~\ref{fig:mh}(c) shows the same type of bifurcation diagram as Figure~\ref{fig:mh}(a) but for thin films on a stepwise-patterned substrate with $A_1=1,\ A_2=1.1$ and $s=L/2$.
The spatial dependence of this disjoining pressure breaks the reflection symmetry and  steady-state droplets centered at $x=0$ and $x=L$ with the same mass now differ in profiles, as illustrated in Figure \ref{fig:mh}(d). 
Compared to the homogeneous case, this symmetry-breaking for the heterogeneous case replaces the pitchfork points with imperfect bifurcations, as seen in \cite{kao2006rupture}, and yields more complicated loop structures.
The outer loops represent branches of solutions with maxima at either $x=0$ or $x=L$, while the inner loops give solutions with interior critical points. Figure~\ref{fig:mh}(e) shows the profile of three distinct steady-state solutions on the inner loop of Figure \ref{fig:mh}(c) with the same mass, $m=0.6$. 
\par
 For convenience, we focus on solutions that are monotone decreasing, describing a half-droplet profile on $x\in [0,L]$ (which can be symmetrically extended to give a single whole drop on $[-L,L]$, as in Fig.~\ref{fig:schematic}). We can write the first integral of \eqref{eq:ss_het_intro} on
$x\in[0,s]$ and $x\in (s,L]$ as
\begin{subequations}
\begin{equation}\lbl{eq:order1_het2}
\frac{dh}{dx}= -\begin{cases}
\sqrt{2R_1(h)} & 0\leq x\leq s,\\
\sqrt{2R_2(h)}& s<x\leq L,
\end{cases}
\end{equation}
where 
\begin{eqnarray}
R_1(h)& =& A_1(U(h)-U(h_{\max}))-\bar{p}(h-h_{\max}),\\
\nonumber
R_2(h)& =& A_2(U(h)-U(h_{\min}))-\bar{p}(h-h_{\min}).
\end{eqnarray}
\end{subequations}
 Equation \eqref{eq:order1_het2} along with the condition \eqref{eq:diff} yields a condition relating the film thickness at the heterogeneity interface, $x=s$, to the extrema for steady-states on stepwise-patterned substrates with $A(x)$ of the form \eqref{eq:myA}, 
\begin{equation}\lbl{eq:matchhx}
(A_1-A_2)U(h(s))+\bar p(h_{\max}-h_{\min})=A_1U(h_{\max})-A_2U(h_{\min}),
\end{equation}
which we will use  for later analysis. Setting both $A$'s to $A_i$ and using $\hmin=\hsi$ reduces \eqref{eq:matchhx} to the first integral equation for $\hmax=\Hmi$.

\section{Classification of branches in the bifurcation diagram in one dimension}\lbl{ssec:1D}
\par
For the remainder of the article, we use a different form of the bifurcation diagram that facilitates describing the effects due to heterogeneous wettability.
Figure~\ref{fig:mh2} shows numerically computed bifurcation diagrams of $\bar{p}$ vs.\ $h_{\max}$ for steady-states on homogeneous and heterogeneous substrates on a domain of length $L=3$ with $\epsilon=0.1$. Here, in the homogeneous case, all flat films are represented by the graph of the disjoining pressure, $\bar{p}=\Pi(\hmax)$ (dotted curve). The branch of nontrivial solutions bifurcating from the flat films in Fig.~\ref{fig:mh2}(a) corresponds to the outer loop from Fig.~\ref{fig:mh}(a). The inset plot shows that the branch bifurcates slightly below $\pmax$.
\par
For the heterogeneous case, we consider a typical problem with $A_2=5$, $s=L/2=1.5$ and similarly plot solutions corresponding to the analogous outer loop from Fig.~\ref{fig:mh}(c). We observe that the family of solutions is continuous and smooth and has fold points separating the curve into six segments, see Figure~\ref{fig:mh2}(b). We will analyze the dependence of solutions in each of these segments with respect to limits for $\epsilon$ and $A_2$.
\par 
Although there is no more flat film solution except for $h(x)\equiv \epsilon$ with $\pbar=0$ for $A_1\neq A_2$, as we will show later in Section~\ref{sssec:s1}, branch 1 and branch 6 yield nearly-flat films that are perturbations of the flat film solutions. Figure~\ref{fig:mh2}(b) includes the graph of $\Pi(\hmax)$ for reference, to show that the heterogeneous bifurcation diagram approaches that curve for the limits of large and small film thickness. From the inset, it is notable that the branch extends to a value of $\pbar$ slightly greater than $\pmax$.
\par
 In the following subsections, we will present our analysis and computation of these steady-state solutions by branch. For each branch, we develop an asymptotic prediction for the steady-state profile and show that the leading order solution for each branch depends on different parameters in $(L,s,A_1,A_2)$, which describe the chemical heterogeneity of the substrates.

Based on the structure of the diagram shown in 
Figure~\ref{fig:mh2}(b), we divide the steady-state solutions that could exist on a heterogeneous substrate with patterning $A(x)$ of the form \eqref{eq:myA} into six different connected branches, as follows:
\begin{itemize}
\itemsep=0in
\item Branch 1: small-thickness films
\item Branch 2: small-width droplets
\item Branch 3: pinned droplets
\item Branch 4: large-width droplets
\item Branch 5: confined droplets
\item Branch 6: large-thickness films
\end{itemize}
In particular, branch 3 is an entirely new branch of solutions characterizing a class of ``pinned'' drops that emerges due to the presence of chemical heterogeneity.

\begin{figure}
\centering
 \begin{subfigure}[b]{0.45\linewidth}
 \centering
  \includegraphics[width=2.5in,height=1.8in]{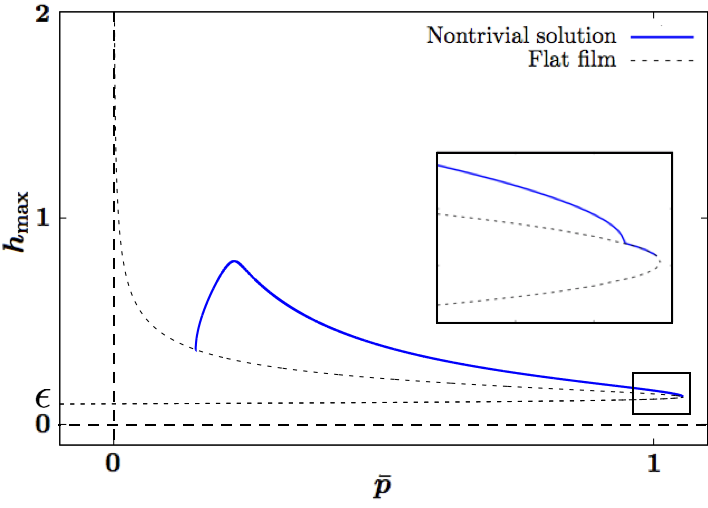}
 \caption{}
\end{subfigure}
 \begin{subfigure}[b]{0.45\linewidth}
 \centering
  \includegraphics[width=2.5in,height=1.8in]{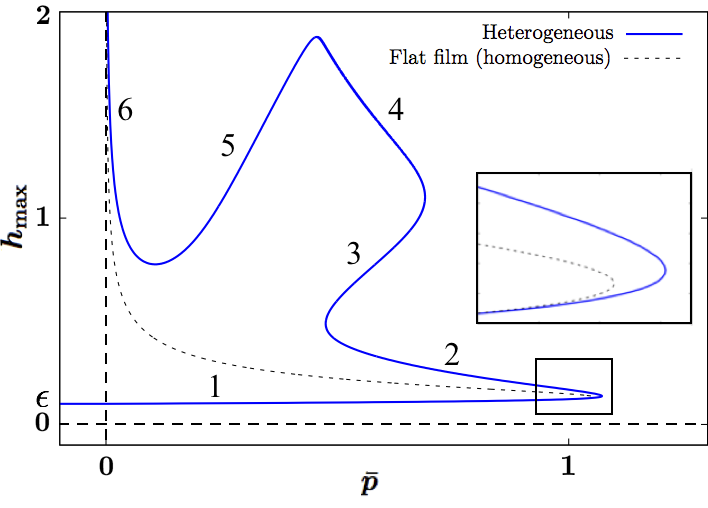}
 \caption{}
\end{subfigure}
\caption{Bifurcation diagram of $\bar{p}$ vs. $h_{\max}$ for steady-states with $\epsilon=0.1$ and $L=3$ on (a) a homogeneous substrate $A(x)\equiv 1$. (b) on a heterogeneous substrate with $A_1=1$, $A_2=5$, $s=1.5$. The solid blue curve represents nontrivial steady-state solutions. The dashed black curve in (a) represents two branches of flat film solutions that merge together with the nontrivial branch at $\bar p=p_{\max}$. In both (a) and (b), the inset plot shows the bifurcation curve zoomed into a small neighborhood near the maximum pressure. }
\lbl{fig:mh2}
\end{figure}


\subsection{Small-thickness and large-thickness nearly-flat films}\lbl{sssec:s1}
In this subsection, we study two types of solutions that are perturbations of flat films. First we study branch 1, which gives steady state solutions with mean thickness $h=O(\epsilon)$. Two examples of steady-state profiles of this type of small-thickness films are given in Figure~\ref{fig:asy1}. Both solutions are characterized by nearly flat films away from the patterning interface $x=s$ and a rapid change in the profile in a small neighborhood of the interface $x=s$.
The rapid change in $h(x)$ near the interface is due to the large change in disjoining pressure for films of thickness $h=O(\epsilon)$. The disjoining pressure $\Pi(h)$ increases rapidly for $h$ in the  range $0<h<4\epsilon/3$ in the limit $\epsilon\to0$. The mean film thickness of branch 1 solutions falls within this range. 
\par 
These solutions can be understood using matched asymptotics for $\epsilon\to 0$. Away from $x=s$, the second derivative in \eqref{eq:ss_het_intro} can be neglected and the outer solutions to all orders are given by the respective saddle points,
\begin{equation}\lbl{eq:b1_h0}
h_{\mathrm{out}}(x)=\begin{cases}
\hsa & 0\leq x\leq s,\\
\hsb & s<x\leq L,
\end{cases}
\end{equation}
where 
\begin{equation}
\hsi=  \epsilon+{\epsilon^2\pbar\over A_i}+O(\eps^3)
\lbl{eq:hsi}
\end{equation} 
with $\pbar=O(1)$. In an $O(\epsilon)$ neighborhood of $x=s^\pm$, the solution satisfies a nonlinear boundary layer equation (balancing the disjoining pressure and the second derivative). However, rather than pursuing this approach to the analysis, we can take advantage of the fact that the range of the solution is small, $\hmax-\hmin=O(\epsilon^2)$, to estimate the local behavior from a linearized analysis.
\begin{figure}
 \begin{subfigure}[b]{0.5\textwidth}
\includegraphics[width=2.5in]{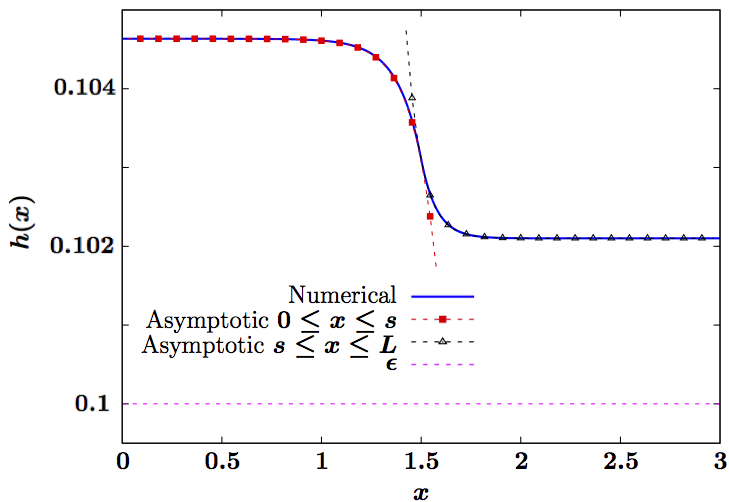}
\caption{}
\end{subfigure}
 \begin{subfigure}[b]{0.5\textwidth}
\includegraphics[width=2.5in]{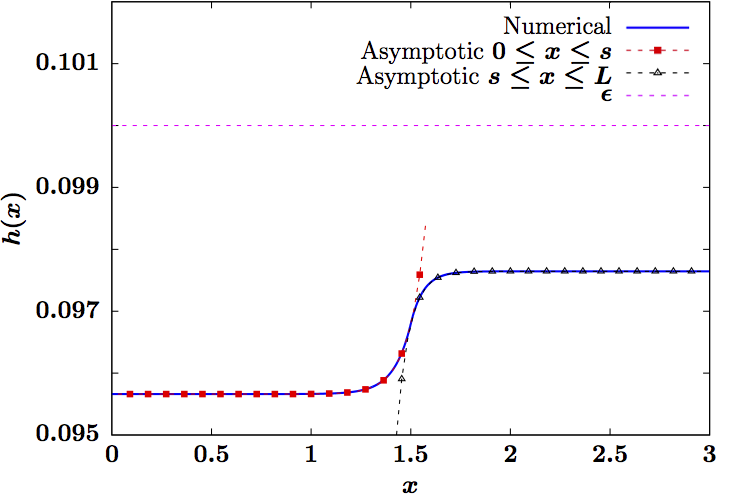}
\caption{}
\end{subfigure}
\caption{Properties of branch~1, small-thickness film solutions, computed with parameters $A_1=1$, $A_2=2$, $s=1.5$, $\eps=0.1$:  (a) A numerical solution (blue) with $\pbar=0.387$ having mean thickness $\hb>\eps$
compared with approximate form \eqref{eq:hprofile_b1} (dotted curves). (b) A computed solution (blue) with $\pbar= -0.518$ yielding mean thickness $\hb<\eps$ also compared with \eqref{eq:hprofile_b1}.}
\lbl{fig:asy1}
\end{figure}
\par 
Linearizing \eqref{eq:ss_het_intro} about each saddle yields the approximate form of the solution as 
\begin{equation}\lbl{eq:hprofile_b1}
h(x) \approx
  \begin{cases}
\hsa+C_1e^{\sqrt{A_1\Pi'(\hsa)}\,(x-s)} & 0\leq x\leq s,\\
\hsb+C_2e^{-\sqrt{A_2\Pi'(\hsb)}\,(x-s)} & s<x\leq L,
 \end{cases}
 \end{equation}
with constants $C_1$, $C_2$ to be determined from conditions \eqref{eq:cont} and \eqref{eq:diff}. 
Solving for $C_i$ shows that for $A_1,A_2=O(1)$, $C_i=O(\epsilon^2\bar p)$ so $C_i\ll \hsi$  as long as $\bar p\ll\epsilon^{-1}$. For $A_1=O(1)$ and $A_2\to\infty$, similarly $C_1=O(\epsilon^2\bar p)\ll \hsa$ and $C_2=O\left({\epsilon^2\bar p}/{\sqrt{A_2}}\right)\ll \hsb$ for $\bar p\ll\epsilon^{-1}$. 
\par
We note that for $\pbar>0$, the saddle points are related by $\hsa>\hsb$ yielding monotone decreasing profiles; this inequality is reversed for $\pbar<0$ (hence the monotone increasing solution in Figure~\ref{fig:asy1}(b)) with the flat-film solution $h\equiv \epsilon$ being the transition state at $\pbar=0$.
\par
Figure~\ref{fig:asy1}(a) shows a small-thickness solution with mean thickness $\hb>\eps$ (corresponding to $\pbar>0$), where we define $\hb=\int_0^L h\,dx/L$.    $m=0.31$ for $\epsilon=0.1, \ L=3$, when $m$ is slightly greater than $\epsilon L=0.3$. Figure~\ref{fig:asy1}(b) shows the profile for another branch~1 solution, with $\pbar<0$ yielding $\hb<\eps$. In both Figure~\ref{fig:asy1}(a) and (b), the boundary layer near $x=s$ can be well approximated by the estimate \eqref{eq:hprofile_b1}.
\par
For $\pbar<0$, equation \eqref{HsHcEqn} has only one root with $\hsi<\epsilon$ and for 
$\bar p\to-\infty$ its leading order behavior is  $\hsi\sim \epsilon^{3/4}(A_i/|\pbar|)^{1/4}$. Consequently the solutions on branch 1 in this limit can still be approximated by the smoothed step profile \eqref{eq:hprofile_b1}, but now the range of the solutions is $\hmax-\hmin=O(|\epsilon^3/\pbar|^{1/4})$ and the width of the interior transition layer is $O(|\pbar|^{-5/8})$.
\par
The limit $\bar p\to0^+$ also describes another class of solutions, characterized by nearly-flat films with large thickness, corresponding to branch 6 in Figure~\ref{fig:mh2}(b). Unlike thin nearly-flat solutions, which have a boundary layer near $x=s$ and approach a step function in the limit $\epsilon\to0$, this class of thick solutions has small amplitude slowly-varying deviations from the mean film thickness. An example of a steady-state profile on this branch is shown in Figure~\ref{fig:asy3}(a).
\par
We write $\hb=m/L$ in terms of the mass of the solution, $m=\int_0^L h\,dx$. 
For $\bar{h}\to\infty$, we write the solution as $h(x)\sim \bar h+\sigma h_1(x)+\sigma^2 h_2(x)$ and we will show that it is convenient to 
define $\sigma=\Pi(\bar{h})$. From \eqref{eq:dis34}, it is clear that the limit $\bar{h}\to\infty$ for any fixed $\epsilon$ is equivalent to $\sigma\to 0$.
\begin{figure}
\centering
 \begin{subfigure}[b]{0.45\linewidth}
 \centering
 \includegraphics[width=2.5in]{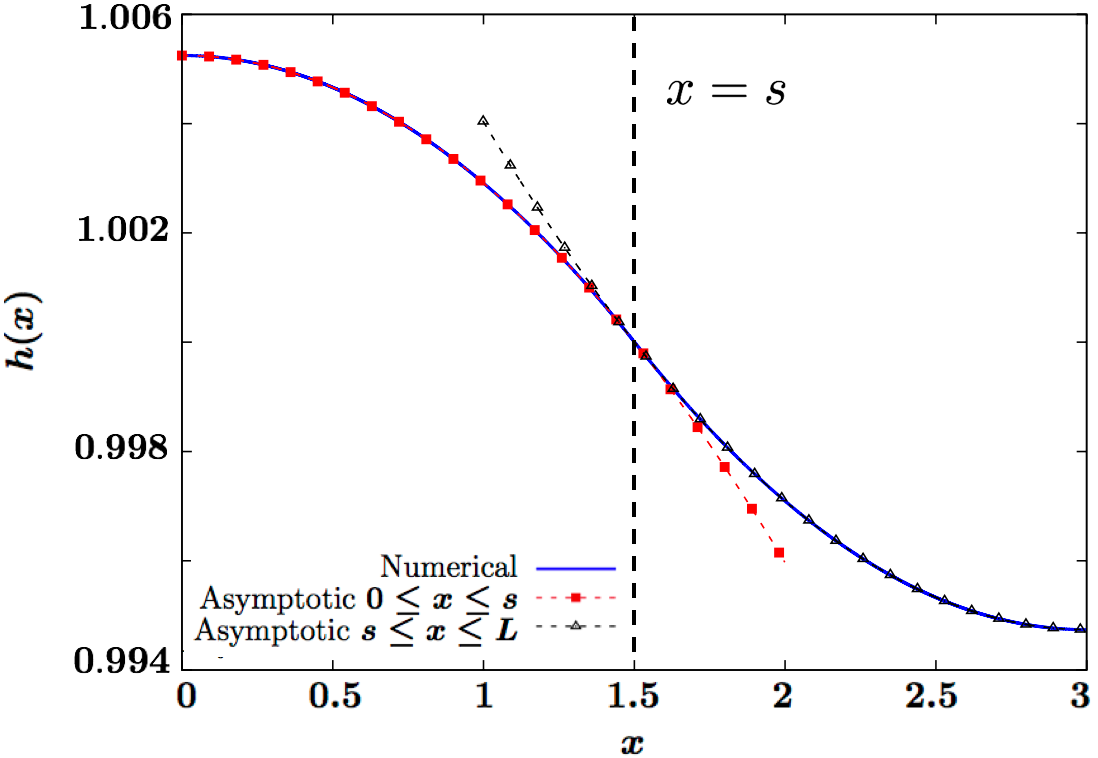}
 \caption{}
\end{subfigure}\qquad
 \begin{subfigure}[b]{0.45\linewidth}
 \centering
\includegraphics[width=2.4in]{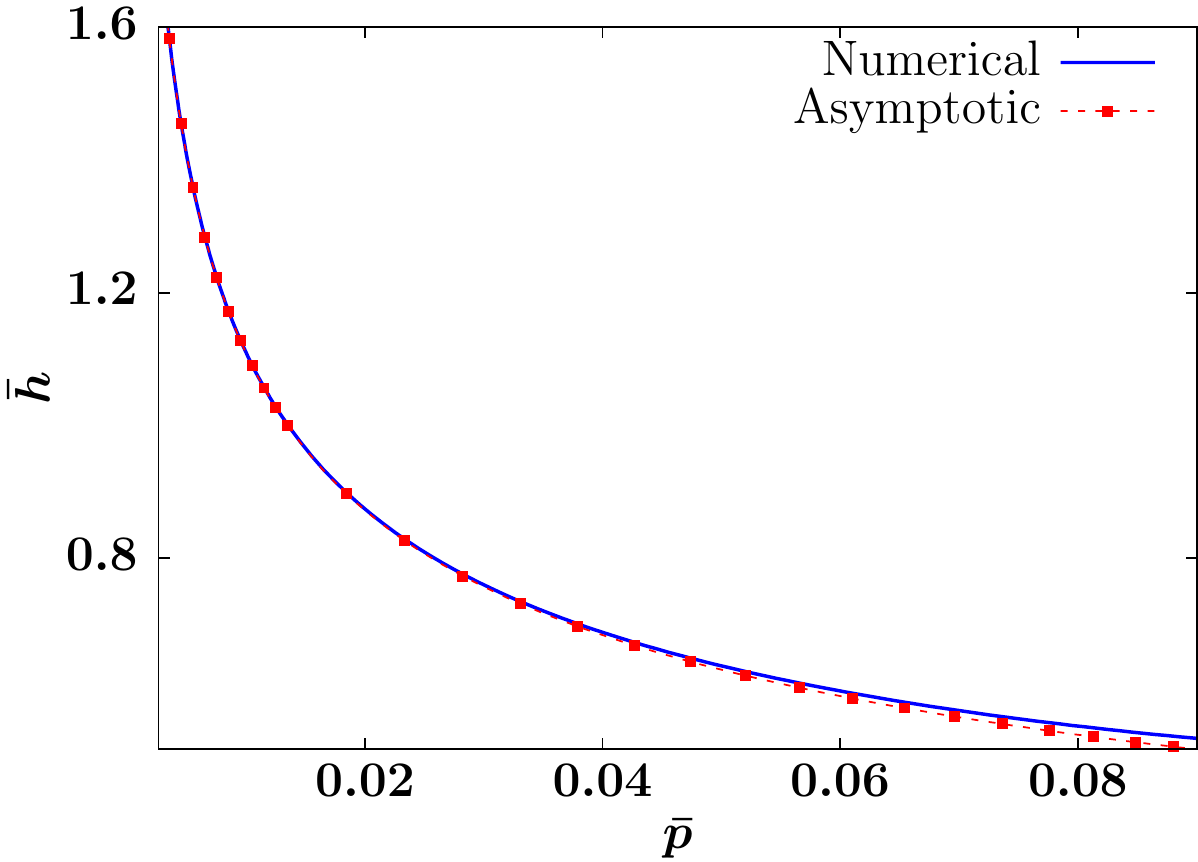}
 \caption{}
\end{subfigure}
\caption{Properties of branch~6, thick film solutions, computed for parameters $\epsilon=0.1$, $A_1=1$, $A_2=2$, $L=3$, $s=1.5$: (a) A numerically computed solution (blue) compared with the asymptotic estimates \eqref{eq:largem} (dotted) at pressure $\pbar=0.0135$. (b) The computed mean thickness $\hb$ (blue) as a function of $\pbar$ compared with the asymptotic prediction \eqref{eq:pm_asy} (red dots).}
\lbl{fig:asy3}
\end{figure}
Substituting this expansion into equation \eqref{eq:ss_het_intro} and expanding $\Pi(h)$ to  $O(\delta)$,  we have
\begin{subequations}
\begin{align}
\sigma h_{1xx}&= \sigma A_1+\sigma A_1\Pi'(\bar h)h_1-\pbar,\qquad h_{1x}(0)=0,\qquad 0<x\leq s,\\
\sigma h_{1xx}&=\sigma A_2+\sigma A_2\Pi'(\bar h)h_1-\pbar,\qquad h_{1x}(L)=0,\qquad s<x\leq L.
\end{align}
\end{subequations}
To balance the equation at $O(\sigma)$, we choose $\pbar=O(\sigma)$ by writing $\pbar\sim \sigma (p_0+\sigma p_1)$ for some $p_0=O(1)$. As $m\to\infty$, $\Pi'(\bar h)<0$. Solving for $h_1(x)$ on $0\leq x\leq s$ and $s<x\leq L$ respectively, we obtain to $O(\sigma)$
\begin{equation}
h(x) \sim
  \begin{cases}
\bar h+\Pi(\hb)\left(C_1\cos(r_1x)+\frac{A_1-p_0}{r_1^2}\right)& 0\leq x \leq s,\\[5pt]
\bar h+\Pi(\hb)\left(C_2\cos(r_2(L-x))+\frac{A_2-p_0}{r_2^2}\right) & s<x \leq L,
  \end{cases}
  \lbl{eq:largem}
\end{equation}
where $r_i=\sqrt{-A_i \Pi'(\bar{h})}$. To determine constants $C_1$ and $C_2$, we use conditions \eqref{eq:cont}-\eqref{eq:diff}. Consequently we find that $C_1$ and $C_2$ are both linear in $p_0$, $C_1=p_0\tilde C_1$ and $C_2=p_0\tilde C_2$.
The definition of the mean thickness $\hb$ yields the condition $\int_0^L h_1(x)\,dx=0$, which we can solve for $p_0$ to obtain 
\begin{equation}\lbl{eq:pm_full}
\bar p\sim\frac{\Pi(\bar h)L}{\frac{s}{A_1}+\frac{L-s}{A_2}+\frac{ \tilde C_1\Pi'(\bar h)}{r_1}\sin(r_1s)+\frac{\tilde C_2\Pi'(\bar h)}{r_2}\sin(r_2(L-s))}.
\end{equation}
Simplifying \eqref{eq:pm_full} further, in the limit of large $\hb$, the pressure can be written as
\begin{equation}\lbl{eq:pm_asy}
\pbar=\left( A_1 {s\over L} + A_2 {L-s\over L}\right)
\left({\epsilon^2\over \hb^3}-{\epsilon^3\over \hb^4}\right)+O(\hb^{-7}),
\end{equation}
This result being in terms of the weighted average of the wetting parameters $A_i$ with respect to domain lengths can be interpreted as giving an effective overall leading order disjoining pressure $\tilde{\Pi}$ for the nearly flat film homogenized at the mean level $\hb$: $\tilde{\Pi}\sim \bar{A}\Pi(\hb)$.
The higher order terms in \eqref{eq:pm_asy} contain factors of $(A_1-A_2)$ and $s(L-s)$ so if the problem was on a homogeneous substrate (via $A_2=A_1$ or $s=0$ or $s=L$) then this trivially reduces to the disjoining pressure for a flat film. 
\par
Figure~\ref{fig:asy3}(a) shows a typical branch~6 thick film solution computed at pressure $\pbar=0.0135$. The asymptotic estimate given by \eqref{eq:largem} agrees very well with the numerical solution, which suggests that $\bar p$ is inversely proportional to ${m^3}$ for large mass. For fixed large mass, $\bar p$ scales linearly in both $A_1$ and $A_2$. Figure~\ref{fig:asy3}(b) shows $\hb$ for numerically computed branch~6 solutions over a range of $\pbar$. The comparison with the analytical predictions show that \eqref{eq:pm_asy} is accurate for the limit of large $\hb$.

\begin{figure}
\centering
\includegraphics[width=2.4in]{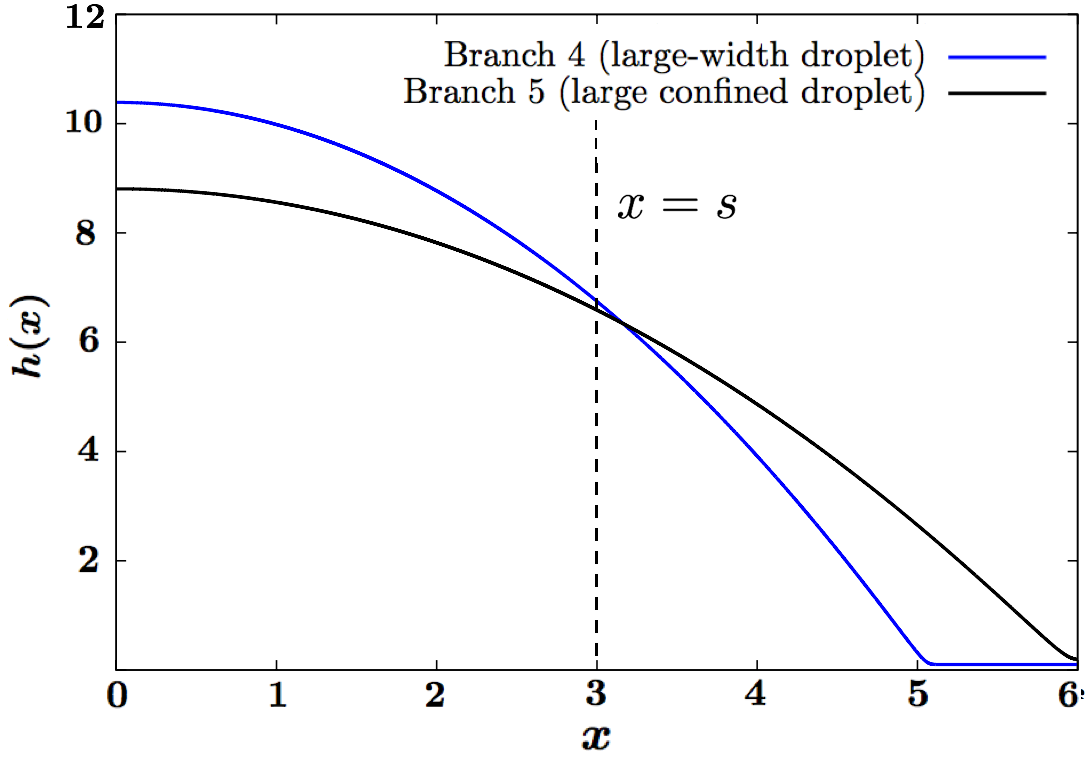}
\caption{Profiles of two droplet solutions with mass $m=35.2$ on the domain $L=6$ for a disjoining pressure with $A_1=1$, $A_2=50$, $L=6$, $s=3$, $\epsilon=0.1$.
A typical branch~4 large-width droplet, with $s<w<L$, is shown by the blue curve for $\pbar=0.493$ and a branch~5 confined droplet, with $w=L$, at pressure $\pbar=0.809$ is given by the black curve.} 
\lbl{fig:profile_b4}
\end{figure}
\subsection{Large-width and confined droplets}\lbl{sssec:s2}
\par
In this subsection, we study branches 4 and 5, which give two families of droplet-type solutions that are similar to droplet solutions on homogeneous substrates. Droplets are states where most of the fluid is concentrated within a region of limited width (or radius) and is surrounded by nearly-uniform very thin films with thickness $h=O(\epsilon)$ set by the disjoining pressure.
\par 
First, we discuss branch 4, which describes a class of large-width droplets with width $s<  w<  L$. 
On the droplet core, we assume $h=O(1)$ on $0\le x< w$, and $h=O(\epsilon)$ outside.
One example of such a steady-state solution is shown by the blue curve in Figure \ref{fig:profile_b4}. We show that in the limit $\epsilon\to0$, the profiles for these droplets can be approximated to leading order by truncations of the homoclinic droplet on the homogeneous substrate with $A(x)\equiv A_2$.
\par
 To obtain an asymptotic estimate of the droplet's maximum ($\hmax=O(1)$), we use \eqref{eq:matchhx} in the limit $\epsilon\to 0$. Since the wetting interface, $x=s$, occurs within the droplet's core, we have $h(s)=O(1)$. For $h=O(1)$, equation \eqref{Ueqn} gives $U(h)=O(\epsilon^2)$. Using this for $\hmax$ and $h(s)$
with $\hmin\sim \epsilon$, equation 
\eqref{eq:matchhx} reduces to 
\begin{equation}
\bar ph_{\max} =-A_2U(h_{\min})+O(\epsilon)
\lbl{reduced215branch4}
\end{equation}
which gives the inverse dependence on the pressure, 
\begin{equation}
h_{\max}=\frac{A_2}{6\bar p}+O(\epsilon).
\lbl{eq:hmax4}
\end{equation}
Note that to leading order this matches $\Hmb$, the maximum of the homoclinic droplet on a homogeneous substrate with $A(x)\equiv A_2$, as shown in \cite{glasner03coarsening}. 
Since $A_2>A_1$, this $\hmax$ describes a droplet larger than the homoclinic for a homogeneous substrate with $A=A_1$.
\par 
For $h=O(1)$, the disjoining pressure scales as $\Pi(h)=O(\epsilon^2)$, so to leading order \eqref{eq:ss_het} on the droplet core reduces to $\frac{d^2h}{dx^2}=-\bar p$, yielding the parabolic profile 
\begin{equation} 
h(x)=\hmax-{1\over 2} \pbar x^2+O(\epsilon).
\lbl{eq:parabolic_soln}
\end{equation} 
The width can then be estimated from $h(w)=O(\epsilon)$ as 
\begin{equation}
w\sim \sqrt{{A_2\over 3\pbar^2}}\, ,
\lbl{eq:width}
\end{equation}
 similar to results in \cite{glasner03coarsening}.
In summary, in the limit $\epsilon\to0$, the leading order profile of a large-width droplet on $[0,L]$ is given by 
\begin{equation}\lbl{eq:profile_eps}
h(x)\sim
\begin{cases}
\frac{A_2}{6\pbar}-\frac{1}{2}\pbar x^2 & 0\leq x<  
w,\\[6pt]
\epsilon &w < x\leq L.
\end{cases}
\end{equation}
The even extension of this profile gives a $2L$-periodic solution and hence its minimum must satisfy
$\hmin> \hsb$.
\par 
In the phase plane, branch 4 solutions lie inside the $A_2$-homogeneous homoclinic orbit, see Fig.~\ref{fig:b4_pp}.
This result is based on two observations for the segments on $x\le s$ and $x>s$. For $x\ge s$ ($h(x)\le h(s)$), this follows directly from the solution's minimum being above the saddle point, $\hmin> \hsb$. For $x\le s$ ($h(x)\ge h(s)$), the trajectory lies outside the $A_1$-homogeneous homoclinic orbit since it starts from $\hmax> \Hma$. To see that this portion lies within the region in the phase plane bounded by the $A_2$-homoclinic, we use \eqref{eq:order1_het2} noting that $R_2(h(s))=R_1(h(s))$ by \eqref{eq:diff} and $R_2(h)> R_1(h)$ for
$h>h(s)$ when $A_2>A_1$, hence $\hmax< \Hmb$.
\begin{figure}
\centering
\includegraphics[width=2.4in]{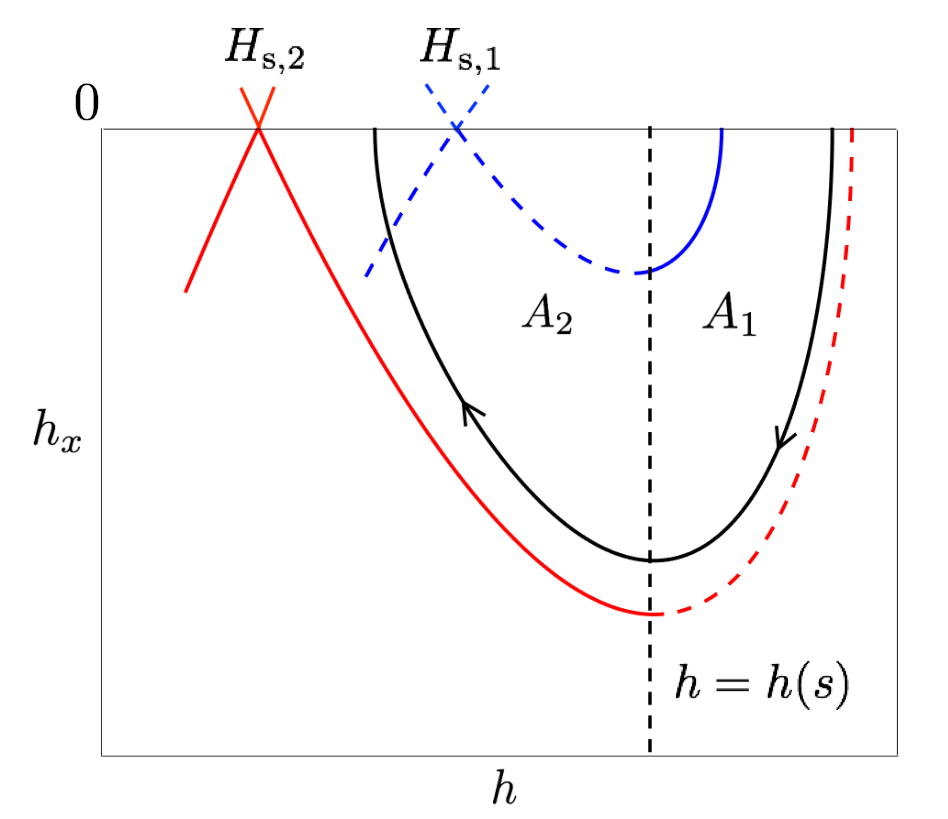}
\caption{Schematic of the phase plane showing the homoclinic orbits for homogeneous substrates, $A(x)\equiv A_1$ (blue) and $A(x)\equiv A_2$ (red) and a branch~4 solution (black), for parameter values $A_1=1$, $A_2=2$, $L=6$, $s=3$, $p=0.211$, $\epsilon=0.1$. The value of $A(x)$ switches across the line $h=h(s)$, the ``in-active'' portions of the homoclinics are drawn with dashed lines.}
\lbl{fig:b4_pp}
\end{figure}
\par
Branch 4 droplet solutions are defined by their widths exceeding the wetting interface position, $w>s$, but not filling the whole domain, $w<L$. 
Using \eqref{eq:width}, this yields the range of pressures for branch 4 as $\sqrt{A_2/(3L^2)}\le\pbar\le \sqrt{A_2/(3s^2)}$. 
At the endpoints, this branch connects to other branches of solutions: at $\pbar^*_{3,4}=\sqrt{A_2/(3s^2)}$ with droplets pinned at the wetting interface (called branch 3, to be described in the next section) and at $\pbar^*_{4,5}=\sqrt{A_2/(3L^2)}$ with droplets limited by the size of the domain (called branch 5, described below).
Figure~\ref{fig:b4_phmax} gives the bifurcation diagram for $\pbar$ vs.\  $\hmax$, showing good agreement of the numerically computed results and compared with the asymptotic predictions. In the derivation we assumed $A_1,A_2=O(1)$; it can be shown that the profile of branch 4 solutions is still described by \eqref{eq:profile_eps} for $A_1=1$ fixed and $A_2\to\infty$, as suggested by Figure~\ref{fig:b4_phmax}(b).
\par
These solutions have mass and width both decreasing with increasing pressure. 
For $\eps\to 0$, the mass of the droplet core is 
\begin{equation}m\sim \int_0^w h\,dx\sim {A_2^{3/2}\over 9\sqrt{3}\,\pbar^2}.
\lbl{eq:mass4drop}
\end{equation}
Note the film mass $m_{4,5}^*\sim \sqrt{3A_2}\,L^2/9$ corresponding to $p^*_{4,5}$ is the maximum  possible mass for a droplet-type solution with domain-size $L$. Above that mass, only nearly-flat film solutions (branch 6) exist. The scaling of this critical mass with $A_2$ shows the importance of the heterogeneous disjoining pressure in controlling droplet structure. 
\par 
Another important physical property characterizing fluid droplets is the contact angle, or angle of inclination at the edge of support. The small aspect ratio assumption essential to lubrication theory justifies use of the small angle approximation, $\tan\theta\sim\theta$, for this context. Consequently the contact angle scales the slope of the droplet profile at the edge of the core, with the constant of proportionality being the aspect ratio. We see that the effective contact angle of all branch 4 droplets is independent of the pressure,
\begin{equation}
\theta\propto |h'(w)| \sim \pbar w= \sqrt{{A_2\over 3}},
\lbl{eq:angle4drop}
\end{equation}
again indicating the controlling influence of the disjoining pressure, as in \cite{glasner2003coarsening}.
It was previously shown in \cite{glasner03coarsening} that large droplets on a homogeneous substrate with $A\equiv 1$ have contact angle given by $|h'(w)|\sim 1/\sqrt{3}$, in agreement with \eqref{eq:angle4drop} when $A_1=A_2=1$. 
\begin{figure}
 \begin{subfigure}[b]{0.45\textwidth}
\includegraphics[width=2.5in]{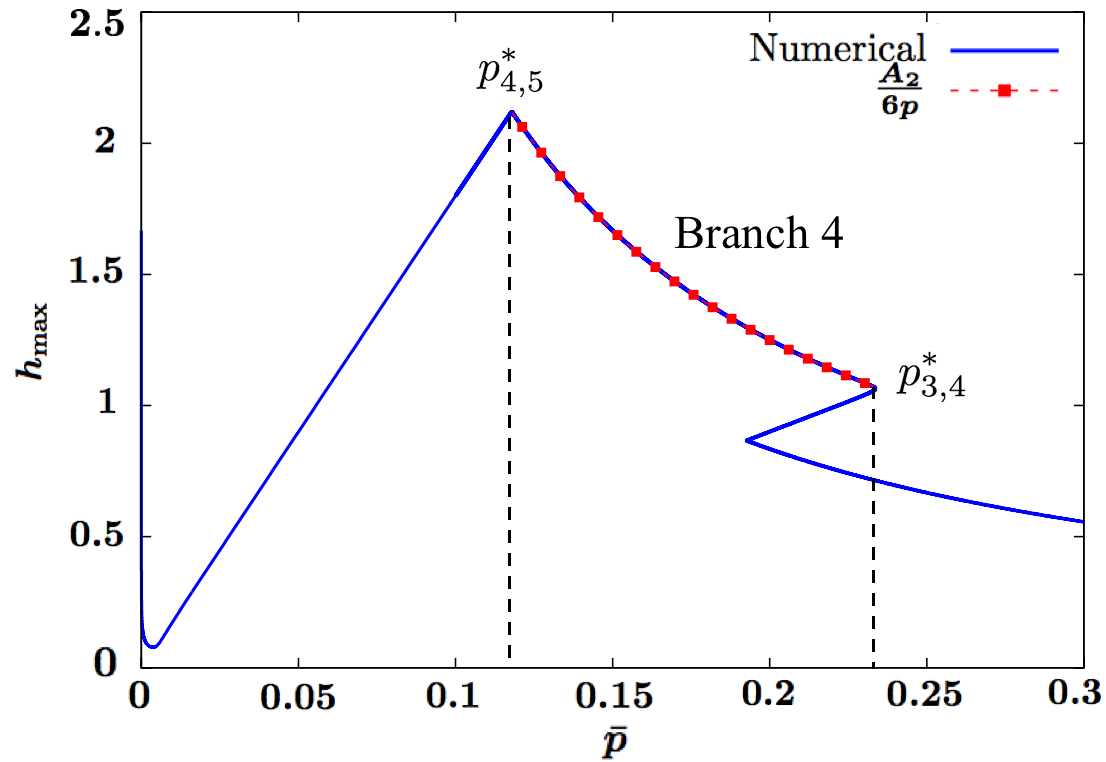}
\caption{}
\end{subfigure}
 \begin{subfigure}[b]{0.45\textwidth}
\includegraphics[width=2.5in]{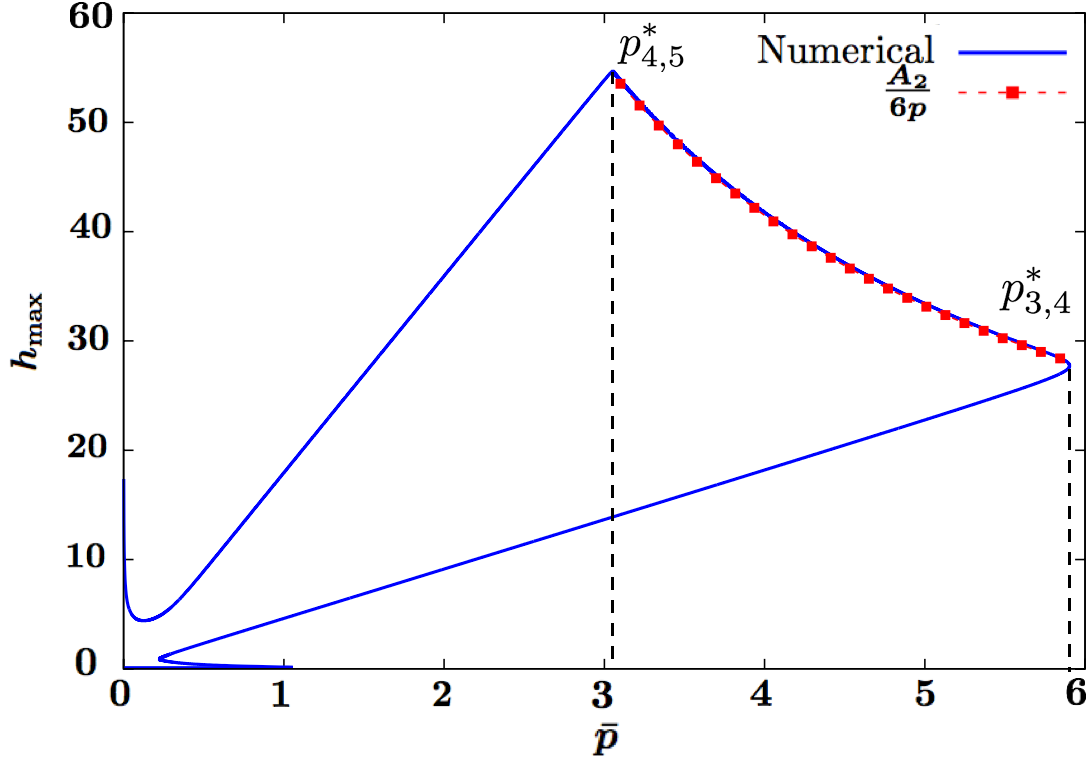}
\caption{}
\end{subfigure}
\caption{Bifurcation diagram $\pbar$ vs. $h_{\max}$ computed numerically and compared with the asymptotic prediction of branch~4 (a) in the limit $\epsilon\to0$, with parameters $A_1=1$, $A_2=1.5$, $\epsilon=0.001$. (b) in the limit $A_2\to\infty$, with parameters $A_1=1$, $A_2=1000$, $\epsilon=0.1$. In both (a) and (b), $L=6$, $s=3$. The blue solid curve represents the numerically computed bifurcation curve. The red dashed and dotted curve represents the asymptotic prediction given by \eqref{eq:hmax4}.}
\lbl{fig:b4_phmax}
\end{figure}
   \begin{figure}
    \centering
    \begin{subfigure}[b]{0.48\textwidth}
        \centering
       \includegraphics[width=2.5in]{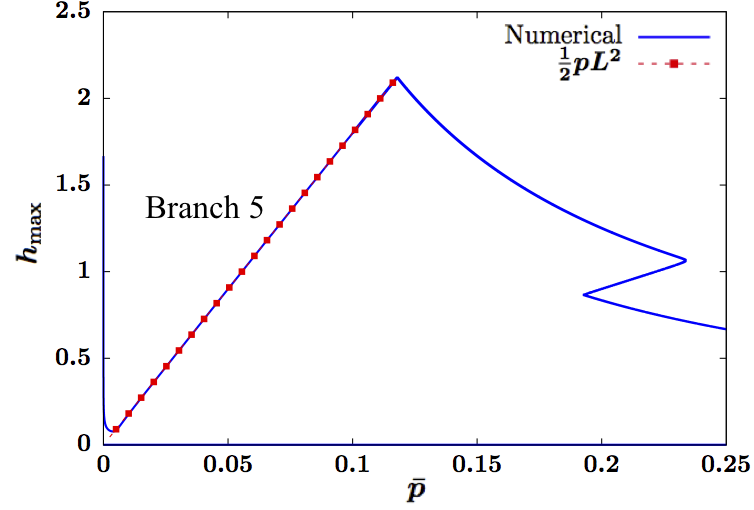}
        \caption{}
    \end{subfigure}%
    \hfill
    \begin{subfigure}[b]{0.48\textwidth}
        \centering
        \includegraphics[width=2.5in]{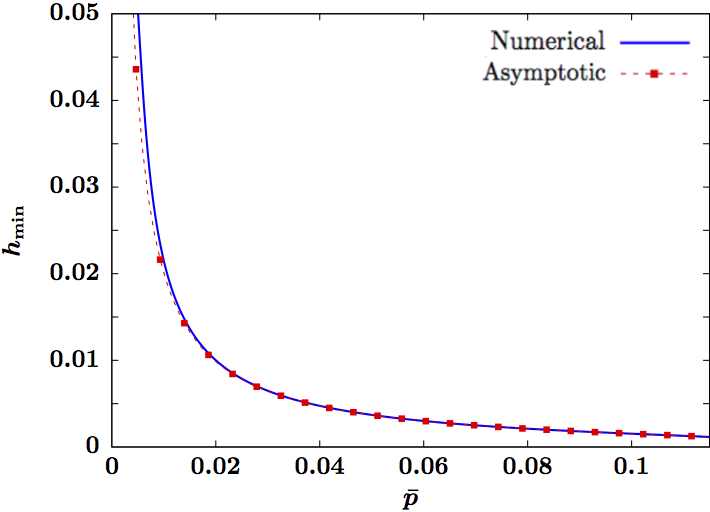}
        \caption{}
    \end{subfigure}
\caption{Properties of branch~5 large drop solutions computed with parameter values $A_1,\ A_2=1.5,\ L=6,\ s=3$, $\epsilon=0.001$:
(a) Branch~5 highlighted in the bifurcation diagram for $\pbar$ vs. $\hmax$ computed numerically (blue solid curve) and compared with the asymptotic estimate \eqref{eq:hmax_b5} (red dotted curve) in the limit of small $\epsilon$. 
(b) The minimum, $\hmin$ as a function of $\pbar$ computed numerically (blue solid curve) and the asymptotic result
(\ref{eq:cubic5}, \ref{eq:hmin5}) (red dotted curve).}
    \lbl{fig:ph_b5_A2small}
\end{figure}
\par 
Branch~5 describes large droplets confined by the domain size so the droplet width is always $w= L$. This set of solutions provides a transition between the large-thickness nearly-flat films (branch~6) and the large-width droplets (branch~4) described above. An example of a solution on branch~5 is shown by the black curve in Figure~\ref{fig:profile_b4}.
\par 
We first investigate the solutions in the limit $\epsilon\to0$ for $A_1, A_2=O(1)$. As with branch~4, on the core region ($0\le x< L$), branch~5 droplets have $h=O(1)$ and the influence of the disjoining pressure can be neglected to yield a parabolic profile, $h(x)=\hmax-{1\over 2} \pbar x^2+O(\eps)$. Here, using $w=L$ gives the drop's maximum as scaling linearly with the pressure, 
\begin{equation}\lbl{eq:hmax_b5}
h_{\max}=\frac{1}{2}\bar pL^2+O(\epsilon)
\end{equation}
To obtain an asymptotic estimate of the minimum film thickness $h_{\min}$, we use \eqref{eq:matchhx}, and assume $h_{\min}=O(\epsilon)$. At leading order, the equation reduces to 
\begin{equation}\label{eq:hmin_b5}
-A_2U(h_{\min})=\bar ph_{\max}\sim\frac{1}{2}\bar p^2L^2,
\end{equation}
similarly to \eqref{reduced215branch4}.
Note that the potential function $U(h)$ has a global minimum at $h=\epsilon$ with $U(\epsilon)=-1/6$. For \eqref{eq:hmin_b5} to have a real solution, we need $\frac{1}{2A_2}\bar p^2L^2\leq \frac{1}{6}$. This upper bound on the pressure on branch~5 coincides with the lower bound for the pressure on branch~4 found above, $\pbar^*_{4,5}=\sqrt{A_2/(3L^2)}$. With $U(h)$ of the form \eqref{Ueqn},
\eqref{eq:hmin_b5} can be written as a cubic polynomial equation,
\begin{equation}
{1\over 2} y^2 - {1\over 3}y^3=  z  \qquad\mbox{with}\qquad y= {\epsilon\over \hmin}\,,\qquad
z={\pbar^2 L^2\over 2A_2}.
\lbl{eq:cubic5}
\end{equation}
The solution for $\hmin$ on $0\le z\le 1/6$ is the smaller of the two positive roots for $y$, given by
\begin{equation} 
y={1\over 2} \left( 1- {1+i\sqrt{3}\over 2} \sigma - {1-i\sqrt{3}\over 2\sigma}\right) \qquad\mbox{with}\qquad
\sigma=\left(  1-12z + 12\sqrt{z^2-z/6}\,\right)^{1/3},
\lbl{eq:hmin5}
\end{equation}
where $\sigma$ is complex-valued yielding
$y\sim 1$ for $z\to 1/6$ and  $y\sim \sqrt{2z}$ as $z\to 0$.
\par 
For $y=O(1)$, it is clear that $\hmin=\epsilon/y=O(\epsilon)$, consistent with our earlier assumption. 
This result holds for solutions on branch~5 with the pressure bounded away from zero,
with $\pbar <\pbar^*_{4,5}$. Branch~6 is approached as $\pbar\to 0$. Figure~\ref{fig:ph_b5_A2small}(a) and (b) show the plots for $\pbar$ vs.\  $\hmax$ and $\pbar$ vs.\ $\hmin$ computed numerically and asymptotically in the limit of small $\epsilon$ for $A_1,A_2=O(1)$. 
\par 
A uniform solution for the droplet can be constructed using matched asymptotics (see \cite{kevorkian}) in the limit of $\eps\to 0$ with the parabolic profile, \eqref{eq:parabolic_soln} with \eqref{eq:hmax_b5}, being the outer solution on $0\le x< L$. To leading order, this outer solution gives the mass and effective contact angle of the droplet as
\begin{equation}
m\sim {1\over 3} \pbar L^3,\qquad  \theta\propto |h'(L)|\sim \pbar L.
\lbl{br5angle}
\end{equation} 
Note that for $\pbar\to 0$, the vanishing contact angle is consistent with the branch of droplets 
transitioning to become the branch of thick films (branch~6) with $\hmin\gg O(\eps)$. 
\par
To satisfy the boundary condition $h'(L)=0$ at the edge of the domain, the solution must have a corner layer to give a rapid transition from the finite contact angle \eqref{br5angle}. 
The local structure for $\eps\to 0$ will actually be a triple deck (see \cite{murdock}) with an inner solution of
the form $h=\eps H(X)$ with $X=(x-L)/\eps$ satisfying
\begin{equation}
{d^2 H\over dX^2} = {A_2\over H^3}\left( 1- {1\over H}\right)-\eps \pbar,
\end{equation}
nested within an intermediate layer $h=\eps^{2/3} \hat{H}(\hat{X})$ with  
$\hat{X}=(x-L)/\eps^{1/3}$ satisfying
\begin{equation}
{d^2 \bar{H}\over d\hat{X}^2} = {A_2\over \hat{H}^3} \left( 1 -{\eps^{1/3}\over \hat{H}}\right) - \pbar.
\lbl{eq321}
\end{equation}
From \eqref{eq321}, the inflection point will occur in the intermediate layer, with $h\sim (\eps^2 A_2/\pbar)^{1/3}$; this could be used to obtain a refined estimate of the contact angle.
We will not go into the details of this construction here. Note that when $\hmin=O(\eps^{2/3})$ the triple deck should reduce to just the intermediate layer and give an estimate for the lower bound on $\pbar$ where the above arguments apply.
\subsection{Pinned droplets}\lbl{sssec:s3}
\par
Solutions on branch~3 are droplets with width pinned by the wetting heterogeneity, $w\sim s$, and
$h=O(\eps)$ for $x\ge s$. Examples of branch~3 profiles for several values of the pressure are shown in Figure~\ref{fig:b3_example}(a). This branch arises as a consequence of the chemical heterogeneity of the substrate. These solutions have several features in common with the confined droplets from branch~5,  differences stem from whether the width is pinned by boundary conditions or the wetting contrast. 
To develop an asymptotic prediction for this type of solutions, we consider the steady-state in the limits $\epsilon\to0$ and $A_2\to\infty$.
\par 
We consider the solution in the limit $\epsilon\to0$ with fixed $A_1,A_2=O(1)$. On $0\le x\le s$ the solution will satisfy the equation
\begin{equation}
    {d^2 h\over dx^2}= {\eps^2 A_1\over h^3}\left( 1- {\eps\over h}\right) -\pbar.
\end{equation}
Similar to branch~5 solutions, the disjoining pressure can be neglected at $O(1)$ and $O(\eps)$  to yield a parabolic profile for
the droplet core, \eqref{eq:parabolic_soln}. Since the droplet has width $w\sim s$, to leading order, the maximum is given by $\hmax\sim\frac{1}{2}\pbar s^2$ and to $O(\eps)$ the solution can be written as
\begin{equation}
    h(x) \sim{1\over 2} \pbar (s^2 -x^2) + \eps C_0\qquad \mbox{on $0\le x< s$.}
\end{equation}
Since the leading order term in this outer solution vanishes as $x\to s^-$, a boundary layer is needed to prevent the divergence of the disjoining pressure contribution there. 
The structure of inner solution at $x=s^-$ follows similarly to the corner layer  at $x=L$ for  confined drops in the previous section except here the coefficient on the disjoining pressure term will be $A_1$ and $\pbar$ will be shown to be $O(1)$ on the whole branch of solutions.
\par
Using equation \eqref{eq:matchhx} with $\hmax=O(1)$ and $\hmin\sim \eps$, at leading order we get a cubic equation for the thickness at the wetting interface,
\begin{equation}\lbl{eq:b3_hs_small}
U(h(s))=\frac{-\frac{A_2}{6}+\frac{1}{2}\bar p^2s^2}{A_2-A_1}
\end{equation}
with $h(s)$ being the real positive root with $h(s)>\epsilon$. Similarly to \eqref{eq:cubic5}, such a solution will exist only if $-1/6< U(h(s))<0$, yielding a condition on the range of pressures for branch~3,
\begin{equation}
    \sqrt{{A_1\over 3 s^2}}< \pbar < \sqrt{{A_2\over 3s^2}}\; ,
    \lbl{br3p234}
\end{equation}
where the upper bound matches $\pbar^*_{3,4}$ for branch~4 solutions, found in section~\ref{sssec:s2}.
\begin{figure}
\centering
\includegraphics[width=2.4in]{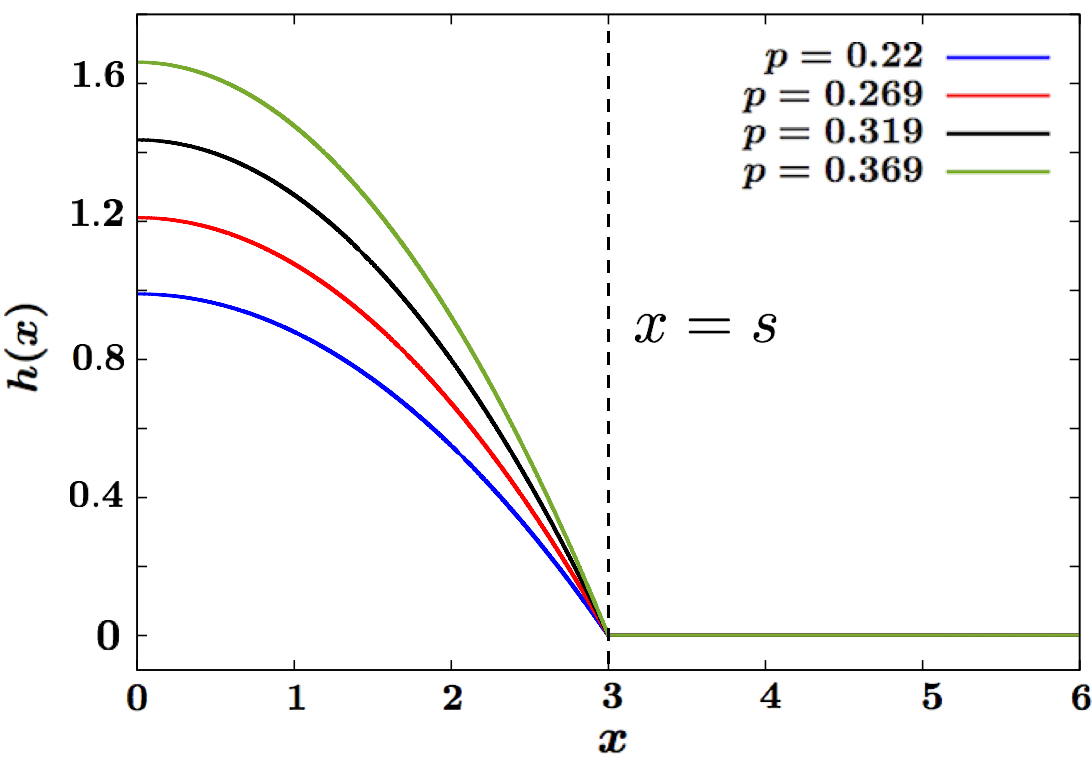}
\caption{Typical pinned droplet branch~3 solutions for $A_1=1, A_2=50, s=3, L=6$ at several values of $\pbar$.}
\lbl{fig:b3_example}
\end{figure}
\par 
To obtain information about the structure of the solution at the contact line, we re-examine the solution in the limit of $A_2\to\infty$. Let $\delta=1/A_2$ then we can write equation \eqref{eq:ss_het_intro} on $s\le x\le L$ as
\begin{equation}
    \delta {d^2 h\over dx^2 }= \Pi(h) -\delta \pbar.
    \lbl{eq:br3ode}
\end{equation}
For $\delta\to 0$ this is a singularly-perturbed problem that can be solved using the method of matched asymptotic expansions in terms of an outer solution and a boundary layer of width $O(\delta^{1/2})$.
The boundary conditions, \eqref{eq:sBCs} and \eqref{eq:ss_het_intro_bc}$_2$,
determine that the boundary layer must be at $x=s^+$.
The outer solution of \eqref{eq:br3ode} for $s<x\le L$ is a constant to all orders, 
\begin{equation} 
h(x)= \eps + \delta \eps^2 \pbar+O(\delta^2);
\lbl{eq:br3outer}
\end{equation} 
this is the $\delta\to 0$ expansion of the saddle point $\hsb$, \eqref{eq:hsi}. Hence, apart from exponentially small terms, the solution's minimum is $\hmin\sim \hsb$.
\begin{figure}
 \begin{subfigure}[b]{0.45\linewidth}
 \centering
   \includegraphics[height=1.8in]{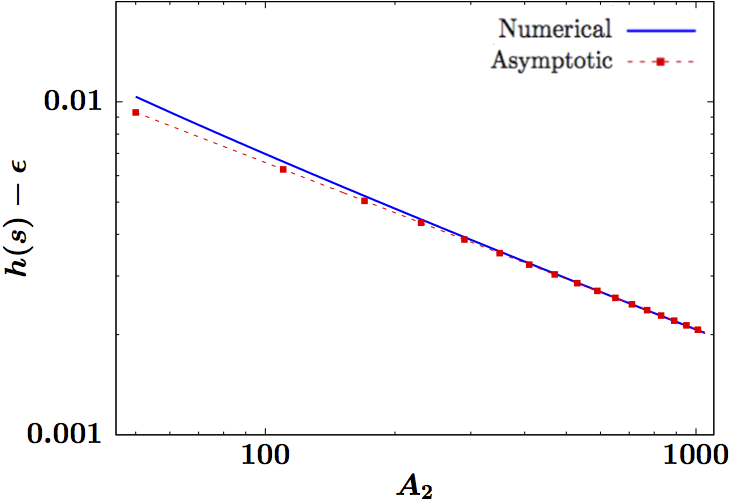}
 \caption{}
\end{subfigure}\qquad
 \begin{subfigure}[b]{0.45\linewidth}
 \centering
  \includegraphics[height=1.8in]{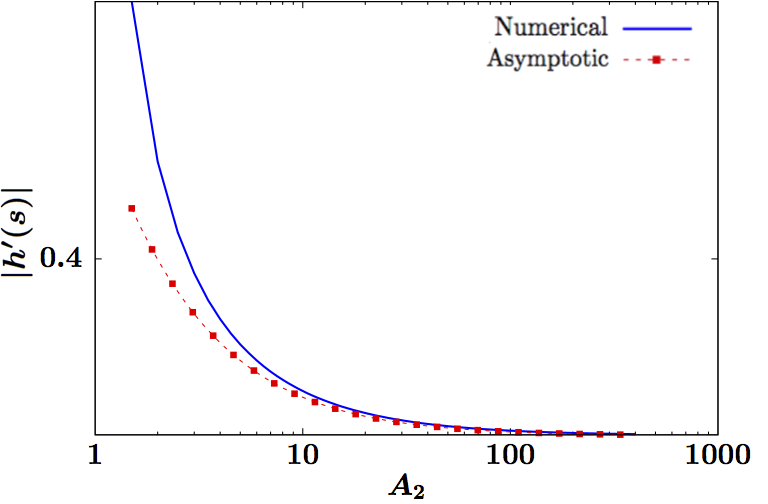}
 \caption{}
\end{subfigure}
\caption{Properties of branch~3 pinned droplet solutions computed with parameter values $A_1=1$, $L=6$, $s=3$, $\epsilon=0.1$: (a) Comparison of $h(s)-\epsilon$ computed numerically with $\pbar=0.292$ (blue) and  from the asymptotic approximation \eqref{eq:hs_pinned} (red dots) for large $A_2$ plotted on log scale. 
(b) the contact angle $|h'(s)|$ vs.\ $A_2$ with fixed pressure $\pbar=0.22$, plotted on log scale, the numerical result (blue) compared with the asymptotic prediction \eqref{eq:hprimes_pinned} (red dots).}
\lbl{fig:b3_leak_scale}
\end{figure}
\par
The form of the inner solution in the boundary layer is $h=\hat{h}(\hat{x})$ 
where $\hat{x}=(x-s)/\delta^{1/2}$ and \eqref{eq:br3ode} becomes 
\begin{equation} {d^2\hat{h}\over d\hat{x}^2}=\Pi(\hat{h})-\delta \pbar.
\lbl{eq:br3ode2}
\end{equation} 
The inner solution must match \eqref{eq:br3outer} for $\hat{x}\to\infty$ and must satisfy \eqref{eq:sBCs} at $\hat{x}=0$. Noting that for $A_2\to\infty$, $\hmax=O(1)$ and $h_x(s^-)=O(1)$ from \eqref{eq:order1_het2}$_1$, there may be concern that the form of $R_2(h)$ suggests that $h_x(s^+)=O(\delta^{-1/2})$. However, from \eqref{eq:diff}, it must be the case that $h_x(s^+)=h_x(s^-)=O(1)$; applied to $R_2$, this forces $U(h)-U(\hmin)=O(\delta)$. Consequently, the expansion of the inner solution must be $\hat{h}(\hat{x})=\eps + \delta^{1/2} \hat{h}_1(\hat{x}) +O(\delta)$ where $\hat{h}_1$ satisfies the linearized equation, $\hat{h}_{1\hat{x}\hat{x}}=\Pi'(\eps)\hat{h}_1$. To satisfy matching, this term must be an exponential decay, $\hat{h}_1(\hat{x})=C_2e^{-\hat{x}/\eps}$,
and overall 
\begin{equation}
    h(x)\sim \eps+\delta^{1/2}C_2 e^{-(x-s)/(\eps\delta^{1/2})}\qquad s\le x< L.
    \lbl{eq:br3a2soln}
\end{equation}
To determine the $C_2$ coefficient, we re-write  \eqref{eq:b3_hs_small} as
\begin{equation}\lbl{eq:delta_eq}
U(h(s)) = { -{1\over 6} + {1\over 2} \delta \pbar^2 s^2\over 1- \delta A_1}
\end{equation}
and plug-in $h(s)\sim \eps+\delta^{1/2}C_2$. Expanding for $\delta\to 0$ we get
\begin{equation}\lbl{eq:hs_pinned}
h(s)\sim \epsilon+\frac{\epsilon}{\sqrt{A_2}}\sqrt{\pbar^2 s^2-\frac{A_1}{3}} \;.
\end{equation}
This result can also be obtained as the leading order approximation from solving \eqref{eq:delta_eq} as a cubic equation as was done with \eqref{eq:cubic5}.
Figure \ref{fig:b3_leak_scale}(a) shows numerically computed values and the asymptotics for $h(s)-\epsilon$ compared with the asymptotic approximation given by \eqref{eq:hs_pinned} for large $A_2$.
\par
Using the asymptotic prediction \eqref{eq:hs_pinned}, we can also derive $h'(s)$, which represents the contact angle of this class of droplets in the limit of large $A_2$. On the $A_1$ region $[0,s]$, as $x\to s^-$, using \eqref{eq:order1_het2}, we have
\begin{equation}\lbl{eq:CA_eqn}
\frac{1}{2}h'(s)^2=A_1U(h(s))-\bar ph(s)+\bar ph_{\max}-A_1U(h_{\max})
\end{equation}
Substituting \eqref{eq:hs_pinned} and $h_{\max}\sim\frac{1}{2}\bar ps^2$ into \eqref{eq:CA_eqn},  we obtain
\begin{equation}\lbl{eq:hprimes_pinned}
h'(s)=-\sqrt{\pbar^2 s^2-\frac{A_1}{3}}\left(1+\frac{A_1}{2A_2}\right)+O(\eps) +O(\eps/A_2).
\end{equation}
This shows how the limiting contact angle is approached as the wettability ratio, $A_2/A_1$, is increased,
see Fig.~\ref{fig:b3_leak_scale}(b). We note that this value is lowered by wettability effects (as represented by the $A_1/3$ term) relative to the contact angle of the confined drop \eqref{br5angle}.
The asymptotic prediction $h_{\max}\sim\frac{1}{2}\pbar s^2$, represented by the red dotted curve in Figure \ref{fig:phmax_b2} is compared with the numerically computed bifurcation curve. We observe that the leading order asymptotic prediction agrees well the numerical results. 
\begin{figure}
 \begin{subfigure}[b]{0.45\linewidth}
 \centering
  \includegraphics[height=1.8in]{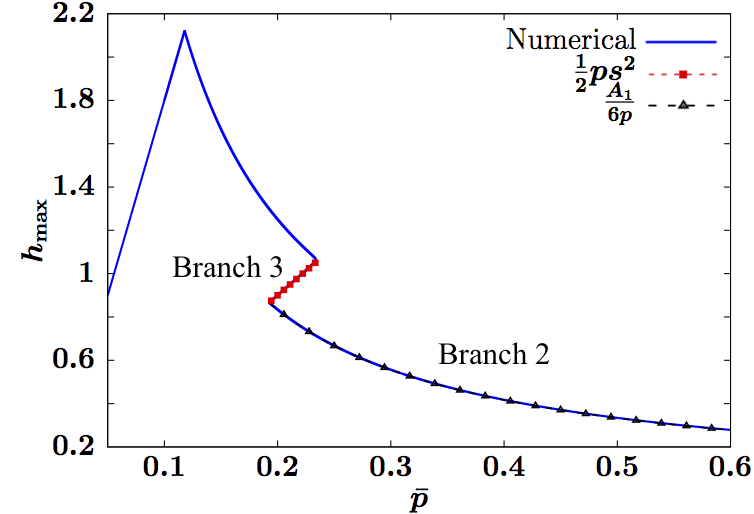}
  \caption{}
\end{subfigure}\qquad
 \begin{subfigure}[b]{0.45\linewidth}
 \centering
  \includegraphics[height=1.8in]{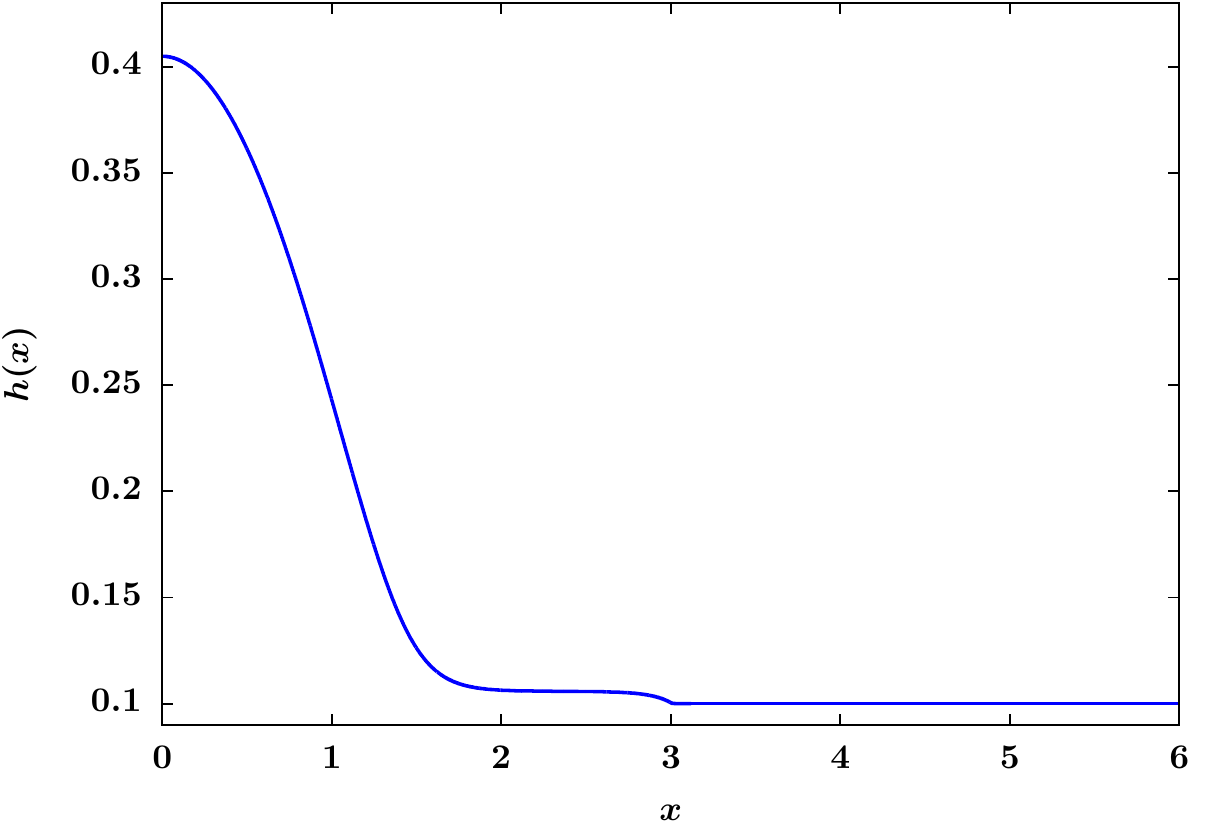}
  \caption{}
\end{subfigure}
\caption{Properties of small and pinned droplets: (a) The bifurcation diagram for $\pbar$ vs.\ $\hmax$ highlighting branches~2 and 3. The solid blue curve gives numerical results. The black and red dotted curves represent the asymptotic prediction of $h_{\max}$ for branch 2 and branch 3 solutions respectively in the limit $\epsilon\to0$, with parameters $A_1=1$, $A_2=1.5$, $L=6$, $s=3$, $\epsilon=0.001$. 
(b) Profile of a steady-state on branch~2, characterized by a droplet on $[0,s]$ and nearly-uniform thin film on $[s,L]$.}
\lbl{fig:phmax_b2}
\end{figure}
\subsection{Small-width droplets}
\lbl{sec:smalldrop}
\par 
Finally, we conclude with branch~2, whose solutions combine features from both droplets and nearly-flat films. 
This branch describes small droplets with an effective width smaller than the size of the hydrophilic domain, $w < s$, and a surrounding nearly-flat film that covers the remainder of the domain, see Fig.~\ref{fig:phmax_b2}(b). 
\par  
Branch~2 folds back from branch~1 in Figure~\ref{fig:mh2}(b) giving droplets whose cores completely reside in the $A_1$ region. Compared to branch 1 solutions which are thin, nearly flat films over the entire domain, in the outer $A_1$ and $A_2$ regions and a boundary layer near $x=s$, solutions on branch 2 are characterized by larger mass so that droplets could form on the $A_1$ region, but not so large as to yield branch 3 or 4 type droplets that fill or extend beyond the $A_1$ region (having widths $w\ge s$). This class of solutions has the smallest mass possible for droplets centered at $x=0$.
 \begin{figure}
  \begin{subfigure}[b]{0.45\linewidth}
 \centering
   \includegraphics[height=1.8in]{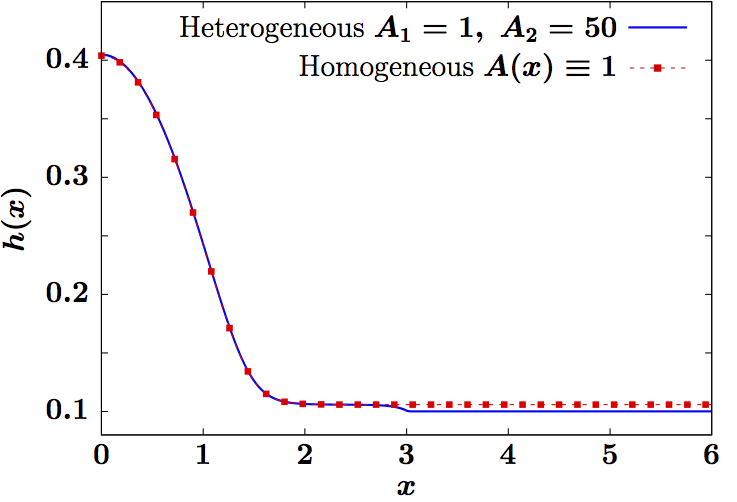}
  \caption{}
\end{subfigure}\qquad
 \begin{subfigure}[b]{0.45\linewidth}
 \centering
  \includegraphics[height=1.8in]{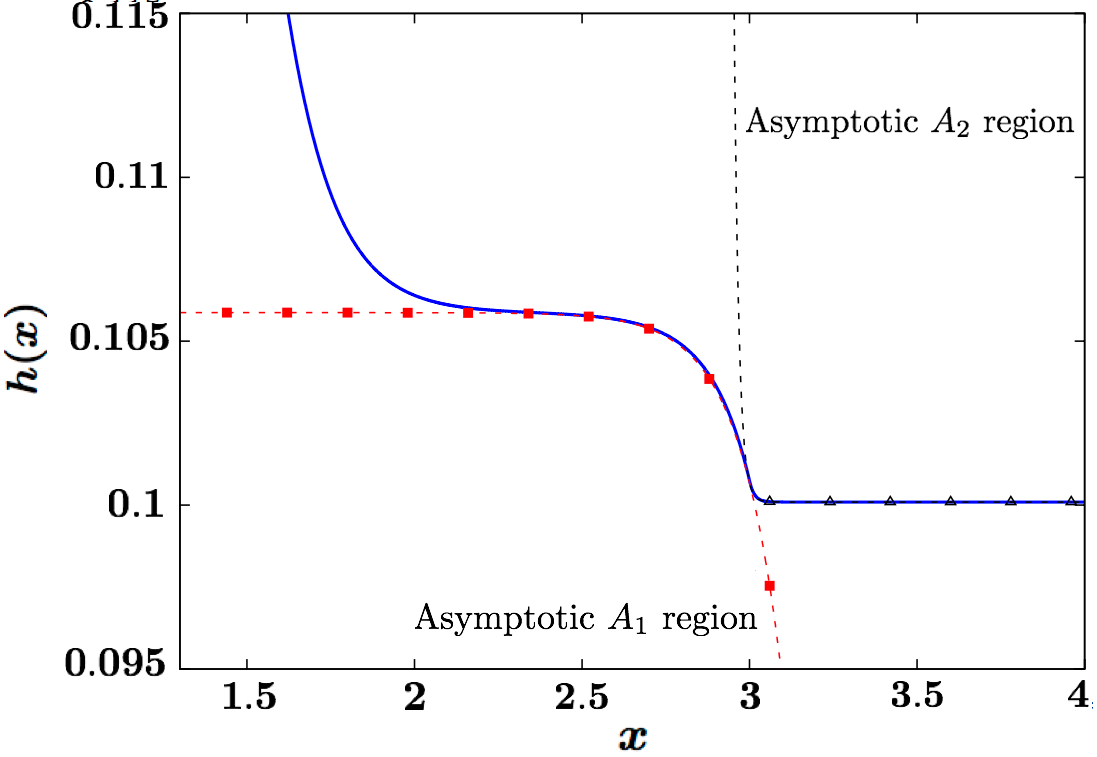}
  \caption{}
\end{subfigure}
\caption{Details for a branch~2 solution computed with parameters $L=6$, $s=3$, $A_1=1$, $A_2=50$, $\epsilon=0.1$, $\bar p=0.467$: (a) Comparison of the numerical solution (blue) on the heterogeneous substrate with its corresponding $A_1$-homogeneous homoclinic (red dots) with the same pressure. (b) Comparison of the numerically computed steady-state (blue)  and the approximations (\ref{eq:h1_A1_eqn},\ref{eq:br3a2soln}) for the structure near the interface $x=s$ (red, black dots). }
\lbl{fig:hprofile_b2}
\end{figure}
\par 
We construct the solution in the limit $A_2\to\infty$ with $A_1$ fixed. On the $A_2$ region, the same matched asymptotics used for branch~3, equations \eqref{eq:br3ode} and \eqref{eq:br3ode2}, similarly yields the solution as \eqref{eq:br3outer} and
\eqref{eq:br3a2soln} with $h(s)=\eps+O(\delta^{1/2})$, but we must use a different argument to determine $C_2$.

\par 
To find $C_2$ in \eqref{eq:br3a2soln}, we consider the steady-state on the $A_1$ region.  To leading order as $\delta\to 0$, the steady-state problem for $h\sim h_0(x)$ on $0\le x\le s$ is given by
\begin{subequations}
\begin{equation}
\frac{d^2h_0}{dx^2}=A_1\Pi(h_0)-\pbar  \lbl{eq:A1Dirichlet}
\end{equation}
\begin{equation}
h_0'(0)=0,\qquad h_0(s)=\epsilon \lbl{eq:DirichletBC}
\end{equation}
\end{subequations}
Noting that the boundary condition $h_0(s)=\eps$ is less than the saddle point $h=\hsa$, 
the trajectory for $h_0(x)$ must lie outside the $A_1$-homoclinic orbit in the phase plane.
Since the solution is monotone decreasing with $h(0)=O(1)$, there must be a point $x_1$ with $0<x_1<s$ where $h_0(x_1)=\hsa$. This will be a non-stationary inflection point of the solution. Linearizing \eqref{eq:A1Dirichlet} about $\hsa$ and using $\eps\ll 1$ the solution on $x_1<  x\le s$ can be approximated by 
\begin{equation}\lbl{eq:h1_A1_eqn}
h(x)\approx \hsa-C_1 e^{\sqrt{A_1\Pi'(\hsa)}\, (x-s)}.
\end{equation}
Applying boundary conditions \eqref{eq:sBCs} to \eqref{eq:h1_A1_eqn} and \eqref{eq:br3a2soln}
yields
\begin{equation}
C_1=\frac{\hsa-\eps}{\epsilon\sqrt{\delta} \sqrt{A_1\Pi'(\hsa)}+1}\,,\qquad
C_2=\frac{\epsilon \sqrt{A_1\Pi'(\hsa)}\, (\hsa-\epsilon)}{\epsilon\sqrt{\delta} \sqrt{A_1\Pi'(\hsa)}+1}\,.
\lbl{eq:F_b2}
\end{equation}
If we take $\epsilon\to0$ and $\delta\to 0$, to leading order we get $C_2\sim \frac{\epsilon^2\bar p}{\sqrt{A_1}}$ and thus
\begin{equation}\lbl{eq:hs_b2_A2large}
h(s)\sim\epsilon+\frac{\epsilon^2\bar p}{\sqrt{A_1 A_2}}.
\end{equation}
This resembles the $H_s$ saddle value with an effective wetting coefficient given by the geometric mean of $A_1$ and $A_2$ and remains less than $\hsa\sim \eps +\eps^2\pbar/A_1$ since  $A_2>A_1$.
\par
Noting \eqref{eq:hs_b2_A2large} and \eqref{eq:br3outer} motivates writing \eqref{eq:matchhx} as
\begin{equation}
A_1 U(h(s)) + \pbar \hmax = \pbar \hmin + A_1 U(\hmax) + A_2 \left( U(h(s))- U(\hmin)\right),\lbl{br2eq215}
\end{equation}
where if $\hmax=O(1)$ then three terms on the right are each $O(\eps)$ or smaller. Consequently balancing terms on the left, at leading order we get $\hmax\sim A_1/(6\pbar)$ for $\eps\to 0$.
This is the leading order approximation of the maximum film thickness of the $A_1$-homoclinic (which can be obtained by solving $R_1(\hsa)=0$). This suggests that as $A_2\to\infty$, the droplet core on $0\le x<x_1<s$ can be approximated to leading order by the $A_1$-homoclinic solution on $0\le x< x_1<s$. Figure \ref{fig:hprofile_b2} shows the profile of a branch~2 solution for $A_2=50$; in \cite{gomba2017} this was called a `D1' solution.
Figure~\ref{fig:hprofile_b2}(a) shows a comparison with the $A_1$-homoclinic solution having the same pressure $\pbar$. 
Figure~\ref{fig:hprofile_b2}(b) shows \eqref{eq:h1_A1_eqn} and \eqref{eq:br3a2soln} compared with the numerical solution on a heterogeneous substrate near the interface of $A_1$ and $A_2$ regions, $x=s$; similarly to the form of the branch~1 solutions \eqref{eq:hprofile_b1}. We note that some of these approximations break down for $\pbar$ near $\pmax$, where $\hmax=O(\eps)$ and $U(\hmax)$ in \eqref{br2eq215} is $O(1)$.
\begin{figure}
\centering
\includegraphics[width=2.4in]{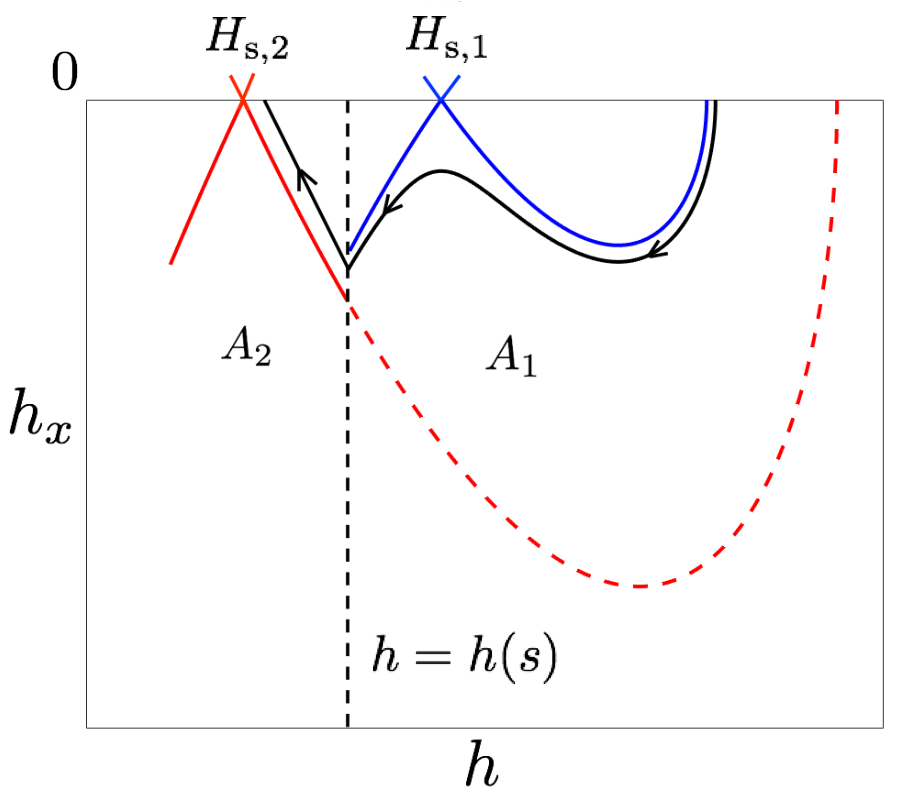}
\caption{Schematic phase plane diagram showing the homoclinic orbits for $A(x)\equiv A_1$ (blue) and $A(x)\equiv A_2$ (red) and the trajectory for a branch~2 solution.}
\lbl{fig:b2_pp}
\end{figure}
\par
Figure~\ref{fig:b2_pp} shows a schematic phase plane for a branch~2 solution (black curve) compared with the homoclinic orbits for the $A_1$-homogeneous and $A_2$-homogeneous problems (blue and red curves respectively). From the arguments connecting to \eqref{eq:A1Dirichlet}, we know that $\hmax>\Hma$ and the branch~2 solution lies outside the $A_1$-homoclinic orbit for $h\ge h(s)$. From the fact that $\hmin > \hsb$ for any finite $L$, we know that the solution must lie within the $A_2$-homoclinic for $h\le h(s)$. The branch~2 solutions have two inflection points, at heights 
$h=\hsa$ and $h=h(s)$, giving them a characteristic `staircase' or `precursor-foot' appearance, see Fig.~\ref{fig:hprofile_b2}.
\par
We also note that corresponding results can also be obtained in the limit $\eps\to 0$  with fixed $A_2=O(1)$, see \cite{liu2019}. In particular for $\eps\to 0$, we can use the parabolic profile \eqref{eq:parabolic_soln} to approximate the droplet portion of branch~2 solutions. Then following similar form \eqref{eq:width} for large drops, we can estimate the effective width of the core from $\hmax$ to yield $w\sim \sqrt{ A_1 /(3\pbar^2)}$. Branch~2 joins branch~3 when the `small' drops attain maximum size as minimal pinned drops with $w\sim s$. This yields $\pbar_{2,3}^*\sim\sqrt{A_1/(3s^2)}$, corresponding to \eqref{br3p234}.
Figure \ref{fig:phmax_b2}(a) shows the bifurcation diagram $\hmax$ vs. $\pbar$ zoomed into a portion of branch 2 and branch 3, computed numerically and asymptotically in the limit of small $\epsilon$.

\subsection{Summary of the steady state branches}
\par
In examining the six branches we have seen that each type of solution is impacted somewhat differently by the presence of the heterogeneous wetting. Distinct from groupings by droplets or film-like states, we can fundamentally separate the solutions into two sets based on phase plane structure:
\begin{itemize}
    \item Solutions on branches 1 and 2 are given by trajectories from Fig~\ref{fig:b2_pp}. They are characterized by having the height at the wetting interface fall between the saddles, $\hsb< h(s) < \hsa$ (for $\pbar<0$ the order of the saddles is reversed). Portions of these solutions follow the stable and unstable manifolds from the $\hsa$ and
    $\hsb$ saddle points.
    \item Solutions on branches 3,4,5 and 6 are given by trajectories from Fig~\ref{fig:b4_pp}. Here the height at the switching point lies above both saddles, $h(s)>\hsa>\hsb$ with branches~4, 5 and 6 having $h(s)=O(1)$ and branch~3 with $h(s)=O(\eps)$.
\end{itemize}

\section{Leakage in the limit of large $A_2$}\label{ssec:leak}
\par
In the limit of large $A_2$, the $A_2$ region effectively becomes increasingly hydrophobic and should give a stronger confining effect on fluid in the $A_1$ region. This behavior holds only for a range of small fluid masses, as wetting effects cannot be expected to influence thick layers of fluids. In terms of the six branches of steady solutions we have analyzed, branches 2 and 3 described small and pinned drops that are effectively confined to the $A_1$ region. From \eqref{eq:mass4drop} with $\pbar^*_{3,4}=\sqrt{A_2/(3s^2)}$, we obtain that the maximum droplet mass that can be confined in the
$A_1$ region is
\begin{equation} 
m_{3,4}^*\sim {s^2 A_2^{1/2}\over 3\sqrt{3}}.\lbl{sec4:maxmass}
\end{equation}
Note that this mass increases when the width of the $A_1$ region ($s$) is increased or $A_2$ is increased.
 In applications where accurate distribution of fluid on solid surfaces are required, it is important to develop a quantitative understanding of the degree of leakage or `spillover' of the fluid from the $A_1$ region into the $A_2$ region. In this section, we present a measure of leakage for branch 2 and 3 solutions and show the leakage is inversely proportional to $A_2$.
\par
In Sections~\ref{sssec:s3} and \ref{sec:smalldrop}, we showed that the film thickness at the heterogeneous interface is $h(s)\sim\epsilon$ as $A_2\to\infty$, see \eqref{eq:hs_pinned} and \eqref{eq:hs_b2_A2large}. We also showed that in the outer $A_2$ region, $h(x)\sim \epsilon$ for $x>s$.
To measure the fluid leakage, we use the fluid mass above $h(x)=\epsilon$ on $x\in[s,L]$, as illustrated by the shaded region in Figure~\ref{fig:mleak_diag}(a). 
\begin{figure}
\begin{subfigure}[b]{0.48\textwidth}
\includegraphics[width=2.5in]{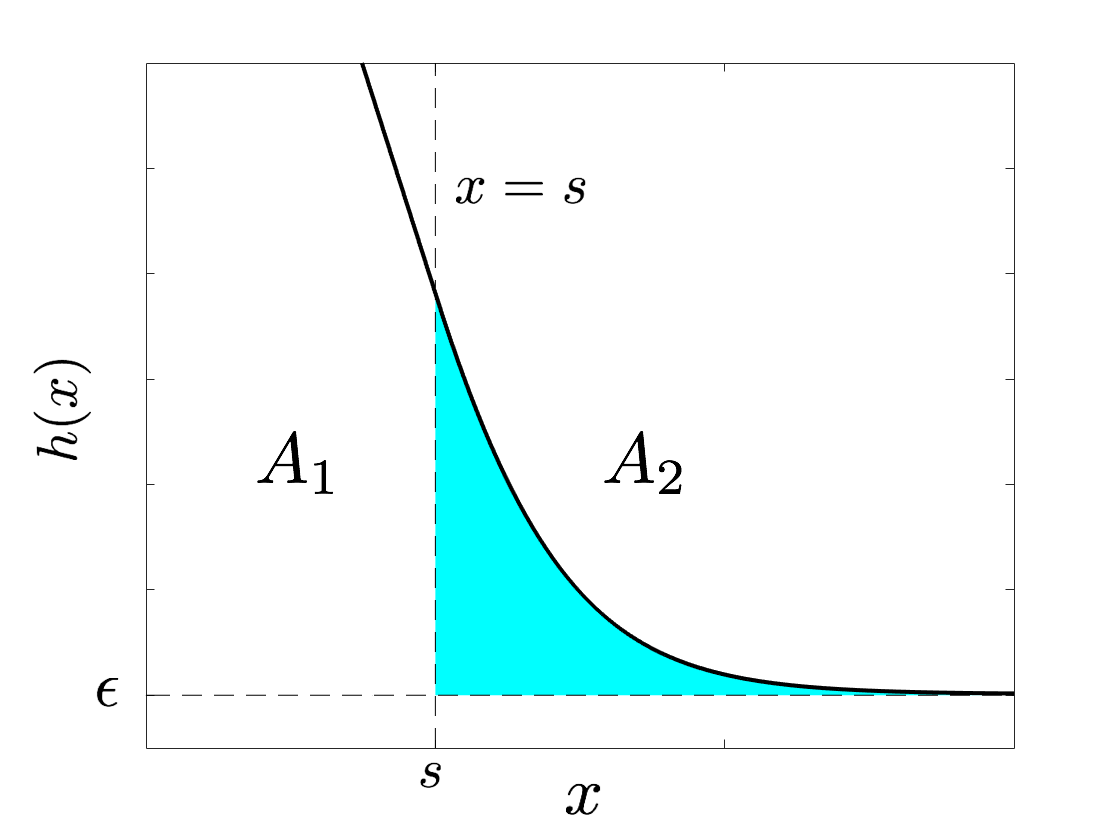}
\caption{}
\end{subfigure}
\begin{subfigure}[b]{0.48\textwidth}
\includegraphics[width=2.5in]{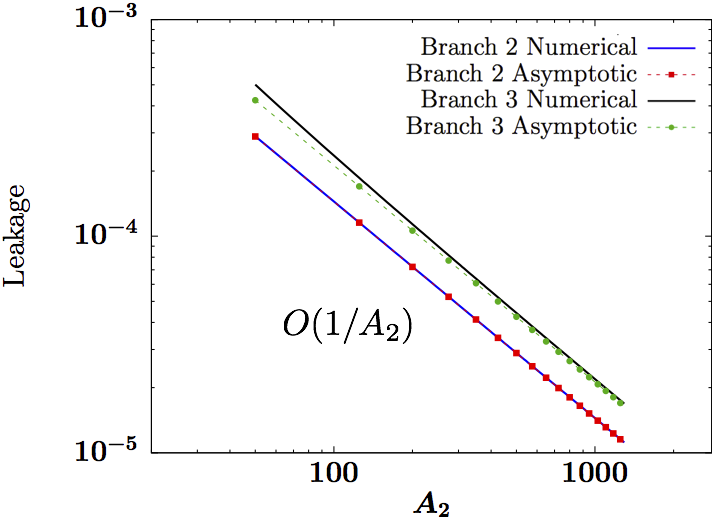}
\caption{}
\end{subfigure}
\caption{(a) Schematic illustration of fluid leakage \eqref{eq:leak_def} (b) Leakage as a function of $A_2$ computed numerically and asymptotically for a branch 2 solution with $\bar p=0.466$ and a branch 3 solution with $\bar p=0.38$ plotted in log scale. The asymptotic prediction for branch 2 and branch 3 is given by \eqref{eq:leak_Asy_b2} and \eqref{eq:leak_b3_final} respectively. The pressure $\bar p$ is fixed as $A_2$ increases with $L=6$, $s=3$, $A_1=1$, $\epsilon=0.1$.}
\lbl{fig:mleak_diag}
\end{figure}
We define the mass of leakage as
\begin{equation}\lbl{eq:leak_def}
{\rm Leakage}=\int_s^L \big[ h(x)-\epsilon \big]\,dx.
\end{equation}
\par
Recalling \eqref{eq:br3a2soln} we can approximate the solution on $s\le x \le L$ as
\begin{equation}
h(x)\sim h_{\min,2} + \delta^{1/2} C_2 e^{-(x-s)/(\eps\delta^{1/2})}.
\end{equation}
Using earlier results, \eqref{eq:br3outer}, we have $h_{\min,2}\sim\epsilon+\delta \epsilon^2\pbar$.
For branch~2 solutions, from \eqref{eq:F_b2}, $C_2\sim \eps^2\pbar/\sqrt{A_1}$ and this gives
\begin{equation}\lbl{eq:leak_Asy_b2}
\mbox{Leakage}_2\sim \left( L-s+\frac{\epsilon}{\sqrt{A_1}}\right)\frac{\eps^2 \pbar}{A_2}
\end{equation}
which gives that at leading order, the fluid leakage of solutions on branch 2 is inversely proportional to $A_2$ for large $A_2$.
Similarly, for branch~3 solutions, we use \eqref{eq:hs_pinned} to obtain
\begin{equation}\lbl{eq:leak_b3_final}
\mbox{Leakage}_3\sim\left(L-s+\sqrt{s^2-\frac{A_1}{3\pbar^2}}\;\right)\frac{\epsilon^2\bar p}{A_2}\;.
\end{equation}
Figure \ref{fig:mleak_diag}(b) shows the fluid leakage computed numerically and compared with the asymptotic estimate for solutions on branches 2 and 3 at fixed pressure over a range of $A_2$, plotted in log scale. The numerical result is obtained by first numerically solving for $h(x)$ and then numerically integrating \eqref{eq:leak_def} using the trapezoid rule. 

\section{Axisymmetric steady state solutions}\lbl{ssec:2D}
\par
We can extend our results for one-dimensional thin films on heterogeneous substrates presented in Section~\ref{ssec:1D} to axisymmetric solutions on two-dimensional heterogeneous substrates with axisymmetric patterning. 
\par
For an axisymmetric film $h(r,t)$ on $0\le r\le L$, the evolution equation \eqref{eq:het_PDE} takes the form
\begin{subequations}
\begin{equation}\label{eq:axis_PDE_full}
\frac{\partial h}{\partial t}=\frac{1}{r}\frac{\partial (rJ)}{\partial r} \qquad\mbox{with}\qquad
J=h^3\frac{\partial}{\partial r}\left( A(r)\Pi(h)-\frac{1}{r}\frac{\partial}{\partial r}\left(r\frac{\partial h}{\partial r}\right)\right)
\end{equation}
where $J\equiv h^3 \partial p/\partial r$ is the radial mass flux.
The boundary conditions corresponding to \eqref{eq:bc_het} are now
\begin{equation}\label{eq:axis_PDE_bc}
\frac{\partial h}{\partial r}\left(0,t\right)=0\qquad
\frac{\partial h}{\partial r}\left(L,t\right)=0,\qquad J(0,t)=0\qquad J(L,t)=0,
\end{equation}
\end{subequations}
and we enforce conditions \eqref{eq:cont} and \eqref{eq:diff} on the smoothness of solutions at the jump in substrate wetting properties, $r=s$. 
The mass of the axisymmetric solutions $h(r;\bar p)$ is given by $m=2\pi\int_0^L hr\,dr$ and the average film height is $\hb=2\int_0^L hr\,dr/L^2$.
\par
The positive steady-states for this problem are still be parametrized by a uniform pressure $p\equiv \pbar$. It follows that the steady-state axisymmetric solutions on $0\le r\le L$ satisfy
\begin{subequations}
\begin{equation}\lbl{eq:ss_axis_eq}
\frac{1}{r}\frac{d}{dr}\left( r\frac{dh}{dr} \right)=A(r)\Pi(h)-\pbar
\end{equation}
\begin{equation}
h'(0)=0\qquad h'(L)=0.\lbl{eq:ss_axis_bc}
\end{equation}
\end{subequations}
Directly corresponding to \eqref{eq:myA}, we take the coefficient of the disjoining pressure to describe an axisymmetrically patterned substrate,
\begin{equation}\lbl{eq:axis_A}
A(r)=
\begin{cases}
A_1& 0\leq r\leq s,\\
A_2& s<r\leq L.
\end{cases}
\end{equation}
Consequently, the axisymmetric equivalent of \eqref{eq:matchhx} is given by
\begin{equation}\lbl{eq:pp_rad}
(A_2-A_1)U(h(s))=A_2U(h_{\min})-A_1U(h_{\max})+\bar p(h_{\max}-h_{\min})-\int_0^L \frac{h'^2}{r}dr
\end{equation}
Since \eqref{eq:ss_axis_eq} is not a piecewise-autonomous equation, the phase plane arguments described in Section~\ref{ssec:1D} do not carry over, but we find that most of the other ideas in the asymptotic constructions do apply similarly.
The steady-state axisymmetric solutions separate into six different branches directly corresponding to the six branches found in Section~\ref{ssec:1D} for one-dimension, see Figure~\ref{fig:rad_bif}. Here we will briefly identify the key steps needed to obtain the axisymmetric solutions and highlight results that we will use further.

\begin{figure}
  \begin{subfigure}[b]{0.45\linewidth}
 \centering
  \includegraphics[height=1.8in]{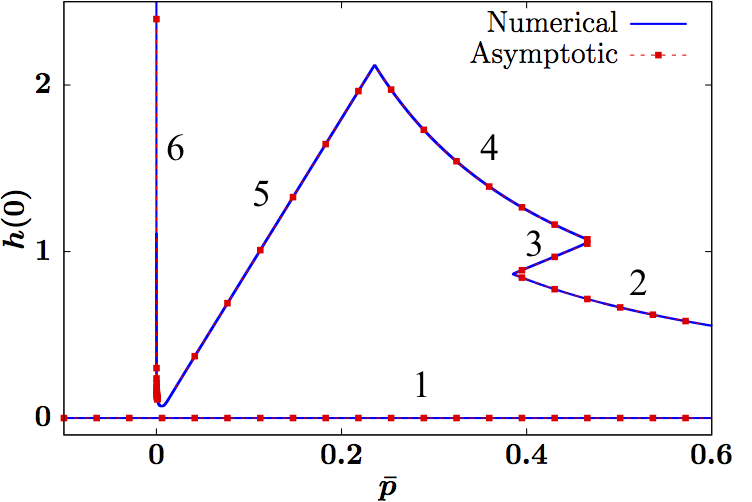}
  \caption{}
\end{subfigure}\qquad
  \begin{subfigure}[b]{0.45\linewidth}
 \centering
  \includegraphics[height=1.8in]{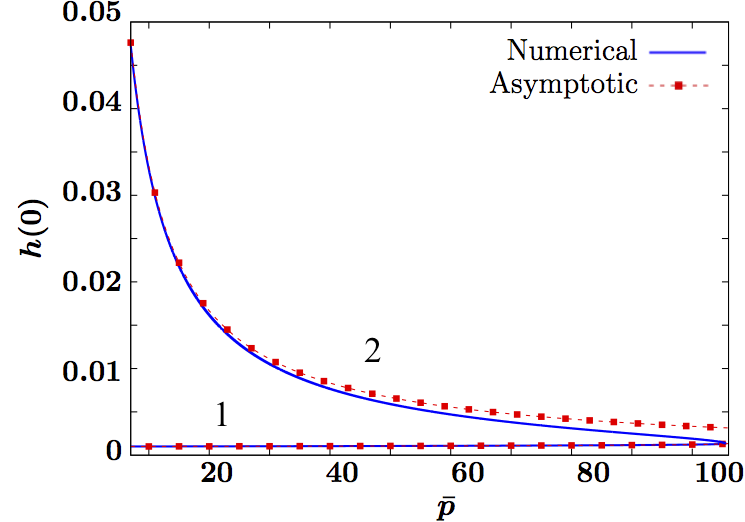}
  \caption{}
\end{subfigure}
\caption{Bifurcation diagram for $h(0)=\hmax$ vs.\ $p$ for (a) small $p$. (b) large $p$. The solid curve denotes the numerically computed bifurcation curve. The dashed and dotted curve denotes the asymptotic prediction of $h_{\max}$ derived for each branch in the limit of small $\epsilon$. In both (a) and (b), $A_1=1,\ A_2=1.5,\ L=6,\ s=3,\ \epsilon=0.001$.}
\lbl{fig:rad_bif}
\end{figure}

\subsection{Small-thickness and large-thickness nearly-flat films}\lbl{sssec:b2_axis}
\par
As in Section~\ref{sssec:s1}, steady nearly-flat solutions generated by the jump in wetting properties can be described by linearizing \eqref{eq:ss_axis_eq}.
\par
For very thin films, the disjoining pressure will balance the uniform pressure, $A_i \Pi(h)=\pbar$, to set piecewise constant heights, $\hsi$, as in \eqref{eq:b1_h0}. Linearizing about these yields a piecewise-defined modified Bessel equation of order zero,
\begin{equation}\lbl{eq:axis_h1_eqn}
h_1''+\frac{1}{r}h_1'=\begin{cases}
A_1\Pi'(\hsa)h_1 & 0\leq r\leq s,\\
A_2\Pi'(\hsb)h_1 & s< r\leq L.
\end{cases}
\end{equation}
Consequently, axisymmetric branch~1 solutions can be approximated by
\begin{equation}\lbl{eq:axis_h1_sol}
h(r)\approx \begin{cases}
\hsa+C_1I_0(\sqrt{A_1\Pi'(\hsa)}\,r) & 0\leq r\leq s,\\
\hsb+C_2K_0(\sqrt{A_2\Pi'(\hsb)}\,r) & s< r\leq L,
\end{cases}
\end{equation}
where $I_0(r)$ and $K_0(r)$ are modified Bessel functions of the first kind and second kind respectively. $C_1$ and $C_2$ are constants to be determined by enforcing the continuity and smoothness of the solution at $r=s$.
\par
For thick films, the disjoining pressure has a weaker influence and the solution can be linearized around a mean height $\hb\gg O(\eps)$.  In the limit $\hb\to \infty$ (having $\Pi'(\hb)<0$), the linearized problem is a regular Bessel equation of order zero, and we can write the axisymmetric branch~6 solutions as
\begin{equation}\lbl{eq:rad_b6_prf}
h(r)\sim
\begin{cases}
\hb+C_1J_0(\sqrt{-A_1\Pi'(\hb)}\,r)-\frac{A_1\Pi(\hb)-\pbar}{A_1\Pi'(\hb)}& 0\leq r\leq s\\
\hb+C_2J_0(\sqrt{-A_2\Pi'(\hb)}\,r)+C_3Y_0(\sqrt{-A_2\Pi'(\hb)}\,r)-\frac{A_2\Pi(\hb)-\pbar}{A_2\Pi'(\hb)}&s<r\leq L
\end{cases}
\end{equation}
where $J_0(r)$ and $Y_0(r)$ are Bessel functions of the first and second kind with constants $C_1, C_2$, and $C_3$ to be determined by the continuity conditions at $r=s$ and the boundary condition at $r=L$.
To determine $\pbar$, we use that \eqref{eq:rad_b6_prf} must satisfy the condition $\int_0^L hr\,dr= \hb L^2/2$. In the limit $\hb\to\infty$ this yields 
\begin{equation}\lbl{eq:pm_b1}
\pbar \sim 
\left( A_1 {s^2 \over L^2} + A_2 {L^2-s^2\over L^2}\right) {\eps^2\over \hb^3}
+O\left( \frac{\epsilon^3}{\hb^4}\right),
\end{equation}
like \eqref{eq:pm_asy}, this pressure is an area-weighted average of the disjoining pressure between the hydrophilic and hydrophobic regions \eqref{eq:axis_A}.

\subsection{Droplet-type axisymmetric solutions}
\par
As was the case for one-dimension, for axisymmetric droplet solutions  are primarily characterized by a core region where $h=O(1)$ as $\eps\to 0$. In the core, to leading order, the uniform pressure balances surface tension with the disjoining pressure being negligible,
\begin{equation}
{d\over dr} \left( r {dh\over dr}\right) \sim -\pbar r.
\end{equation}
This yields a parabolic profile analogous to \eqref{eq:parabolic_soln} but with a modified coefficient,
\begin{equation}\lbl{eq:b2_rad_prf}
h(r)=\hmax -\frac{1}{4}\bar pr^2+O(\epsilon),
\end{equation}
and $h=O(\eps)$ outside the core. 
The width $w$, where $h(w)=O(\eps)$, now represents the effective radius of the core,
\begin{equation}
w\sim \sqrt{{4\hmax\over \pbar}}\;,
\lbl{eq:axi_width}
\end{equation}
and \eqref{eq:b2_rad_prf} can also be written as $h\sim {1\over 4}\pbar (w^2-r^2)$ yielding a mass $m\sim 2\pi \hmax^2/\pbar$, also see \cite{glasnerEJAM}.
What remains to define the different branches of droplet solutions is to make use of information on the contact line position and the far-field of the droplet through \eqref{eq:pp_rad}.
\par
Using \eqref{eq:b2_rad_prf} and \eqref{eq:axi_width}
we can approximate the integral term in \eqref{eq:pp_rad} as
$$
\int_0^L {h'(r)^2\over r}\,dr\sim \int_0^w {h'(r)^2\over r}\,dr\sim
\int_0^w {1\over 4} \pbar^2 r\,dr\sim {1\over 2} \pbar\hmax\,.
$$
Consequently \eqref{eq:pp_rad} for droplet solutions can be approximated by
\begin{equation}\lbl{eq:pp_rad1}
(A_2-A_1)U(h(s))=A_2U(h_{\min})-A_1U(h_{\max})+{1\over 2}\pbar h_{\max}- \pbar h_{\min},
\end{equation}
then the axisymmetric droplet solutions follow using analogous arguments from Section~\ref{ssec:1D}
\begin{itemize}
\item
Branch~2: small radii droplets, $w<s$ with
$$
\hmax\sim {A_1\over 3\pbar} \qquad 
w\sim \sqrt{{4A_1\over 3\pbar^2}}\qquad h(s)\sim\eps\qquad \mbox{for}\qquad 
\pbar> \sqrt{{4A_1\over 3 s^2}}.
$$
\item
Branch~3: pinned droplets, $w\sim s$ with
$$\hmax\sim {1\over 4} \pbar s^2\qquad 
h(s)\sim\epsilon+\frac{\epsilon}{\sqrt{A_2}}\sqrt{\frac{1}{4}\bar p^2s^2-{A_1\over 3}}\qquad 
h'(s)\sim -\sqrt{\frac{1}{4}\pbar^2s^2-{A_1\over 3}}
$$
for $\sqrt{4A_1/(3s^2)}<\pbar< \sqrt{4A_2/(3s^2)}$.
Note that these results differ from the 1-D results \eqref{eq:hs_pinned} and \eqref{eq:hprimes_pinned} only by a coefficient and the contact angle is lowered relative to estimate based on the droplet core, $|h'(w)|\sim {1\over 2} \pbar w$.
\item
Branch~4: large radii droplets, $s<w<L$ with
$$
\hmax\sim {A_2\over 3\pbar} \qquad 
w\sim \sqrt{{4A_2\over 3\pbar^2}}\qquad \mbox{for} \qquad \sqrt{{4A_2\over 3 L^2}}<\pbar<  \sqrt{{4A_2\over 3 s^2}}.
$$
on $\sqrt{4A_2/(3L^2)} <\pbar< \sqrt{4A_2/(3s^2)}$.
\item
Branch~5: confined droplets, $w\sim L$ with
$$\hmax\sim {1\over 4} \pbar L^2\qquad 
h'(L)\sim -{1\over 2} \pbar L\qquad\mbox{for}\qquad \pbar<  \sqrt{{4A_2\over 3 L^2}}.
$$
\end{itemize}
Figure~\ref{fig:rad_bif}(a,b) show the bifurcation diagram $h(0)=\hmax$ vs. $\pbar$ computed for small and large $\pbar$ respectively compared with the asymptotic estimates given above.

\section{Stability of the steady-state solutions}\lbl{ssec:stab_het}
\par
We use linear stability analysis on the one-dimensional steady-state solutions described in Section~\ref{ssec:1D}. 
Writing the steady states as $\hstab(x)=h(x;\pbar)$, 
we express perturbed solutions as $h(x,t)=\hstab(x)+\delta h_1(x,t)$ for $\delta\ll 1$. Plugging into the full evolution equation \eqref{eq:het_PDE} and linearizing, at $O(\delta)$, we obtain
\begin{equation}
\frac{\partial h_1}{\partial t}=\Lh\, h_1,
\end{equation}
where the linear operator $\Lh$ is given by
\begin{subequations}
\begin{equation}
\Lh\, g\equiv {\partial \over \partial x}\left(\hstab^3(x) {\partial\over \partial x}\left[A(x)\Pi'(\hstab)g-{\partial^2 g\over \partial x^2}\right]\right)
\end{equation}
\begin{equation}
\frac{\partial g}{\partial x}(0,t)=0\qquad \frac{\partial g}{\partial x}(L,t)=0, \qquad \frac{\partial^3 g}{\partial x^3}(0,t)=0\qquad \frac{\partial^3 g}{\partial x^3}(L,t)=0
\end{equation}
\end{subequations}
By separation of variables, we can write $h_1(x,t)=\sum_{n}c_ng_n(x)e^{\lambda_nt}$ where $(g_n(x),\lambda_n)$ are
eigenmodes of
\begin{equation}\lbl{eq:eig_het}
\Lh\,g=\lambda g\, .
\end{equation}
The steady state $\hb(x)$ is then linearly stable if all $\mbox{Re}(\lambda_n)<0$.
\begin{figure}
\begin{subfigure}[b]{0.48\textwidth}
\centering
\includegraphics[width=2.4in]{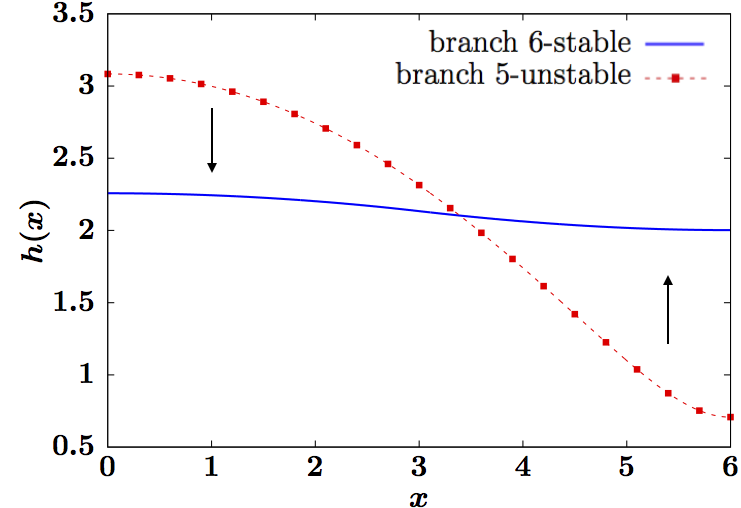}
\caption{}
\end{subfigure}\qquad
\begin{subfigure}[b]{0.48\textwidth}
\centering
\includegraphics[width=2.4in]{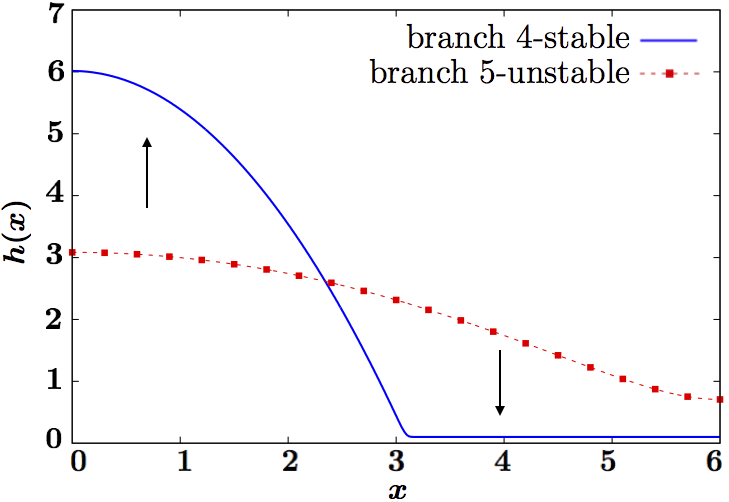}
\caption{}
\end{subfigure}
\caption{Bi-stable dynamics with respect to perturbations of an unstable branch~5 solution $\hstab(x)$. (a) Initial conditions (red dotted curve) $h(x,0)=\hstab(x)+\delta g_1(x)$, with $g_1(x)$ being the unstable eigenmode of $\hstab$ and
small $\delta>0$, evolving to the stable branch~6 solution (blue curve). (b) Initial conditions $h(x,0)=\hstab(x)-\delta g_1(x)$ evolving to the stable branch~4 solution.}
\lbl{fig:evolve}
\end{figure}
\par
To investigate the linear stability of the six different branches of solutions discussed in Section~\ref{ssec:1D} we solve \eqref{eq:eig_het} with typical parameters $A_1=1, \ A_2=50,\ L=6, \ s=3, \ \epsilon=0.1$ using the eigenvalue solver in MATLAB. By continuation in pressure $\pbar$, we find that of all the six different branches discussed above, branch~5 is the only unstable branch, while other branches all characterize stable steady-state solutions.
\begin{figure}
\begin{subfigure}[b]{0.48\textwidth}
\centering
\includegraphics[width=2.4in]{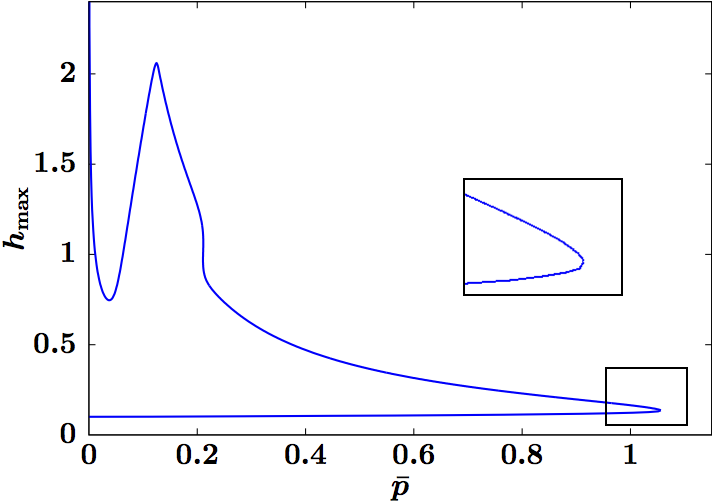}
\caption{}
\end{subfigure}\qquad
\begin{subfigure}[b]{0.48\textwidth}
\centering
\includegraphics[width=2.5in]{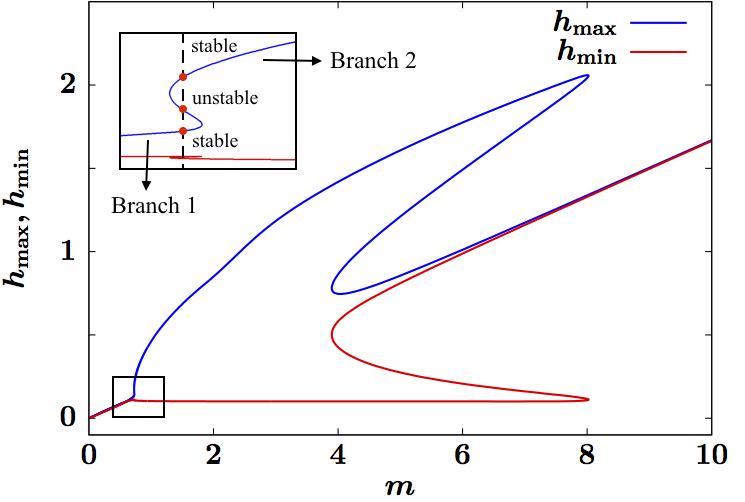}
\caption{}
\end{subfigure}
\caption{Bifurcation diagrams for parameters $L=6,\ s=3, \ A_1=1,\ A_2=1.5,\ \eps=0.1$: (a) bifurcation diagram for $\bar p$ vs. $h_{\max}$ (b) bifurcation diagram for $m$ vs. $h_{\max}, h_{\min}$}
\lbl{fig:stability_2}
\end{figure}
\par
Depending on the choice of parameters, a part of branch~2 near the connection with branch~1 may also be unstable. Figure~\ref{fig:stability_2} shows bifurcation diagrams for $\pbar$ vs. $\hmax$ and $m$ vs. $\hmax, \hmin$. The inset plot in Figure~\ref{fig:stability_2}(a) zooms into the end of branch~2 that connects with branch~1. This corresponds to the inset plot shown in Figure~\ref{fig:stability_2}(b), showing the same solution branches yield an S-shaped curve plotted using $m$ vs. $h_{\max},h_{\min}$, indicating saddle-node bifurcations. There is a small range of mass for which three different steady-states exist with the same mass. Two of the steady-states are branch~2 solutions. Of the two branch 2 solutions, the solution with the smaller mass is unstable. The third steady-state is a branch~1 solution. As will be discussed further, increasing $A_2$ has the effect of stabilizing branch 2 solutions. We found that for $L=6,\ s=3, \eps=0.1$ fixed, as $A_2$ increases, the unstable part of branch~2 vanishes.
\par 
It was shown that steady states on branch~5 are parametrized by a finite range of pressures $\pbar$. Corresponding to a finite range of masses, in Fig.~\ref{fig:mh}(c) this is $1.2<m<1.6$. Branches 4 and 6 are defined over the same range of masses, suggesting mass-conserving bi-stable dynamics of \eqref{het_prob} separated by branch~5.
Figure~\ref{fig:evolve} confirms this description by showing the two different stable equilibria approached by the solution at large times starting from initial data given by  a branch~5 solution with small perturbations of opposite sign.
Figure~\ref{fig:evolve}(a) shows when the initial condition is given by $h(x,0)=\hstab(x)+\delta g_1(x)$, the stable branch~6 solution is approached. In contrast, in Figure~\ref{fig:evolve}(b) starting from the initial condition $h(x,0)=\hstab(x)-\delta g_1(x)$, the dynamics lead to the branch~4 solution with the same mass, $m\approx 12.8$.
\begin{figure}
\begin{subfigure}[b]{0.48\textwidth}
\centering
\includegraphics[width=2.4in]{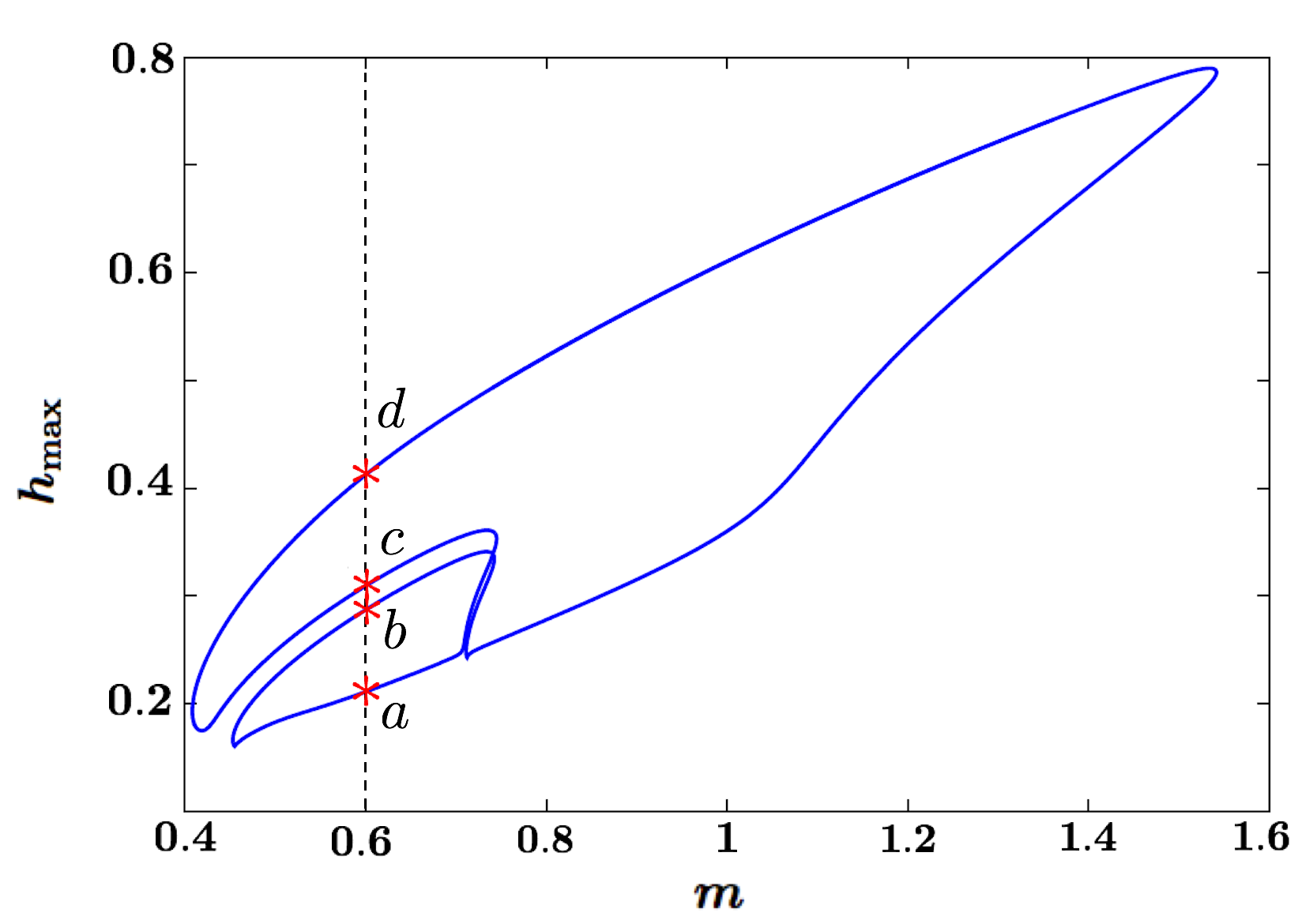}
\caption{}
\end{subfigure}\qquad
\begin{subfigure}[b]{0.48\textwidth}
\centering
\includegraphics[width=2.4in]{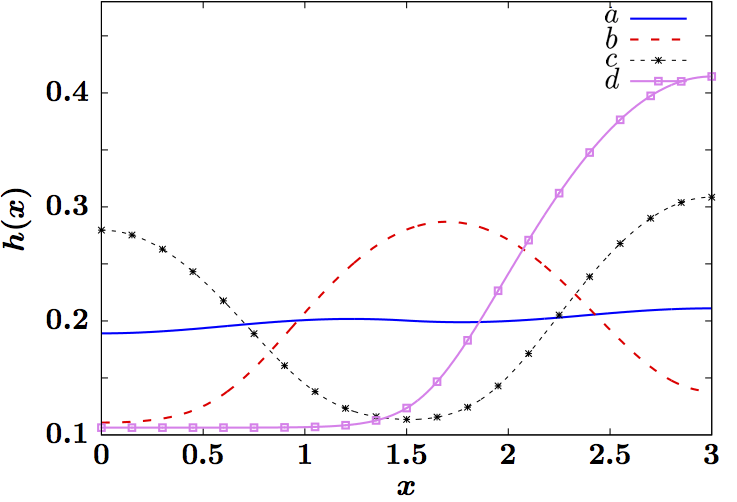}
\caption{}
\end{subfigure}
\caption{(a) Part of the bifurcation diagram from Fig.~\ref{fig:mh}(c) highlighting  four distinct non-primary steady-states with mass $m=0.6$. (b) The corresponding height profiles of the four solutions, of which $a-c$ were shown in Fig.~\ref{fig:mh}(e) and $d$ is a large droplet centered at $x=L$.}
\lbl{fig:big_stab_small}
\end{figure}
\par
In Section 2, we noted the existence of further branches of solutions in the bifurcation diagram,  besides the primary (outer-most) loop (Fig.~\ref{fig:mh2}(b)) that we have been studying, for heterogeneous substrates with sufficiently large domain size $L$. 
We use the values of the system parameters from Fig.~\ref{fig:mh}(c) and consider the stability of solutions off the primary loop.
We compute the eigenvalues for four solutions, all with mass $m=0.6$, marked by asterisks in the bifurcation diagram shown in Figure~\ref{fig:big_stab_small}(a). Figure~\ref{fig:big_stab_small}(b) shows the corresponding profiles of the four solutions. Linear stability analysis suggests that of these four steady-states, only solution $d$, which is an outer loop solution representing a large droplet centered at $x=L$ is stable; solutions $a$-$c$ are unstable. The dominant eigenvalue for solution $d$, $\lambda_1\approx -0.065$, is smaller in amplitude compared to $\lambda_1\approx -0.08$ for the stable droplet centered at $x=0$. This suggests that while both droplets are stable to infinitesimal perturbations, the droplet in the hydrophilic region may be the attracting state for dynamics starting from most generic initial conditions at this mass. Such stability considerations led us to focus on the outer loop of solutions.
\par
In Sections~\ref{sssec:s3} and \ref{ssec:leak}, we quantified the pinning effect of an increasing wettability contrast on branch~2 and 3 droplets. Here, we show that increasing $A_2$ can increase the relative stability of a branch~2 droplet at a  fixed mass. Figure~\ref{fig:eig_A2} shows the largest eigenvalue of a steady-state branch~2 droplet with mass $m=3.5$ as a function of $A_2$. As $A_2$ increases, the leading eigenvalue $\lambda_1$ becomes more negative, making the steady-state more stable with small perturbations decaying faster. We will see further influences of large $A_2$ on the dynamics in the next section.  
\begin{figure}
\centering
\includegraphics[width=2.5in]{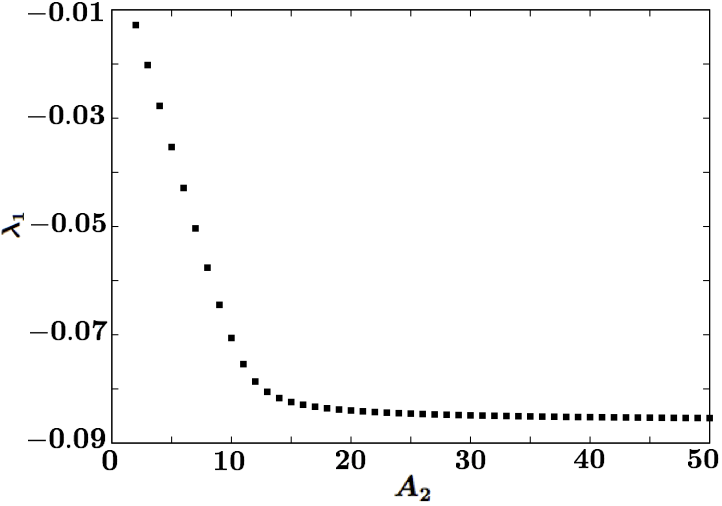}
\caption{The largest eigenvalue $\lambda_1$ of a steady-state branch~2 droplet with fixed mass on a substrate with increasing $A_2$ for parameters $m=3.5$, $L=10$, $s=5$, $A_1=1$, $\eps=0.1$.}
\lbl{fig:eig_A2}
\end{figure}

\begin{figure}[ht!]
\noindent
\mbox{\includegraphics[height=1.6in]{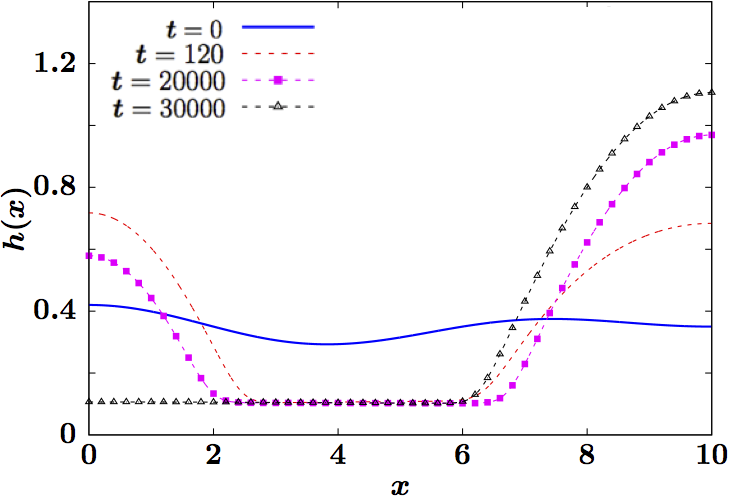}\qquad\includegraphics[height=1.55in]{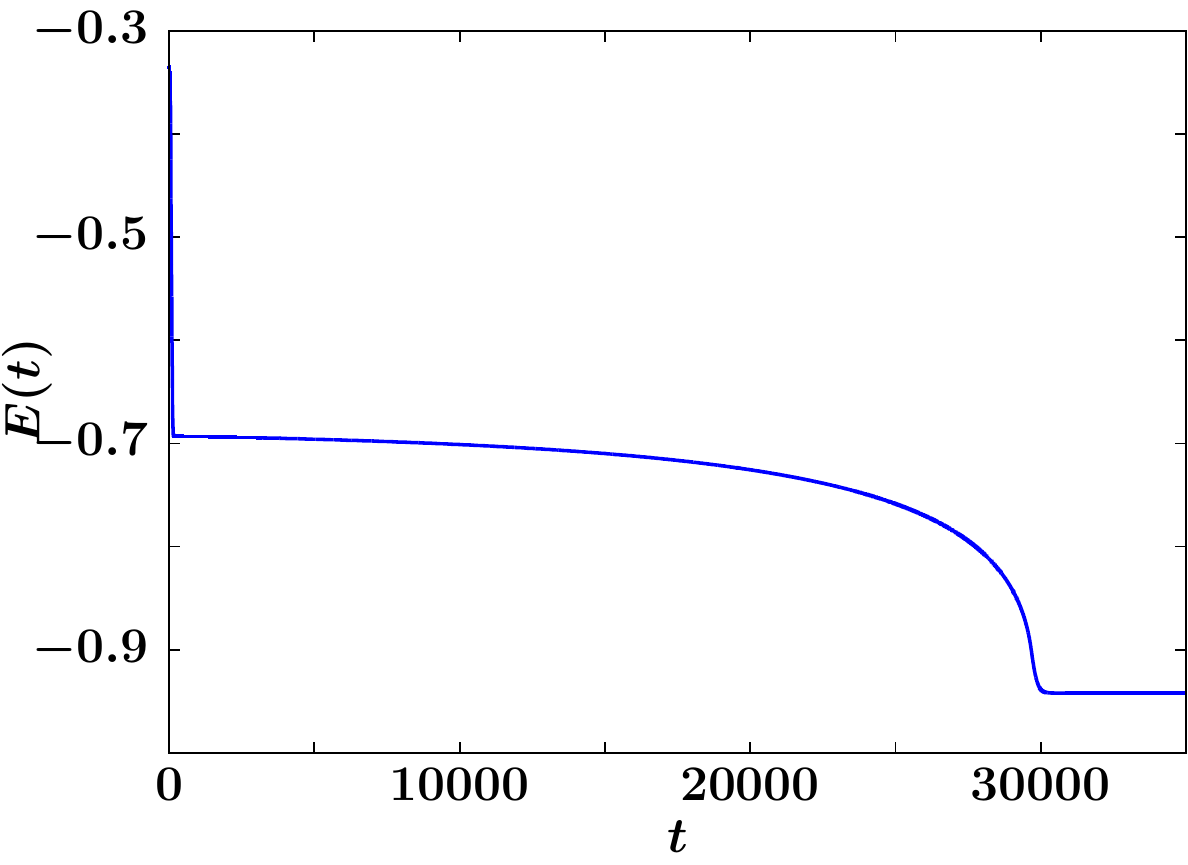}}\\[5pt]
\mbox{\includegraphics[height=1.6in]{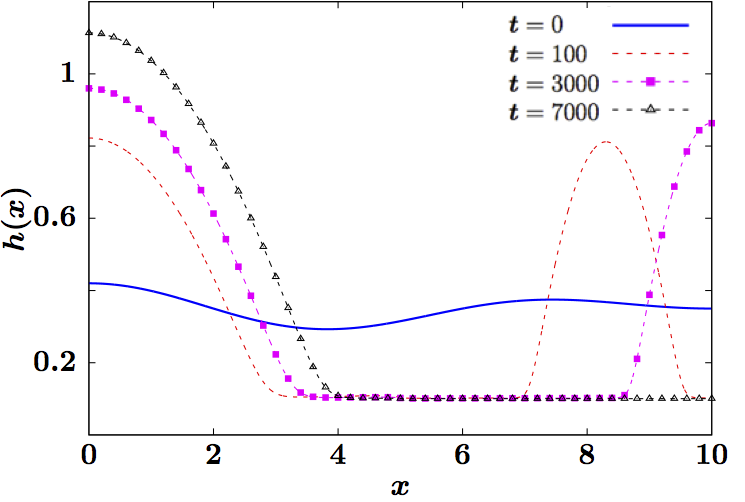}\qquad\includegraphics[height=1.55in]{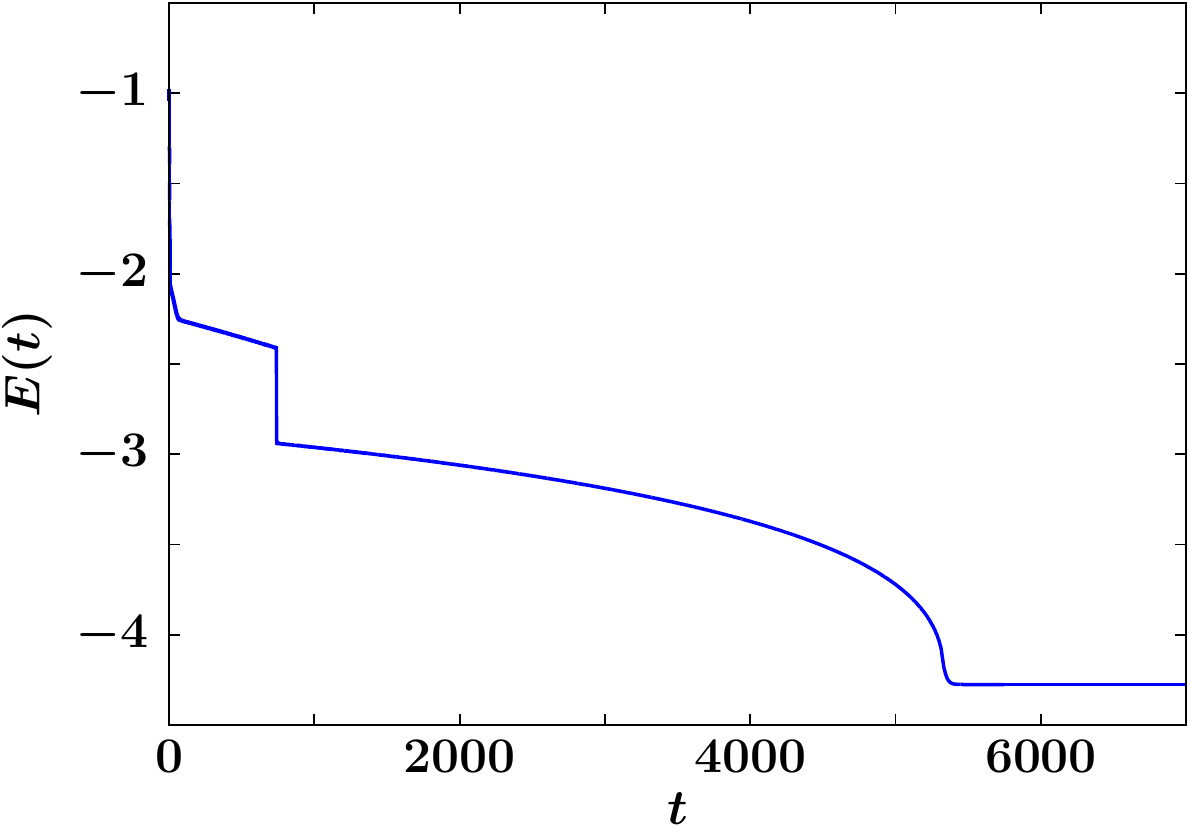}}\\[5pt]
\mbox{\includegraphics[height=1.6in]{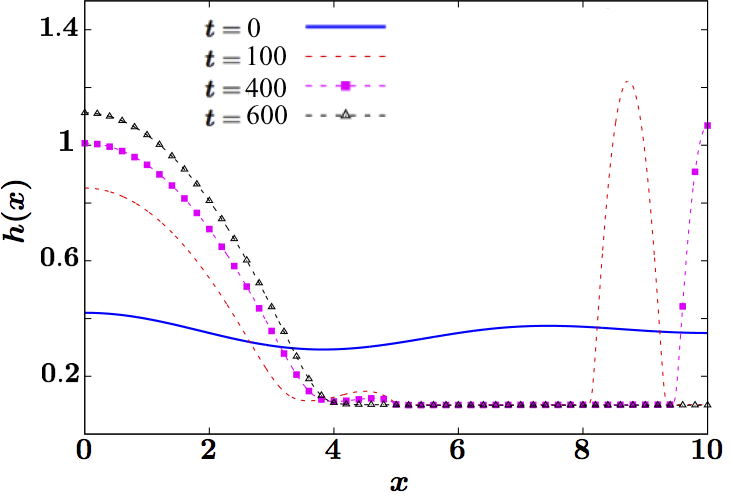}\qquad\includegraphics[height=1.55in]{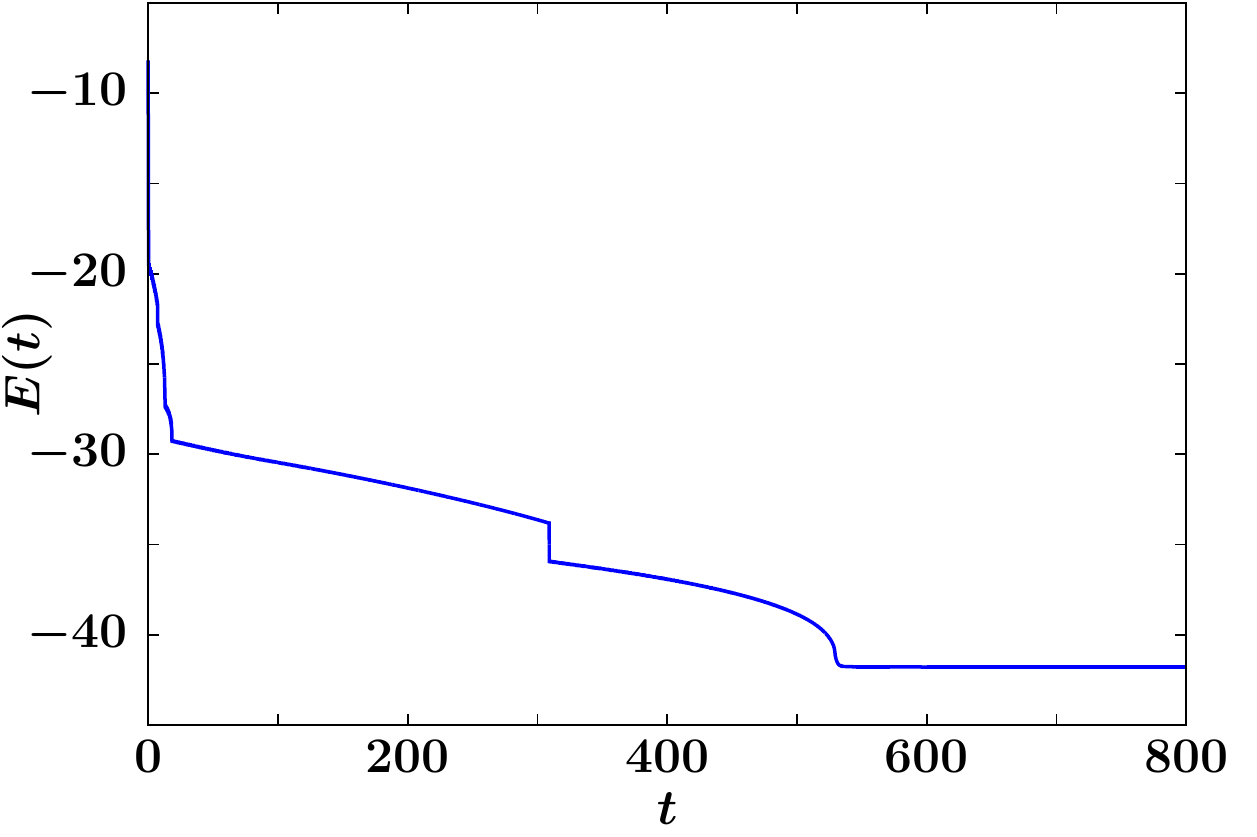}}
\caption{Dewetting dynamics of thin films over time starting from the same initial profile on $0\le x\le 10$ with 
    $\eps=0.1$. Each row shows height profiles $h(x,t)$ at selected times (left) and the evolution of the energy \eqref{eq:E} (right). (Top) Evolution on a homogeneous substrate, $A(x)\equiv 1$, (Middle) evolution on a heterogeneous substrate with $A_2=5$ on $5\le x\le 10$, (Bottom) evolution on a heterogeneous substrate with $A_2=50$ on $5\le x\le 10$. Note the differences in time-scales and the differences in the final states of the solutions.}  
    \lbl{fig:evolve_het1}
 \end{figure}
\section{Dynamics of one-dimensional solutions}\lbl{ssec:dynamics}
\par
The dewetting dynamics of thin films on hydrophobic substrates involves many regimes starting from linear instabilities of perturbed films, leading to pattern formation and long-time break up into droplets connected by thin precursor films, see for example \cite{thiele2001dewetting}. An important step in showing that model \eqref{het_prob} can represent these dynamics for homogeneous substrates ($A(x)\equiv 1$) was the proof in \cite{bertozzi2001dewetting} that film thicknesses remain positive for all times. In the Appendix here we extend their proof to apply to \eqref{het_prob} with heterogeneous wetting given by \eqref{eq:myA}. Given that result, here we briefly address the influence of heterogeneous wetting properties on the time-scales of the dewetting dynamics.
 \par
  Figure~\ref{fig:evolve_het1} compares the evolution of a thin film on substrates with homogeneous and heterogeneous wetting properties, \eqref{eq:dis34} with $\eps=0.1$, on a domain with $L=10$. The initial condition is given by a perturbed thin film  $h(x,0)=0.35[1+0.1\cos\left(\frac{2\pi x}{L}\right)+0.1\cos\left(\frac{3\pi x}{L}\right)]$ with mass $m=3.5$. In each of three simulations, we illustrate the dynamics by showing height profiles at selected times along with plotting the evolution of the energy \eqref{eq:E}.
  \par
On the homogeneous substrate (see Fig.~\ref{fig:evolve_het1}(top)), the thin film de-stabilizes very quickly to form two droplets of different sizes centered at $x=0$ and $x=L$. This is accompanied by a rapid decrease in the energy from the initial
value of $E_0\approx -0.334$. Thereafter, the drops slowly evolve.
The droplet at $x=L$ slowly gains mass as time increases, eventually leading to an equilibrium with one large droplet centered at $x=L$.
\par
  Figure \ref{fig:evolve_het1}(middle) shows the evolution starting from the same initial film on a stepwise-patterned substrate with $s=5$, $A_1=1$ and  $A_2=5$. While the film also breaks up to form two droplets in this case, the right droplet initially develops at an interior position, at some $x<L$. As time increases, the right droplet moves towards $x=L$ and loses mass, eventually leading to one single equilibrium droplet centered at $x=0$. In this evolution, the energy of the thin film has  two stages of rapid decrease, first forming two drops from the film followed by the movement of the interior droplet to the edge of the domain. The two edge droplets then slowly evolve until a single-drop equilibrium is approached, as shown in Fig.~\ref{fig:evolve_het1}(middle). Note that compared to the homogeneous substrate case, the final droplet formed on the other side of the domain, and the timescale to reach this near-equilibrium phase was reduced by a factor of five.
  \par
    Fig.~\ref{fig:evolve_het1}(bottom) shows the evolution of the thin film profile on the patterned substrate with $A_2=50$. The evolution of the thin film goes through a similar dewetting process. However, the droplet formed at the right boundary has a smaller width compared to the $A_2=5$ case and the stages of dynamics occurred roughly ten times faster. This is consistent with the stabilizing effect of increasing $A_2$ evidenced by the eigenvalue calculation shown earlier in Fig.~\ref{fig:eig_A2}. Further work is needed to better understand the significant influence of substrate heterogeneity on the overall dynamics of thin film evolution.

\begin{figure}[h]
\centering
 \includegraphics[width=3in]{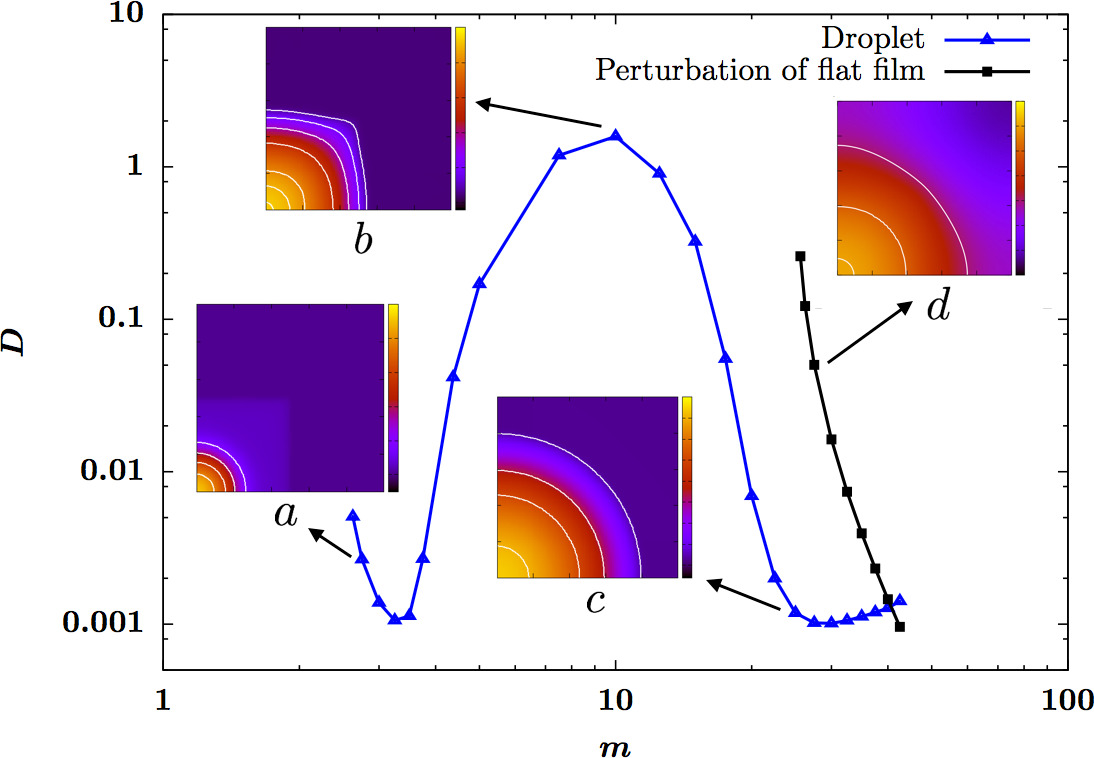}\\[2pt]
  \mbox{
 \begin{subfigure}[b]{0.3\textwidth}
  \includegraphics[width=2in]{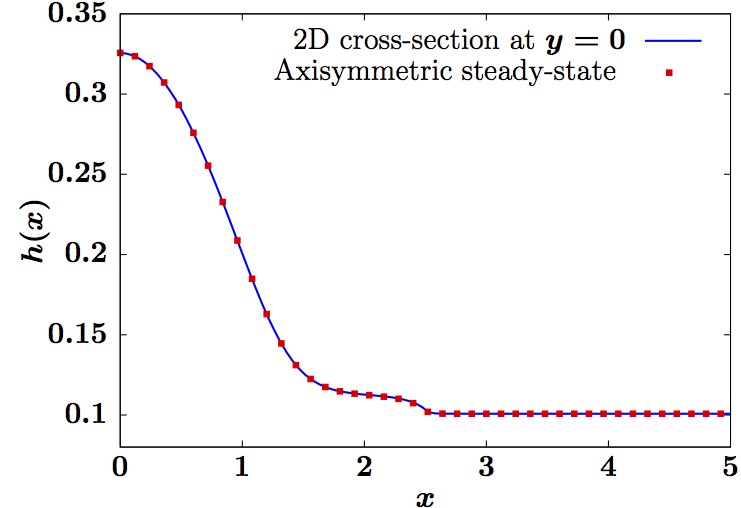}
 \caption{}
\end{subfigure}\qquad
 \begin{subfigure}[b]{0.3\textwidth}
\includegraphics[width=2in]{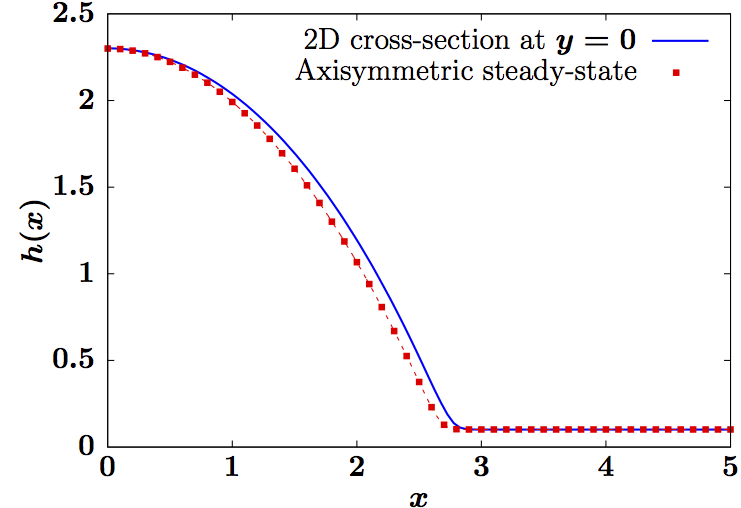}
 \caption{}
\end{subfigure}\qquad
 \begin{subfigure}[b]{0.3\textwidth}
\includegraphics[width=2in]{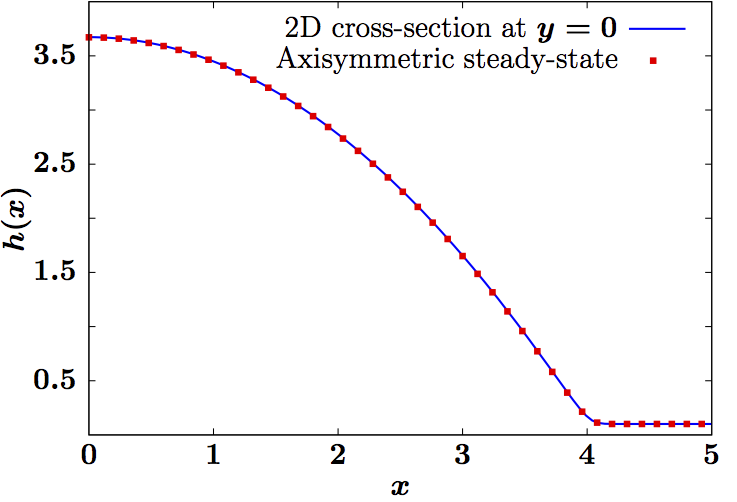}
 \caption{}
\end{subfigure}}
 \caption{Thin films on a hydrophilic square patch, $-s\le x,y \le s$: (top) Film mass $m$ vs. $D$ plotted in log scale for both droplet-type solutions and perturbations of thick flat films, with parameters $L=L_x=L_y=5$, $A_1=1$, $A_2=10$, $s=L/2$. The insets show color contour maps of four selected solutions (on one quarter of the domain, $0\le x,y\le L$, reduced by symmetry).
 (a,b,c) Cross-section of the two-dimensional solution at $y=0$ compared with the axisymmetric steady-state with the same maximum film thickness for (a) droplet $a$ (b) droplet $b$ (c) droplet $c$. No cross-section profile is shown for  the nearly-uniform thick film marked by inset $d$.}
\lbl{fig:mD}
\end{figure}
\section{Steady-state thin films on two-dimensional substrates}
\par
So far, we have mainly focused on solutions for the simple one-dimensional and axisymmetric cases.
However, the chemical patterning of surfaces used in many microfluidic applications is generally much more complicated, see for example \cite{darhuber2001,darhuber2005,kavspar2016confinement}.
Here, we study thin films on two-dimensional heterogeneous surfaces with square and stripe patterning and show that the cross-sections of some two-dimensional steady-state solutions on such surfaces can be approximated by one-dimensional and axisymmetric solutions.
\par
The generalization of \eqref{eq:het_PDE} to two dimensions for the evolution of $h(x,y,t)$ is 
\begin{equation}
{\partial h\over \partial t} = \nabla \cdot\left(h^3 \nabla \left[ A(x,y) \Pi(h) -\nabla^2 h\right]\right),
\lbl{2dpde}
\end{equation}
and steady-states are characterized by having constant pressures, $p=\pbar$, yielding the semilinear elliptic partial differential equation problem 
\begin{equation}
\pbar=A(x,y)\Pi(h)-\nabla^2 h.\lbl{2dsteady}
\end{equation}
We computationally obtain stable steady states by applying efficient numerical schemes for \eqref{2dpde}  (see \cite{Witelski03}) and evolving the solution to sufficiently long times starting from initial conditions over a range of masses.
\par
As in the one-dimensional and axisymmetric cases, we focus on the droplet solutions centered at the origin.
First, we study drops on a heterogeneous substrate with a relatively hydrophilic $A_1$-square patch in the center, surrounded by a hydrophobic $A_2$ region on a square domain with the Hamaker coefficient modeled by
\begin{equation}\lbl{eq:A_rec}
A(x,y)=
\begin{cases}
A_1 & 0\leq x\leq s \ {\rm and}\ 0\leq y\leq s,\\
A_2 & {\rm otherwise}.
\end{cases}
\end{equation}
With this geometry, we can take advantage of four-fold symmetry to get the solutions in terms of computing just the first quadrant. Our expectations are that small droplets, whose core fits well-inside the the $A_1$ square should be close to axisymmetric, as should large drops that overflow the $A_1$ square but are not so large as to be strongly influenced by the confining effects of the finite domain size. These correspond to branches 2 and 4 of the axisymmetric solutions found in Section~\ref{ssec:2D}. Between these cases should be two-dimensional pinned drops whose structure depends significantly on the shape of the hydrophilic region.
\par
 To quantitatively compare the computed solutions on this substrate with the axisymmetric steady-state solutions we define a measure for the difference of $h=h(x,y)$ from being an axisymmetric form, $h=h(r)$, as
\begin{equation}
D\equiv \int_0^{L_y}\int_0^{L_x} \bigg| x\frac{\partial h}{\partial y}-y\frac{\partial h}{\partial x}\bigg|^2 \, dx\,dy
\end{equation}
Note that written in polar coordinates, $D=\big\lVert \frac{\partial h}{\partial \theta}\big \rVert_{L_2}^2$, hence if a solution is axisymmetric, then $D=0$. 
\par
Figure~\ref{fig:mD} shows 
film mass $m=\iint h\,dx\,dy$ vs.\ $D$ plotted on log scale over a range of fluid masses on a square hydrophilic patch \eqref{eq:A_rec} with $A_2/A_1=10$.
Droplet-type solutions, represented by blue triangular data points, correspond to pinned and unpinned droplets similar to those studied in Sections~\ref{sssec:s2} and \ref{sssec:s3}. We observe that the maximum $D$ occurs at a pinned steady-state droplet with droplet width $w\approx s$. The contour map of the surface of the solution labeled $b$ is also shown in Figure~\ref{fig:mD}. For solutions with mass larger than droplet $b$, the droplet becomes a large-radii unpinned droplet like a branch~4 solution, shown by the contour map labeled $c$. In this process, $D$ gradually decreases. For masses smaller than droplet $b$, droplets gradually transition to being small-radii droplets like a branch~2 droplet with a smaller $D$, shown by the contour map of droplet $a$. Fig.~\ref{fig:mD}(a,b,c) confirm the excellent agreement of the computed solution with the axisymmetric height profiles for cases $a,c$ and the noticeable difference with the anisotropic pinned droplet $b$. Above a certain mass, the wettability contrast is not strong enough to maintain droplets and the solution will take the form of a nearly-uniform thick film. A branch of these solutions is also shown in the figure (indicated with black dots); as should be expected from earlier results for branch~6 solutions, the influence of the form of $A(x,y)$ decreases with increasing thickness.
\par
Processes in many applications involve depositing liquids on periodic striped wettability patterns, see \cite{ajaev2016stability,brasjen2013dewetting,honisch2015instabilities,kargupta2002morphological}. In particular, several different regimes for liquid droplets on substrates with stripe-like patterns have been identified in \cite{honisch2015instabilities}. Here, we show that depending on the regime,  cross-sectional profiles of the two-dimensional droplet can be predicted using the axisymmetric or one-dimensional steady-states. To simulate the deposition of liquids on a substrate with stripe-like patterns, we consider $A(x,y)$ of the form
\begin{equation}\lbl{eq:A_stripe}
A(x,y)=
\begin{cases}
A_1 & 0\leq x\leq s,\\
A_2 & \mbox{otherwise.}
\end{cases}
\end{equation}
We focus on one-quarter of a droplet whose maximum film thickness occurs at $(0,0)$, in the center of the stripe. Figure~\ref{fig:2d_stripe} shows $m$ vs.\ $D$ plotted on log scale for droplets on striped substrates. When the fluid mass is small, with the droplet core fitting well inside the width of the stripe, the influence of the chemical heterogeneity on the droplets is limited. The droplets are closer to axisymmetric solutions with small $D$, as shown by the color map of the surface of droplet $a$ and droplet $b$ highlighted in Figure~\ref{fig:2d_stripe}. As mass increases, the fluid grows in the $y$-direction and becomes increasingly non-axisymmetric, as shown by droplet $c$ labeled in Figure~\ref{fig:2d_stripe}.Figure~\ref{fig:2d_stripe}(a)-(c) show the cross-section of the two-dimensional computed solution at $y=0$ compared with the axisymmetric or one-dimensional steady-states with the same maximum film thickness for droplets $a$-$c$. We observe that the cross-section of droplet $a$ and droplet $b$ can be well approximated by the axisymmetric solution with the same maximum film thickness. As the fluid mass increases, $D$ increases. The one-dimensional steady-state gives a better prediction of the cross-sectional profile at $y=0$ for the elongated pinned droplet $c$.
\begin{figure}[ht!]
\centering
 \includegraphics[width=3in]{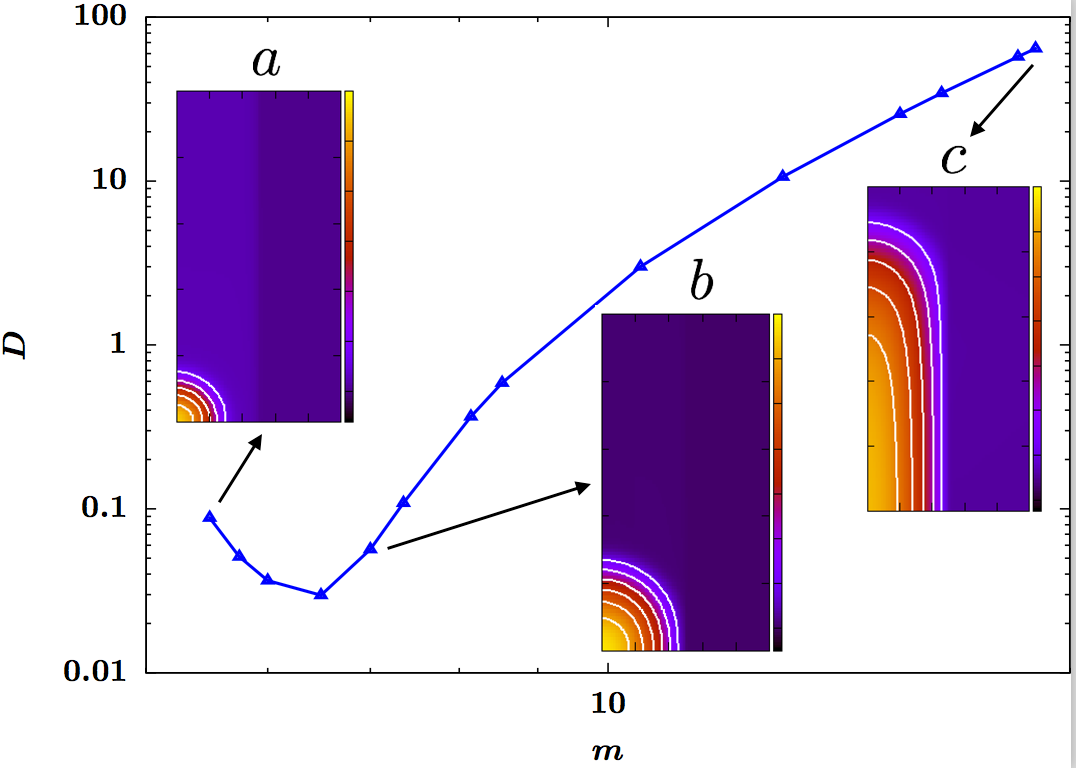}
 \mbox{
 \begin{subfigure}[b]{0.3\textwidth}
 \includegraphics[width=2in]{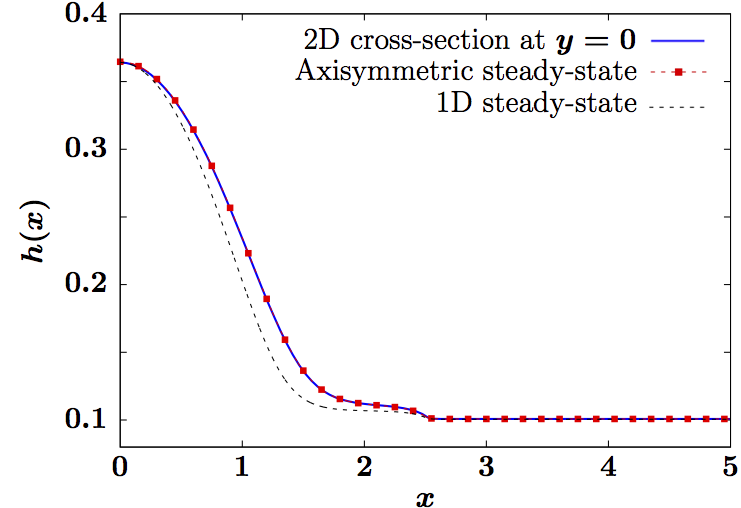}
 \caption{}
\end{subfigure}\qquad
 \begin{subfigure}[b]{0.3\textwidth}
\includegraphics[width=2in]{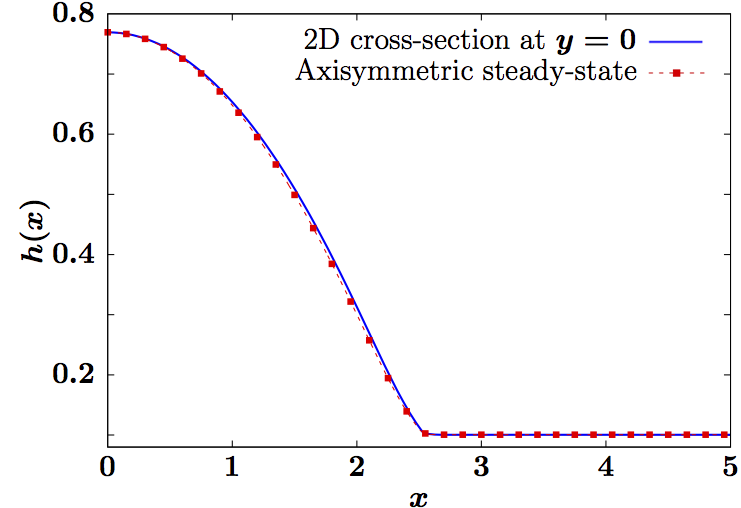}
 \caption{}
\end{subfigure}\qquad
 \begin{subfigure}[b]{0.3\textwidth}
\includegraphics[width=2in]{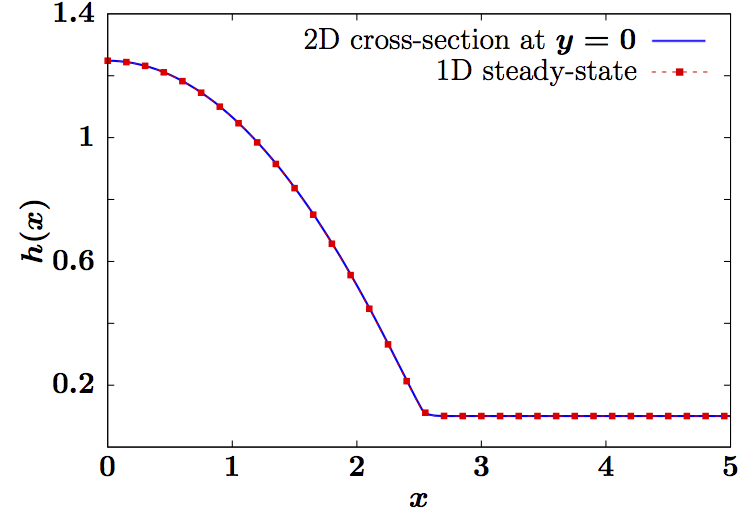}
 \caption{}
\end{subfigure}}
 \caption{Thin films on a hydrophilic stripe, $-s\le x\le x$: (top) Film mass $m$ vs.\ $D$ plotted in log scale for both droplet-type solutions and perturbations of thick flat films, with parameters $L_x=5$, $L_y=10$, $A_1=1$, $A_2=10$, $s=L_x/2$. The insets show color contour maps of three selected solutions (on one quarter of the domain, reduced by symmetry).
 (a,b,c) Cross-section of the two-dimensional solutions at $y=0$ compared with the axisymmetric and one-dimensional steady-states with the same maximum film thickness for droplets $a,b,c$ respectively.}
\lbl{fig:2d_stripe}
\end{figure}

\section{Conclusions}
This paper has considered the steady-state thin films on a finite chemically heterogeneous substrate with stepwise patterning. We have classified the primary steady-state solutions in one dimension into six different branches, for which we presented asymptotic analysis of solutions and have considered two limits, the small $\epsilon$ limit and the large wettability contrast limit. In particular, we investigated two new types of pinned droplet solutions that arise completely due to the heterogeneity of the substrate. We identified that an increasing $A_2$ has a confining effect on these two pinned droplets. Through asymptotic analysis, we quantified the degree of confinement and leakage of fluid film in terms of the wettability contrast.
\par
We showed that the results of the asymptotic analysis derived for one-dimensional solutions can be directly extended to axisymmetric solutions. In addition, we discussed the stability of these steady-state solutions using linear stability analysis. We also extended a proof of positivity of solutions on homogeneous substrates to the case of heterogeneous substrates. Last, we explored the effect of heterogeneity on the dynamics of thin film evolution in one-dimension and in square and striped geometries in two dimensions.  
\par
There are many interesting questions for further study suggested by this work including understanding the structure of the higher-order branches in one-dimension and approaches for systematically simplifying solutions of the two-dimensional elliptic problem \eqref{2dsteady} in simple geometries like those studied in  \cite{brasjen2013dewetting}. Further work is needed to compare our results for branch~3 pinned drops with the results for pinned drops on square patches given in \cite{kavspar2016confinement}. Much more work is also needed to better understand the influence of heterogeneous wetting on dewetting and coarsening dynamics on larger domains as in  \cite{brusch2002dewetting,thiele2003modelling,asgari2012coarsening}. 

\newpage



\bibliographystyle{imamat}
\bibliography{Author_tex}
%


\appendix
\section{Proof of positivity and global existence of solutions}\lbl{app_proof}
\par
Here we extend the proof given in 
\cite{bertozzi2001dewetting} showing the global existence of positive solutions to \eqref{eq:het_PDE} from the homogeneous case ($A(x)\equiv 1$) to apply to heterogeneous substrates with positive $A(x)$ bounded from above.
\begin{theorem}
{\rm Consider initial data for \eqref{het_prob} satisfying $h_0(x)>0$ with $h_0\in H^1([0,L])$ and $E(h_0)<\infty$, then the solution $h(x,t)$ is positive for all $t>0$.}
\end{theorem}

\begin{proof}
We derive {\it a priori} pointwise upper and lower bounds for the solution. The energy $E$, as given by \eqref{eq:E} is monotonically decreasing following \eqref{eq:dEdt}. It follows that at any time $T>0$,
\begin{equation}\lbl{eq:bound}
\displaystyle\frac{1}{2}\int_0^L\bigg| \frac{\partial h}{\partial x}(T)\bigg|^2\,dx\leq \frac{1}{2}\int_0^L\bigg| \frac{\partial h_0}{\partial x}\bigg|^2\,dx+\int_0^LA(x)U(h_0)\,dx-\int_0^LA(x)U(h(T))\,dx
\end{equation}
Using that $A(x)$ is bounded and $-U(h)$ has an {\it a priori} upper bound independent of $h$ (from \eqref{Ueqn}, $U(h)\ge 1/6$ for all $h>0$), implies that $\int |\partial_xh(x,T)|^2\,dx$ is bounded. Hence, $h(x,T)\in H^1([0,L])$. Then $h(x,T)$ has both  {\it a priori} pointwise and $C^{0,1/2}$ upper bounds by the Sobolev embedding theorem.
\par
Note that \eqref{eq:bound} along with the boundedness of $A(x)$ implies $\int_0^LU(h(x,T))\,dx<C$. Suppose $h(x,T)$ attains its minimum $h_{\min}$ at $x=x_0$. By Holder continuity, $h(x)\leq h_{\min}+C_h|x-x_0|^{1/2}$. Therefore,
\begin{equation}
C>\int_0^L U(h(x,T))\,dx\geq\int_0^L\left(\frac{\epsilon^3}{3(h_{\min}+C_h|x-x_0|^{1/2})^3}-\frac{\epsilon^2}{2h_{\min}^2}\right)\,dx
\geq \frac{C_2(\epsilon, L)}{h_{\min}}+O(1) 
\end{equation}
Hence, the solution cannot go below a positive threshold for any $T>0$.
\end{proof}

\end{document}